\documentclass[twocolumn,superscriptaddress,amsmath,amssymb,floatfix,aps,notitlepage,nokeys,prl,longbibliography]{revtex4-1}
\usepackage{mathrsfs, hyperref}
\usepackage{amssymb, amsbsy, amsmath, latexsym, dsfont, array, layout, graphics,mathrsfs,braket,amsfonts,amsthm,
  amssymb,graphicx,subfigure, youngtab,color,bm,mathtools,braket,verbatim,url,cleveref,natbib,hypernat,extarrows,inputenc,multirow,bbm,dsfont}%,multicol
\usepackage{graphicx,placeins}% Include figure files and provide \FloatBarrier
\usepackage{dcolumn}% Align table columns on decimal point
\usepackage{bm}% bold math
\usepackage[dvipsnames]{xcolor}
\usepackage[normalem]{ulem}
\usepackage[T1]{fontenc}
\hypersetup{
     colorlinks = true,
     linkcolor = magenta,
     citecolor = magenta
}
\graphicspath{ {./images/} }

\allowdisplaybreaks

\begin{document}\let\supplementaddcontentsline\addcontentsline\renewcommand{\addcontentsline}[3]{}

\title{Similarity of fractional Chern insulators to fractional quantum Hall
    states: Moir\'e-enhanced gaps and excitation-spectrum correspondence}

\author{Siddhartha Sarkar}\thanks{These two authors contributed equally}
\affiliation{Max Planck Institute for the Physics of Complex Systems, N\"othnitzer Stra\ss e 38, 01187 Dresden, Germany}

\author{Yitong Zhang}\thanks{These two authors contributed equally}
\affiliation{Department of Physics, University of Michigan, Ann Arbor, MI 48109, USA}

\author{Kai Sun}
\email{sunkai@umich.edu}
\affiliation{Department of Physics, University of Michigan, Ann Arbor, MI 48109, USA}

\begin{abstract}
Fractional Chern insulators (FCIs) realize fractional quantum Hall topology in lattice bands, but their excitation spectra remain far less understood than their ground states. Here we establish a theoretical principle relating the periodic electron-density modulations of flat Chern bands to the many-body gap and excitation spectrum of FCIs. Contrary to the conventional view that such density modulations are detrimental to fractional topology, we show that different reciprocal-lattice Fourier components play sharply distinct roles: components at smaller reciprocal lattice vectors suppress the FCI gap, whereas components at larger reciprocal lattice vectors enhance it. By suppressing the harmful small-wave-vector components and amplifying the beneficial large-wave-vector components, the gap enhancement can, in principle, be made arbitrarily large within the projected flat-band theory. Moreover, the same enhancement factor rescales the full low-energy spectrum, making the FCI excitation spectrum predictable from the corresponding Landau-level problem. We further generalize this correspondence to non-Abelian states. Applying this principle to moir\'e Chern bands, we identify these reciprocal-lattice density components as practical diagnostics for robust FCIs.
\end{abstract}
\maketitle

\noindent\textit{Introduction.}---Fractional topological phases are characterized not only by quantized response but also by the exotic excitations above their ground states. In the fractional quantum Hall (FQH) effect~\cite{stormer1999fractional,laughlin1999nobel}, fractionally charged quasiparticles, quasiholes, and collective modes form a well-understood excitation spectrum that controls the stability of the phase and the energetics of anyon manipulation. Fractional Chern insulators (FCIs) provide a lattice route to analogous fractional topological order without external magnetic fields, and have emerged as a central platform for fractionalized quantum matter~\cite{tang2011high,sun2011nearly,neupert2011fractional,sheng2011fractional,Regnault2011fractional,xiao2011interface,bernevig2012emergent,wu2012fractional,parameswaran2013fractional,roy2014band,repellin2014single,wu2015fractional,repellin2020chern,ledwith2020fractional,simon2020contrasting,liu2021gate,mera2021engineering,li2021spontaneous,Devakul2021Magic,ledwith2021strong,wang2021exact,xie2021fractional,cai2023signatures,zeng2023thermodynamic,park2023observation,xu2023observation,ledwith2023vortexability,wu2024quantum,lu2024fractional,xie2025tunable}. Yet while FCIs can realize the same ground-state topology as FQH states, a fundamental question remains open: what determines their excitation spectrum?

This question is intrinsically challenging because the celebrated correspondence between FCIs and FQH states is a ground-state correspondence: an FCI ground state can be adiabatically connected to its FQH counterpart without closing the many-body gap~\cite{wu2012adiabatic}. However, this correspondence is not expected to extend generally to excited states. In contrast to Landau levels, where quasiparticles, quasiholes, and collective modes can be systematically characterized, FCI excitation spectra remain far less understood and are often accessible only numerically. Numerical studies indicate that quasiparticles and quasiholes in FCIs can acquire strong residual interactions, producing softened charge-neutral excitations, nontrivial quasiparticle and quasihole dispersions, and spectra that differ strongly from their FQH counterparts~\cite{wu2012adiabatic,repellin2014single,lu2024interaction,schleith2025anyon,yan2025anyon,iyer2026dispersion}. These effects are not merely microscopic details: the excitation gap controls the robustness of the fractional phase, while the anyon dispersion and full excitation spectrum control the energetics of anyon manipulation and may play a key role in exotic descendant phases such as anyon superconductivity.

\begin{figure*}[t]
    \centering
    \includegraphics[width=\textwidth]{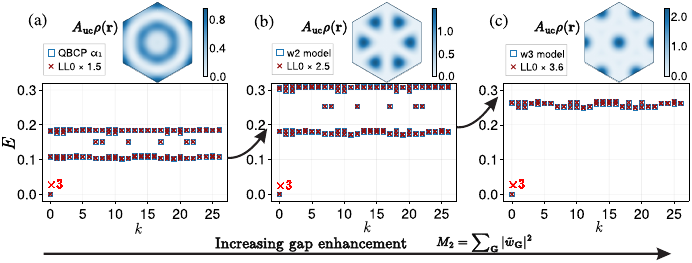}
\caption{{\bf Charge neutral gap enhancement of $\nu=1/3$ FCI in ideal bands over LLL in short-range interaction limit $d_s/\ell_B = 0.01$.} Panels (a)-(c) correspond to models with increasing lattice modulation strength $M_2=\sum_{\mathbf{G}}|\tilde{w}_{\mathbf{G}}|^2$, as indicated. The vertical axes in the energy spectra are normalized by $V_\text{int}(d_s/\ell_B)^3$, where $V_\text{int} = e^2/(\epsilon \ell_B)$. Blue boxes denote the spectra of fractional Chern insulators, while red crosses denote the spectra of the corresponding fractional quantum Hall states rescaled by factors of $1.5$, $2.5$, and $3.6$, respectively. Remarkably, after this simple rescaling, the FCI and FQH spectra collapse onto the same spectral structure, despite being realized in very different microscopic systems. The insets show the corresponding electron density $\rho(\mathbf{r})$ of the FCI ground state within a moir\'e unit cell. In direct contrast to FQH states, the FCI ground states are spatially inhomogeneous. The density modulation becomes stronger with increasing $M_2$, whose value precisely matches the scaling factor relating the FCI and FQH spectra. Details of the ideal bands considered here are given in Appendix A. Exact diagonalization calculations were performed for a system with $N_k=27$ unit cells and $C_6$ symmetry (SM~\cite{SM2026}).
}
\label{fig:1}
\end{figure*}

A central origin of the difference between FCIs and FQH states is the lattice itself. Unlike a continuum Landau level, a Chern band lives in a periodic environment, so its electronic density can acquire periodic modulations. These modulations can be characterized by the reciprocal-lattice Fourier components of electron density:
$w_{\mathbf G}=
\frac{1}{A_{\rm uc}}
\int_{\rm uc} d^2{\mathbf r}\,
\rho(\mathbf r)e^{-i\mathbf G\cdot \mathbf r}$ of the filled band. Here $\mathbf G$ is a reciprocal lattice vector, $A_{\rm uc}$ is the unit-cell area, and $w_{\mathbf G}$ measures the amplitude of the density modulation at the Bragg wave vector $\mathbf G$. Nonzero harmonics with $\mathbf G\neq 0$ are commonly viewed as detrimental to fractional topology: they generate Umklapp scatterings, weaken the Landau-level analogy, reduce the many-body gap, and can even destabilize the FCI state~\cite{wu2012fractional,parameswaran2013fractional,roy2014band,wu2015fractional,ledwith2020fractional,ledwith2021strong,wang2021exact,mera2021engineering,ledwith2023vortexability,kousa2025theory,shi2025effects,liu2025broken,moitra2025instability}. This conventional view raises a natural question: are such electron density modulations ($w_{\mathbf G\neq 0}$) necessarily enemies of fractional topological order?

Here we show that the answer is no. Using the framework of vortexable and higher-vortexable bands, we establish a theoretical principle connecting density modulation $w_{\mathbf{G}}$ to the many-body gap and excitation spectrum of FCIs. Contrary to conventional intuition, we find that different Fourier components $w_{\mathbf{G}}$ play sharply distinct roles: the lowest harmonics (first ${\mathbf{G}}$ shell) induce strong Berry-curvature fluctuations in momentum space and suppress the FCI gap, whereas higher harmonics (large ${\mathbf{G}}$) enhance the gap without introducing such fluctuations. In the regime where the harmful low harmonics are small, we prove that for short-range repulsive interactions, FCI Laughlin states acquire an analytic gap-enhancement factor
\begin{align}
M_2=\sum_{\mathbf G}|\tilde w_{\mathbf G}|^2=1+\sum_{\mathbf G\ne0}| w_{\mathbf G}/w_{\mathbf{0}}|^2.
\end{align}
Under otherwise identical conditions, the FCI gap is enhanced according to $\Delta_{\rm FCI}=M_2\Delta_{\rm FQH}$, where $\Delta_{\rm FQH}$ is the gap of the corresponding FQH state in a Landau level. Because the higher-harmonic weight is non-negative and has no intrinsic upper bound within the projected flat-band theory, the resulting gap enhancement is, in principle, unbounded. For longer-range interactions, the enhancement factor is reduced, while the same qualitative result still holds.

Remarkably, this enhancement applies not only to the gap but also to the full low-energy spectrum. By constructing topological flat bands with suppressed low Bragg harmonics, we find that the FCI spectrum is nearly identical to the corresponding FQH spectrum up to the same overall enhancement factor $M_2$. Thus, the excitation spectrum of these FCIs can be predicted from the Landau-level problem, while the lattice provides a mechanism to enhance all excitation energies, which is another key result of this study. Finally, using standard continuum models for twisted bilayer graphene and twisted bilayer MoTe$_2$ (tMoTe$_2$), we show that the harmonic structure of Chern-band wavefunctions provides a practical diagnostic for identifying and engineering robust FCIs in realistic moir\'e materials.

\noindent\textit{Vortexable/ideal bands.}--We begin with a vortexable Chern band with $|C|=1$, whose single-particle wavefunction can always be written as~\cite{wang2021exact}
\begin{equation}\label{eq:vortexablebandWF}
    \psi_\mathbf{k}(\mathbf{r}) = \mathcal{N}_\mathbf{k}\psi^\text{LLL}_\mathbf{k}(\mathbf{r})h(\mathbf{r}),
\end{equation}
where $\mathcal{N}_\mathbf{k}$ is the normalization factor, $\psi^\text{LLL}_\mathbf{k}(\mathbf{r})$ is the (normalized) LLL wavefunction in a homogeneous magnetic field $B_0$, obeying magnetic
Bloch periodicity: $\psi^\text{LLL}_\mathbf{k}(\mathbf{r}+\mathbf{a}_i) = e^{i\mathbf{a}_i\times \mathbf{r}/2\ell_B^2}e^{i\mathbf{k}\cdot\mathbf{a}_i}\psi^\text{LLL}_\mathbf{k}(\mathbf{r})$, with magnetic length $\ell_B=\sqrt{\hbar/eB_0}$ and lattice vectors $\mathbf{a}_i$ such that every unit cell contains one flux quantum $\Phi_0 = 2\pi\hbar/e$ or, equivalently, the unit cell area satisfies $A_\text{uc}=\mathbf{a}_1\times \mathbf{a}_2 = 2\pi\ell_B^2$. The function $h(\mathbf{r})$ is a $\mathbf{k}$-independent (quasi-)periodic (moir\'e) lattice modulation~\cite{aharonov1979ground,guerci2025layer}. 
Bands described by these wavefunctions are termed ``ideal'' as they satisfy the ideal trace condition $\mathrm{tr}(g(\mathbf{k})) = |F_{xy}(\mathbf{k})|$ relating the quantum metric $g(\mathbf{k})$ and Berry curvature $F_{xy}(\mathbf{k})$. These bands are termed ``vortexable'' because multiplication of the wavefunctions $\psi_\mathbf{k}(\mathbf{r})$ by an arbitrary holomorphic function $f(z)$, with $z=x+iy$, yields a state $f(z)\psi_\mathbf{k}(\mathbf{r})$ that remains fully within the same band ~\footnote{Note that a generalization of Eq.~\eqref{eq:vortexablebandWF} where a unit cell contains multiple flux quanta is also possible~\cite{sarkar2023symmetry,sarkar2025unconventional}, leading to multiple flat bands and unconventional fractional phases}. 
Since $|h(\mathbf{r})|^2$ is (moir\'e) lattice periodic, expanding it in Fourier series $|h(\mathbf{r})|^2= \sum_\mathbf{G}w_\mathbf{G}e^{i\mathbf{G}\cdot\mathbf{r}}$ (where $\mathbf{G}$ are the reciprocal lattice vectors), the extent of lattice modulation can be measured by the $n$-th order (normalized) raw moment of $|h(\mathbf{r})|^2$:  $M_n=\langle|h(\mathbf{r})|^{2n}\rangle_\text{uc}/\langle|h(\mathbf{r})|^2\rangle^n_\text{uc} = \langle|\tilde{h}(\mathbf{r})|^{2n}\rangle_\text{uc}=\sum_{\mathbf{G}_1,\dots,\mathbf{G}_n}\delta_{\sum_{i=1}^n \mathbf{G}_i,\mathbf{0}}\prod_{i=1}^n\tilde{w}_{\mathbf{G}_i}$, where $\langle.\rangle_\text{uc}$ stands for average over unit cell, $|\tilde{h}(\mathbf{r})|^2 = |h(\mathbf{r})|^2/\langle|h(\mathbf{r})|^2\rangle_\text{uc}$, and $\tilde{w}_\mathbf{G}\equiv w_\mathbf{G}/w_\mathbf{0}$. For a LLL, $M_n =1$ for all positive integer $n$; whereas any lattice modulation makes $M_n>1$. Note also that the electron density of a filled ideal band is $\rho(\mathbf{r}) \approx A_\text{uc}^{-1}|\tilde{h}(\mathbf{r})|^2$ (see Supplemental Material (SM)~\cite{SM2026}), and hence $\tilde{w}_\mathbf{G}$ are approximately Fourier components of $\rho(\mathbf{r})$.

\noindent\textit{Enhancement of the many-body gap of $\nu=1/3$ FCI with lattice modulation.}--To understand how the moir\'e modulation affects FCI stability, we consider the projected repulsive interaction:
\begin{equation}
\begin{split}
H_\text{int} = \frac{1}{2A} \sum_{\mathbf{k}_1\mathbf{k}_2\mathbf{k}_3\mathbf{k}_4}^\text{BZ}& c^\dagger_{\mathbf{k}_1}c^\dagger_{\mathbf{k}_3}c_{\mathbf{k}_4}c_{\mathbf{k}_2}\times\\
&\sum_\mathbf{q}V(\mathbf{q})\,\lambda_\mathbf{q}(\mathbf{k}_1,\mathbf{k}_2)\lambda_{-\mathbf{q}}(\mathbf{k}_3,\mathbf{k}_4)
\end{split}
\end{equation}
where $A=N_k A_\text{uc}$ is the system area containing $N_k$ unit cells,
$V(\mathbf{q}) = 2\pi e^2 \tanh(d_s |\mathbf{q}|)/(\epsilon |\mathbf{q}|)$ 
is the screened Coulomb interaction, $d_s$ is the separation between the screening electrodes, and $\lambda_\mathbf{q}(\mathbf{k},\mathbf{k}') = \langle\psi_\mathbf{k}|e^{-i\mathbf{q}\cdot\mathbf{r}}|\psi_{\mathbf{k}'}\rangle$ 
is the form factor through which the particular properties of the band enter into $H_\text{int}$. We choose various $|C|=1$ ideal bands with a varying lattice modulation parameter $1.5\lesssim M_2\lesssim3.6$ but a small lowest harmonic $\tilde{w}_\mathbf{G}$;  they all exhibit highly uniform Berry curvature distribution (standard deviation below 4\% as shown in Appendix A). We plot the charge neutral spectrum at filling fraction $\nu=1/3$ in such bands obtained via numerical exact diagonalization (ED) in Fig.~\ref{fig:1} for short-range interaction $d_s\ll \ell_B$. {\it We find that the spectrum remains essentially identical to that of FQH in LLL, but all energy scales, and consequently the charge neutral gap, increase approximately linearly with $M_2$. We also find the same trend in the anyon excitation spectrum as shown in SM~\cite{SM2026}. However, we find that the electron density of the FCI ground state is $A_\text{u.c.}\rho(\mathbf{r}) \approx \nu |\tilde{h}(\mathbf{r})|^2$, and hence becomes increasingly inhomogeneous as $M_2$ increases (Fig.~\ref{fig:1}).} We emphasize that this effect is beyond the quantum geometric properties since all these bands are ideal and have uniform Berry curvature. Furthermore, in SM~\cite{SM2026}, we show that for all bands considered in Fig.~\ref{fig:1} the many-body spectra still look identical to LLL as $d_s$ increases, but the gap enhancement factor $\Delta/\Delta_\text{LLL}$, defined as the ratio of charge neutral gaps corresponding to ideal band ($\Delta$) and  LLL ($\Delta_\text{LLL}$) evaluated at the same $d_s$, reduces with increasing $d_s$ from $M_2$ at $d_s\rightarrow 0$ to $\sim1$ for $d_s\gtrsim \ell_B$. We see a similar trend for $\nu=1/5$ FCI as well (SM~\cite{SM2026}).

\noindent\textit{$M_2$ renormalizes pseudopotential and enhances the many-body gap.}--To understand the mechanism responsible for the enhancement, we analyze how the extra $h(\mathbf{r})$ in the single-particle wavefunction modifies the form factor $\lambda_{\mathbf{q}}(\mathbf{k},\mathbf{k}') = \int d^2\mathbf{r} e^{-i\mathbf{q}\cdot\mathbf{r}}\psi_\mathbf{k}^*(\mathbf{r})\psi_{\mathbf{k}'}(\mathbf{r})$ away from LLL limit. 
Upon insertion of Eq.~\eqref{eq:vortexablebandWF}, the form factor reads
\begin{equation}\label{eq:idealbandLLLFFrelation}
    \lambda_{\mathbf{q}}(\mathbf{k},\mathbf{k}')=\mathcal{N}_\mathbf{k}\mathcal{N}_{\mathbf{k}'} \sum_{\mathbf{G}}w_{\mathbf{G}} \lambda^\text{LLL}_{\mathbf{q}-\mathbf{G}}(\mathbf{k},\mathbf{k}').
\end{equation}
The normalization factor $\mathcal{N}_\mathbf{k}^{-2} = \sum_{\mathbf{G}}\eta_{\mathbf{G}}w_{\mathbf{G}} e^{(i\mathbf{G}\times\mathbf{k}-\frac{1}{4}|\mathbf{G}|^2)\ell_B^2}$ ($\eta_\mathbf{G}$ is $+1$ if $\mathbf{G}/2$ is a reciprocal lattice vector and $-1$ otherwise) encodes the quantum geometry: $F_{xy}(\mathbf{k}) = \ell_B^2-\nabla_\mathbf{k}^2 \log\mathcal{N}_\mathbf{k}$~\cite{wang2021exact}. Importantly, since the summand in $\mathcal{N}_\mathbf{k}^{-2}$ has the Gaussian factor $e^{-\frac{1}{4}|\mathbf{G}|^2\ell_B^2}$, the effect of higher harmonics of $w_\mathbf{G}$ is suppressed in $\mathcal{N}_\mathbf{k}$ and consequently in $F_{xy}(\mathbf{k})$. For example, for a triangular lattice, the first and second harmonic $w_\mathbf{G}$'s get Gaussian suppression factors of $0.16$ and $0.04$, respectively. Hence, as long as the first harmonic $\tilde{w}_\mathbf{G}$ is small, the zeroth harmonic (the LLL piece) dominates, and $\mathcal{N}_\mathbf{k}\approx w_\mathbf{0}^{-1/2}$ as well as the Berry curvature distribution is very uniform in BZ (assuming $\tilde{w}_\mathbf{G}$'s reduce with increasing $|\mathbf{G}|$ for smooth $|h(\mathbf{r})|^2$). This is exactly the reason behind the bands considered in Fig.~\ref{fig:1} having uniform Berry curvature. The approximation $\mathcal{N}_\mathbf{k}\approx w_\mathbf{0}^{-1/2}$ simplifies $\lambda_{\mathbf{q}}(\mathbf{k},\mathbf{k}')\approx \sum_{\mathbf{G}}\tilde{w}_{\mathbf{G}} \lambda^\text{LLL}_{\mathbf{q}-\mathbf{G}}(\mathbf{k},\mathbf{k}')$ making the interaction potential $V(\mathbf{r}_1-\mathbf{r}_2)$ projected to the ideal band equivalent to a modified interaction $\tilde{V}(\mathbf{r}_1,\mathbf{r}_2)=V(\mathbf{r}_1-\mathbf{r}_2)|\tilde{h}(\mathbf{r}_1)|^2|\tilde{h}(\mathbf{r}_2)|^2$ projected to LLL~\cite{wang2021exact}. 
Expanding this interaction, we obtain
\begin{equation}\label{eq:ProjIntLLL}
    \tilde{V}= \sum_\mathbf{q}\sum_{\mathbf{G}_1,\mathbf{G}_2}\frac{V(\mathbf{q})}{A}\tilde{w}_{\mathbf{G}_1}\tilde{w}_{\mathbf{G}_2} e^{i((\mathbf{G}_1+\mathbf{q})\cdot\mathbf{r}_1+(\mathbf{G}_2-\mathbf{q})\cdot\mathbf{r}_2)}. %(\mathbf{r}_1,\mathbf{r}_2) 
\end{equation}
Clearly, $\tilde{w}_\mathbf{G}$'s are responsible for Umklapp scattering. Next, we separate $\tilde{V}$ into purely relative motion and center-of-mass (COM) dependent pieces $\tilde{V}_\text{rel}(\mathbf{r}_1-\mathbf{r}_2)$ and $\tilde{V}_\text{COM}(\mathbf{r}_1,\mathbf{r}_2)$, respectively. Importantly, as we show in End Matter, the projection of $\tilde{V}_\text{COM}(\mathbf{r}_1,\mathbf{r}_2)$ to LLL is small; it is always suppressed by the Gaussian factors in the form factors. The remaining relative interaction is
\begin{equation}
    \tilde{V}_\text{rel}(\mathbf{r}_1-\mathbf{r}_2) = \sum_\mathbf{q}\sum_{\mathbf{G}}\frac{V(\mathbf{q})}{A} |\tilde{w}_{\mathbf{G}}|^2 e^{i(\mathbf{G}+\mathbf{q})\cdot(\mathbf{r}_1-\mathbf{r}_2)}.
\end{equation}
By changing the order of the summation in the above equation and performing a $\mathbf{q}\to \mathbf{q}-\mathbf{G}$ shift, $\tilde{V}_\text{rel}(\mathbf{r}_1-\mathbf{r}_2)$ can be written as
\begin{equation}\label{eq:Vrel}
    \tilde{V}_\text{rel}(\mathbf{r}_1-\mathbf{r}_2) = \frac{1}{A}\sum_\mathbf{G}|\tilde{w}_{\mathbf{G}}|^2  \sum_{\mathbf{q}}V(\mathbf{q}-\mathbf{G}) e^{i\mathbf{q}\cdot(\mathbf{r}_1-\mathbf{r}_2)}.
\end{equation}
When $d_s\ll \ell_B$, the dominant contribution in $V(\mathbf{q})$ is the $V_1$ pseudopotential $\propto -d_s^3 q^2$ (the contact potential $\propto d_s$ does not contribute)~\cite{haldane1983fractional,trugman1985exact,Haldane1987QHE}. The shift $\mathbf{q}\to\mathbf{q}-\mathbf{G}$ leaves the interaction unchanged up to terms that do not couple to fermions ($\mathbf{q}\cdot\mathbf{G}$ and $G^2$). Hence, for $d_s\ll \ell_B$ and small first harmonic $\tilde{w}_\mathbf{G}$'s, the interaction projected to an ideal Chern band is equivalent to
\begin{equation}\label{eq:GapEnhancement}
\begin{split}
    \tilde{V}(\mathbf{r}_1,\mathbf{r}_2) &\approx \tilde{V}_\text{rel}(\mathbf{r}_1-\mathbf{r}_2)\\
    &\approx -\left(\sum_\mathbf{G}|\tilde{w}_{\mathbf{G}}|^2\right)  \sum_{\mathbf{q}} \frac{2\pi e^2}{\epsilon A}d_s^3q^2 e^{i\mathbf{q}\cdot(\mathbf{r}_1-\mathbf{r}_2)}\\
    &\approx \left(\sum_\mathbf{G}|\tilde{w}_{\mathbf{G}}|^2\right) V(\mathbf{r}_1-\mathbf{r}_2)
\end{split}
\end{equation}
projected to LLL (up to terms that do not couple to electrons). Since $\sum_\mathbf{G}|\tilde{w}_{\mathbf{G}}|^2 = M_2$, we conclude that the many-body spectrum (charge neutral, anyon or charge sector) of the ideal band with small first harmonic $\tilde{w}_\mathbf{G}$'s must have approximately identical many-body spectrum as LLL with enhancement of all energy scales by $M_2$~\footnote{For pseudopotentials, the FCI ground state wavefunction is known to be the Laughlin wavefunction modulated by $\prod_{i=1}^{N_e} h(\mathbf{r}_i)$~\cite{ledwith2020fractional}}. 

\begin{figure}[t]
    \centering
\includegraphics[width=0.482\textwidth]{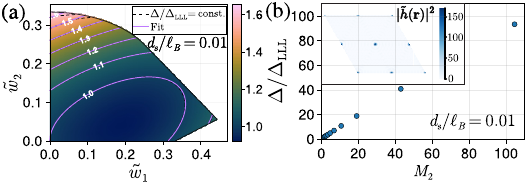}
\caption{{\bf Role of lattice harmonics in gap enhancement of Abelian FCIs for 
short-range interactions.}  {(a) Charge-neutral gap ratio $\Delta/\Delta_{\mathrm{LLL}}$ 
of the $\nu=1/3$ FCI in ideal bands with triangular-lattice geometry and $C_6$ symmetry versus the 
first two harmonics $\tilde w_1,\tilde w_2$. Black dashed and purple contours show ED and the fit through 
third order in the Gaussian-suppression hierarchy, respectively; the solid black curve is the positivity 
boundary $|\tilde h(\mathbf r)|^2\geq0$. The fit captures both the leading first-shell suppression and 
the weaker second-shell correction present at $\tilde w_1=0$. (b) ED gap enhancement ratio
$\Delta/\Delta_{\mathrm{LLL}}$ versus $M_2$ for the Gaussian-peak models (see Supplemental Material (SM)~\cite{SM2026}) with different
peak widths at $d_s/\ell_B=0.01$. The inset shows a representative real-space profile of
$|\tilde h(\mathbf r)|^2$. This construction shows that the $M_2$ enhancement can become very large
within the single-band theory.}}
\label{fig:2}
\end{figure}

Note that while the positivity of $|\tilde{h}(\mathbf{r})|^2 \ge 0$ puts constraints on the Fourier coefficients $|\tilde{w}_{\mathbf{G}}|$, it does not bound the enhancement factor: we show in SM~\cite{SM2026} that $M_2$ can be made arbitrarily large under this constraint. 
We also provide an explicit model in SM~\cite{SM2026}, where by concentrating $|\tilde{h}(\mathbf r)|^2$ into increasing smaller fraction of the unit-cell area, we can achieve any large value of $M_2$, and by performing ED at filling $\nu=1/3$ at each value of $M_2$, we find that $\Delta/\Delta_\text{LLL}\approx M_2$ up to $M_2 \approx 100$ as shown in Fig.~\ref{fig:2}(b)~\footnote{In ED numerics, we could not go beyond $M_2\approx 100$ since larger values of $M_2$ requires very large number of $\\mathbf{G}$-shells to be kept in the ED calculation, and would make it increasing numerically expensive}. However, when $M_2$ becomes sufficiently large that the many-body gap is comparable to the single-particle band gap, the projected-interaction approximation used here breaks down, and band-mixing effects must be taken into account.

Furthermore, using the formalism developed here, in SM~\cite{SM2026} we show that (1) $V_3$ pseudopotential is also renormalized by $M_2$, and consequently the $\nu=1/5$ gap is also enhanced by $M_2$, and (2) the electron density of the FCI ground states in these ideal bands are indeed $A_\text{uc}\rho(\mathbf{r}) \approx \nu |\tilde{h}(\mathbf{r})|^2$.

In the opposite limit,  $d_s \gg \ell_B$, Eq.~(\ref{eq:Vrel}) yields $V(\mathbf{q}-\mathbf{G}) \propto 1/|\mathbf{q}-\mathbf{G}|$, which suppresses contributions from $|\mathbf{G}|\neq 0$. As a result,
$\tilde{V}(\mathbf{r}_1,\mathbf{r}_2) \approx V(\mathbf{r}_1-\mathbf{r}_2)$,
and the interaction projected to the ideal band reduces to that of the LLL. This explains the absence of gap enhancement at large $d_s$. 

A physically intuitive way of understanding results for short- and long-range interactions is the following. As discussed earlier, projecting $V(\mathbf r_1-\mathbf r_2)$ to the ideal band amounts to projecting
$V(\mathbf r_1-\mathbf r_2)|\tilde h(\mathbf r_1)|^2|\tilde h(\mathbf r_2)|^2$ to the LLL. For short-range interactions, $\mathbf r_1\simeq \mathbf r_2$, so the interaction samples the local weight $|\tilde h(\mathbf r)|^4$, whose unit-cell average is $M_2$. In other words, when an excitation creates a short-distance pair, that pair feels a stronger local repulsion because both particles are more likely to sit in the high-density region of the unit cell. This enhances the interaction energy scale. For longer-range interactions, this enhancement is reduced toward 1 because for longer-range interactions, electron locations in real space become less important.

{Recall that the simple \(M_2\) scaling in Eq.~\eqref{eq:GapEnhancement} relies on two approximations: neglecting the momentum dependence of the normalization, \(\mathcal N_{\mathbf k}\approx w_{\mathbf 0}^{-1/2}\), and neglecting the COM-dependent interaction \(\tilde V_{\mathrm{COM}}\). Both approximations generate corrections to \(\Delta/\Delta_{\mathrm{LLL}}\), which are strongest for the lowest reciprocal-lattice harmonics. As shown in SM~\cite{SM2026}, these corrections can be organized in a Gaussian hierarchy in powers of
$\zeta \equiv e^{-b^2\ell_B^2/8}$,
where \(b\) is the magnitude of the smallest reciprocal-lattice vector. The leading correction arises from the first shell contribution from $\tilde{V}_\text{COM}$ at order \(\zeta\), while normalization and higher-shell COM effects enter at higher powers of \(\zeta\). Thus, even when the first harmonic vanishes, small residual corrections remain, whereas a finite first harmonic generally produces the dominant deviation from the ideal \(M_2\) scaling.
To show this concretely, we quantify the corrections for a \(C_6\)-symmetric triangular-lattice model with first- and second-harmonic amplitudes \(\tilde w_1\) and \(\tilde w_2\). Figure~\ref{fig:2}(a) maps the resulting gap ratio over the \(\tilde w_1\)-\(\tilde w_2\) plane. For \((N_k,N_e)=(27,9)\), the third-order fit gives
\[
\begin{split}
\frac{\Delta}{\Delta_{\mathrm{LLL}}}\simeq M_2
&-0.775\tilde w_1-4.195\tilde w_1^2-1.931\tilde w_1\tilde w_2\\
&+2.106\tilde w_1^3-3.402\tilde w_1^2\tilde w_2
+0.231\tilde w_1\tilde w_2^2\\
&+0.005\tilde w_2-0.119\tilde w_2^2,
\end{split}
\]
with \(M_2=1+6\tilde w_1^2+6\tilde w_2^2\) and \(R^2=0.99998\). For \(\tilde w_1=0\), the deviation from \(\Delta/\Delta_{\mathrm{LLL}}=M_2\) is therefore very small, while finite \(\tilde w_1\) produces the dominant gap-suppressing corrections (the linear terms in $\tilde{w}_i$ was studied in~\cite{shi2025effects}). A sufficiently large \(\tilde w_2\), however, restores \(\Delta/\Delta_{\mathrm{LLL}}>1\). The full hierarchy of corrections and its finite-size analysis are given in SM~\cite{SM2026}.}

\noindent\textit{Gap enhancement in Abelian FCIs in higher-vortexable bands.}--In relation to gap enhancement in moir\'e flat bands that mimic Landau levels, it was shown in~\cite{zhang2025beyond} that higher vortexable bands with multi-component wavefunctions $\psi^\text{MC}_\mathbf{k} = \{\sqrt{t}\mathcal{N}_{1\mathbf{k}}\psi_\mathbf{k}^\text{LL1}(\mathbf{r}),\sqrt{1-t}\mathcal{N}_{0\mathbf{k}}\psi_\mathbf{k}^\text{LLL}(\mathbf{r})\}h(\mathbf{r})$ can enhance the charge-neutral gap of $\nu =1/3$ FCI by several orders of magnitude relative to pure LLL at the $d_s\ll \ell_B$ limit with maximal enhancement near $t =1/2$ (here $\psi_\mathbf{k}^{\text{LL}n}(\mathbf{r})$ is the single particle wavefunction of the $n$th LL). This enhancement originates from the multicomponent structure of the wavefunction. In contrast to pure LLL-type bands, where only the term $V(\mathbf{q})\propto -d_s^3 q^2$ contributes to the effective $V_1$ pseudopotential, multicomponent wavefunctions allow the contact term $V(\mathbf{q})\propto d_s$ to contribute as well. Consequently, in the small-$d_s$ limit the enhancement scales as $1/d_s^2$. More importantly, Ref.~\cite{zhang2025beyond} numerically observed that moir\'e bands with $h(\mathbf{r})\neq1$ and nearly uniform Berry curvature exhibit even larger gap enhancement than the corresponding hybridized Landau-level bands with $h(\mathbf{r})=1$, although the origin of this additional enhancement remained unclear. We reproduce both effects in Appendix C Fig.~\ref{fig:end_ll_mix} The additional enhancement in the moir\'e case follows naturally from the analysis above. Repeating the derivation leading to Eq.~\eqref{eq:Vrel}, but now using the dominant interaction $V(\mathbf{q})\propto d_s$, again yields an overall renormalization of the projected interaction by
$
M_2
=
\sum_{\mathbf{G}}
|\tilde{w}_{\mathbf{G}}|^2
$
(see SM~\cite{SM2026} for full derivation). Thus, lattice modulation enhances the many-body gap in higher-vortexable bands through the same mechanism identified above for ideal LLL-type bands.

\begin{figure}[t]
    \centering
\includegraphics[width=0.482\textwidth]{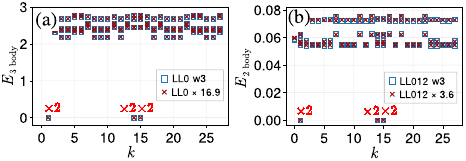}
\caption{{\bf Gap enhancement in non-Abelian FCIs.} (a) ED spectra at filling $\nu=1/2$ for the three-body Moore-Read pseudopotential projected to an ideal band with nearly uniform Berry curvature (blue boxes) and to the LLL(red crosses). The LLL spectrum has been rescaled by a factor $16.9$. The spectra remain nearly identical in structure, while the many-body gap is enhanced by the factor $M_3$ in the ideal band. (b) ED spectra at filling $\nu=1/2$ for the screened Coulomb interaction projected to higher-vortexable bands with multicomponent wavefunctions, comparing moir\'e bands with lattice modulation (blue boxes) and hybridized LL bands with $h(\mathbf{r})=1$ (red crosses). The latter spectrum has been rescaled by $M_2 = 3.6$. In both panels, the red numbers indicate the ground-state degeneracies in the corresponding momentum sectors. The details of the lattice modulation of the (higher-) vortexable bands considered here are given in  Appendix A. ED calculations were performed on a system containing $N_k=28$ unit cells (SM~\cite{SM2026}).
}
\label{fig:3}
\end{figure}

\noindent\textit{Gap enhancement in non-Abelian FCIs.—}
A similar enhancement mechanism also applies to non-Abelian FCIs. In the LLL at filling $\nu=1/2$, the Moore-Read state is the exact zero-energy ground state of the three-body pseudopotential
$
V(\mathbf{r}_1-\mathbf{r}_2,\mathbf{r}_2-\mathbf{r}_3)
=
\nabla_1^4\nabla_2^2
\delta(\mathbf{r}_1-\mathbf{r}_2)
\delta(\mathbf{r}_2-\mathbf{r}_3)$
\cite{rezayi2000incompressible,reddy2024non}. Following the derivation above, projection of this interaction to an ideal band with nearly uniform Berry curvature is equivalent to projecting
$
V(\mathbf{r}_1-\mathbf{r}_2,\mathbf{r}_2-\mathbf{r}_3)
|\tilde{h}(\mathbf{r}_1)|^2
|\tilde{h}(\mathbf{r}_2)|^2
|\tilde{h}(\mathbf{r}_3)|^2
$
to the LLL. As in the two-body case, the center-of-mass dependent contributions are strongly suppressed by Gaussian form factors and can be neglected. The remaining purely relative interaction acquires an overall renormalization factor~\footnote{In momentum space, the three-body pseudopotential is
$V(\mathbf{q}_1,\mathbf{q}_2)\propto q_1^4 q_2^2$.
Under shifts $\mathbf{q}\to\mathbf{q}-\mathbf{G}$ analogous to Eq.~\eqref{eq:Vrel}, one generates anisotropic terms and lower-order contributions such as $q_1^4$, $q_1^2q_2^2$, $q_2^4$, $q_1^2$, and $q_2^2$. In real space these correspond to derivatives acting on the two delta functions $\delta(\mathbf{r}_1-\mathbf{r}_2)\delta(\mathbf{r}_2-\mathbf{r}_3)$. Their matrix elements vanish identically in the LLL because fermionic three-particle wavefunctions necessarily contain the factor $(z_1-z_2)(z_2-z_3)(z_3-z_1)$.}
$
M_3
=
\sum_{\mathbf{G}_1,\mathbf{G}_2}
\tilde{w}_{\mathbf{G}_1}
\tilde{w}_{\mathbf{G}_2}
\tilde{w}_{-\mathbf{G}_1-\mathbf{G}_2}.$
In Fig.~\ref{fig:3}(a), we compare the ED spectra of the three-body interaction projected to the ideal band and to the LLL at $\nu=1/2$. We find that the spectra remain nearly identical in structure, while the many-body gap in the ideal band is enhanced by the factor $M_3$, consistent with the above analysis. 

We further investigate the possibility of enhancing Moore-Read gaps in higher-vortexable bands using only two-body interactions. It is known that in pure LL1, very short-range interactions do not robustly stabilize the Moore-Read state~\cite{zhang2025beyond}, with the charge-neutral gap exhibiting strong finite-size dependence in ED. Motivated by this, we consider higher-vortexable bands with multicomponent wavefunctions
$
\psi^{\text{MC}}_{\mathbf{k}}
=
\left\{
\sqrt{t_0}\,\mathcal{N}_{0\mathbf{k}}\psi^{\text{LL0}}_{\mathbf{k}}(\mathbf{r}),
\sqrt{t_1}\,\mathcal{N}_{1\mathbf{k}}\psi^{\text{LL1}}_{\mathbf{k}}(\mathbf{r}),
\sqrt{t_2}\,\mathcal{N}_{2\mathbf{k}}\psi^{\text{LL2}}_{\mathbf{k}}(\mathbf{r})
\right\}
h(\mathbf{r}).
$ We find that the screened Coulomb interaction
$
V(\mathbf{q})
=
\frac{2\pi e^2}{\epsilon |\mathbf{q}|}
\tanh(d_s |\mathbf{q}|)
$
is sufficient to stabilize the Moore-Read state in these bands. In the limit $d_s\ll \ell_B$, the charge-neutral gap scales linearly with $d_s$ (see SM~\cite{SM2026}), indicating that the dominant contribution arises from the contact interaction
$
V(\mathbf{r}_1-\mathbf{r}_2)
\propto
\delta(\mathbf{r}_1-\mathbf{r}_2).
$
From the analysis above, we then expect lattice modulation that preserves nearly uniform Berry curvature to enhance the gap by the same factor
$
M_2.
$
An explicit example of this enhancement is shown in Fig.~\ref{fig:3}(b).

\begin{figure}[t]
    \centering
\includegraphics[width=0.482\textwidth]{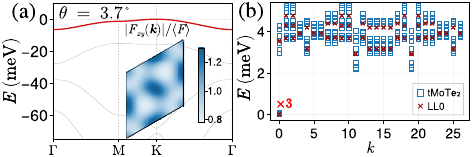}
\caption{{\bf FCI in twisted bilayer MoTe$_2$ at twist angle $\theta=3.7^\circ$.} (a) Band structure showing the top four valence bands in the moir\'e BZ (top valence band is marked in red). The band structure only shows the bands in a single valley. The inset shows the normalized Berry-curvature distribution $|F_{xy}(\mathbf{k})|/\langle F_{xy}\rangle_{\mathrm{BZ}}$ of the top valence band. The continuum Hamiltonian as well as parameters are taken from~\cite{wang2024fractional}. (b) ED spectra at $\nu_h=2/3$ filling of tMoTe$_2$ (blue boxes) and $\nu=2/3$ filling of LLL (red crosses) for screened Coulomb interaction with parameters $d_s = 30\text{ nm, } \epsilon = 15.0$ and $N_k =27$ unit cells. We assume valley polarization for tMoTe$_2$ calculations.
}
\label{fig:4}
\end{figure}

\noindent\textit{Application to tMoTe$_2$.}--To connect our results to a realistic material platform, we study a continuum model of tMoTe$_2$ at twist angle $\theta=3.7^\circ$, where FCI has been experimentally observed at hole filling $\nu_h=2/3$~\cite{cai2023signatures,zeng2023thermodynamic,park2023observation,xu2023observation} and persists up to temperatures of a few Kelvin. The top valence band of tMoTe$_2$ is known to be nearly vortexable~\cite{dong2023composite,morales2023pressure}, with only a small deviation from the ideal condition, $\langle\text{tr}(g)\rangle_\text{BZ}/\langle F_{xy}\rangle_\text{BZ}\approx 1.15$ and relatively weak Berry-curvature fluctuations ($\sim 11.5\%$) [Fig.~\ref{fig:4}(a)]. Numerically approximating the band wavefunctions to the ideal limit~\footnote{The approximation scheme is akin to~\cite{gao2023untwisting}, see~\cite{SM2026} for details}, we find that the first harmonic $|\tilde{w}_\mathbf{G}|\approx 0.12$ is small (compared to twisted bilayer graphene flat bands, where the first harmonic $|\tilde{w}_\mathbf{G}|\approx 0.24$~\cite{wang2021exact}). {(The idealization for tMoTe$_2$ is only to extract $\tilde w_{\mathbf G}$, all the EDs below are performed based on the original continuum model.)} The analytical framework developed above is therefore expected to apply approximately to this system. In the experimentally relevant regime $d_s/a\gtrsim1$ ($a$ is the moir\'e lattice constant), our theory predicts that the charge-neutral spectrum of the FCI in an ideal band with nearly uniform Berry curvature should closely resemble that of the FQH problem in the LLL~\footnote{Note that previously an adiabatic approximation to the continuum Hamiltonian of tMoTe$_2$ was proposed~\cite{morales2024magic,shi2024adiabatic}, which effectively maps the top valence band to Aharonov-Casher band. Based on this, tMoTe$_2$ ED spectra have been compared with ED spectra of Dirac particles under inhomogeneous magnetic field~\cite{li2026abelian}. Here we are directly comparing with LLL under a uniform magnetic field}. To test this expectation, in Fig.~\ref{fig:4}(b) we compare the charge-neutral spectra of tMoTe$_2$ and the LLL at filling $\nu_h=2/3$. The two spectra are indeed remarkably similar, with only small quantitative deviations~\footnote{In experiments on twisted MoTe\(_2\), the FCI is most prominent near \(\nu_h=2/3\), while the FCI is weaker at \(\nu_h=1/3\)~\cite{pan2026optical}. In SM~\cite{SM2026} we show that for this continuum model, the charge neutral gap at $\nu=1/3$ is smaller than $\nu=2/3$ due to nonzero bandwidth and particle-hole asymmetry.}. These results suggest that the near-vortexable nature of the top valence band and its unusually small first harmonic $|\tilde{w}_{\mathbf G}|$ are key ingredients underlying the remarkable robustness of the FCI phase observed experimentally in tMoTe$_2$. 

Interestingly, the slight nonideality of the tMoTe$_2$ valence band can become quantitatively important for $d_s\ll a$. In SM~\cite{SM2026} we show that it then produces a gap enhancement factor of $\sim100$ relative to the LLL, analogous to the enhancement mechanism found in higher-vortexable bands.

\noindent\textit{Discussion.}--Our results identify lattice modulation not merely as a perturbation to Landau-level physics, but as a design principle for stabilizing fractional topology. The lattice modulations encoded in $w_{\mathbf G}$ directly control the FCI gap and excitation spectrum, and therefore serve as practical diagnostics and design knobs for achieving robust FCIs. This makes it possible to engineer FCIs with gaps substantially larger than those of their Landau-level counterparts while retaining a predictable FQH-like spectrum. Such control is important not only for improving the robustness of FCIs at finite temperature and against disorder but also for enabling phases built from fractionalized excitations, such as anyon fluids and anyon superconductors.

\begin{acknowledgments}
\noindent \textit{Acknowledgements}.---We thank Jennifer Cano, Zhao Liu, Ady Stern, and Bo Yang for insightful comments and suggestions. This work was supported in part by Air Force Office of Scientific Research MURI FA9550-23-1-0334 and by the Gordon and Betty Moore Foundation Grant No. GBMF10694 (YZ, KS).
\end{acknowledgments}

\bibliographystyle{apsrev4-1}
\bibliography{ref}

\appendix

\setcounter{equation}{0}  %  this will re-count eq from 1
\setcounter{figure}{0}
\renewcommand{\theequation}{A\arabic{equation}}
\renewcommand{\thefigure}{A\arabic{figure}}
% \newpage
%\begin{widetext}
%\end{widetext}

\section{End Matter}

\subsection{Appendix A.  Details of the (higher)-vortexable bands studied in main text}
\begin{figure}[t]
    \centering
    \includegraphics[width=0.482\textwidth]{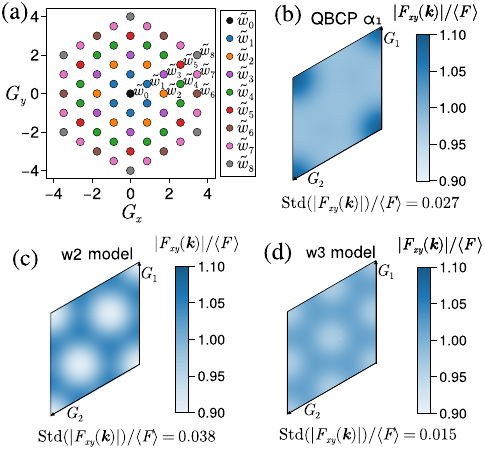}
    \caption{
    \textbf{Fourier components and Berry curvature distributions.}
    (a) Reciprocal-lattice shells used to parameterize the Fourier coefficients
    of the modulation,
    \(|h(\mathbf r)|^2=\sum_{\mathbf G}w_{\mathbf G}e^{i\mathbf G\cdot\mathbf r}\).
    Each point denotes a reciprocal lattice vector \(\mathbf G\).  We impose
    \(C_6\) symmetry, so all vectors in the same shell have the same real coefficient,
    labeled \(w_1,w_2,\ldots\) according to their distance from the origin.
    The  model used in Fig.~\ref{fig:1}(b), called w2 model, is specified by the normalized
    shell coefficients
    \((\tilde w_1,\ldots,\tilde w_8)
    =(0,-1/3,0,0,1/3,-1/6,0,0)\), and where the model in Figs.~\ref{fig:1}(c)  and~\ref{fig:3}(a), called w3 model is given by
    \((\tilde w_1,\ldots,\tilde w_8)
    =(0,0,3/5,0,0,1/4,0,1/8)\).
    (b-d) Normalized Berry curvature \(|F_{xy}(\mathbf k)|/\langle F\rangle\) for QBCP $\alpha_1$, w2, and w3 models. The QBCP $\alpha_1$ model refers to the ideal band achieved in a single layer system with a quadratic band crossing point (at the zone center $\Gamma$) under a moir\'e periodic strain field with magic strain strength $\alpha \approx 2.13 \equiv \alpha_1$; see Ref.~\cite{wan2023topological} for the details of the continuum Hamiltonian and details.
    }
    \label{fig:end1}
\end{figure}

Here we give details of the conventions as well as the single-particle bands we used for the ED results shown in the main text. 

Fig.~\ref{fig:end1}(a) shows how the Fourier coefficients \(w_{\mathbf G}\) are organized upon reciprocal lattice shells. 

The ideal band used in Fig.~\ref{fig:1}(a) comes from a single particle continuum model Hamiltonian which describes a quadratic band crossing point (QBCP) at a time reversal invariant momentum coupled to a moir\'e periodic strain field~\cite{wan2023topological,sarkar2023symmetry,sarkar2025unconventional},
\begin{equation}\label{eq:hamiltonian}
    \mathcal{H}(\mathbf{r})
    = \begin{pmatrix}0 &\text{h.c.}\\ -4 \overline{\partial_{z}}^2 +\tilde{A}(\mathbf{r}) & 0\end{pmatrix}
\end{equation}
Here $z=x+iy$ is the complex coordinate, $\tilde{A}(\mathbf{r}) = A_x(\mathbf{r})+iA_y(\mathbf{r})$ where $A_x=u_{xx}-u_{yy}$ and $A_y=u_{xy}$ are moir\'e periodic shear strain fields. The $C_{6z}$ symmetric moir\'e strain field used here is given by $\tilde{A}(\mathbf{r})= -\frac{\alpha}{2}\sum_{n=1}^{3}e^{i(4-n)\phi} \cos\left(\mathbf{G}_n \cdot \mathbf{r} \right)$, where $\mathbf{G}_n = \frac{2\sqrt{\pi}}{3^{1/4}\ell_B}(-\sin(2\pi (n-1)/3),\cos(2\pi (n-1)/3))$ are primitive moir\'e reciprocal lattice vectors, and $\phi=2\pi/3$. Note that this Hamiltonian is chiral/sublattice symmetric $\{\sigma_z,\mathcal{H}(\mathbf{r})\}=0$. For a magic strain amplitude $\alpha \approx 2.13 |\mathbf{G}_1|^2\equiv \alpha_1$, the band structure contains two exact ideal flat bands at Fermi level. These two flatbands are polarized to opposite sublattices and are related by time reversal symmetry. For filling fractions $\nu\leq 1$, exchange interaction drives all electrons to one of the sublattice polarized bands, and thus we consider only one of the two bands in our calculations. Notably, these sublattice polarized bands have very uniform Berry curvature (with a standard deviation of 0.027) as shown in Fig.~\ref{fig:end1}(c). This is because the first harmonic $\tilde{w}_1\approx 0.017$ is small (and in fact smaller than the second harmonic $\tilde{w}_2\approx 0.076$; see SM~\cite{SM2026} for more details).

The harmonics $\tilde{w}_\mathbf{G}$ of the bands referred to as w2 and w3 models are given in Fig.~\ref{fig:end1}. These bands can be thought of as bands that appear in the Aharonov-Casher model~\cite{aharonov1979ground} with a magnetic field $B(\mathbf{r}) = B_0(1-\ell_B^2\nabla^2\log |h(\mathbf{r})|)$. Indeed, the magnetic fields corresponding to the w2 and w3 models are very inhomogeneous, yet they have very uniform Berry curvature (standard deviations 0.038 and 0.015, respectively) since the first harmonic $\tilde{w}_1 = 0$, and as we show in the main text, the ED spectra are identical to LLL ED spectra in structure.

\subsection{Appendix B. COM dependent part of the interaction is small}
In the main text we argued that interaction potential $V(\mathbf{r}_1-\mathbf{r}_2)$ projected to the ideal band with near uniform Berry curvature distribution is equivalent to a modified interaction $\tilde{V}(\mathbf{r}_1,\mathbf{r}_2)=V(\mathbf{r}_1-\mathbf{r}_2)|\tilde{h}(\mathbf{r}_1)|^2|\tilde{h}(\mathbf{r}_2)|^2$ projected to LLL, and expanded $\tilde{V}(\mathbf{r}_1,\mathbf{r}_2)$ in Fourier space as Eq.~\eqref{eq:ProjIntLLL}. Splitting this expression into purely relative motion and COM dependent pieces, we obtain
\begin{equation}
\begin{split}
    \tilde{V}_\text{rel}(\mathbf{r}_1-\mathbf{r}_2) &= \sum_\mathbf{q}\sum_{\mathbf{G}}\frac{V(\mathbf{q})}{A} |\tilde{w}_{\mathbf{G}}|^2 e^{i(\mathbf{G}+\mathbf{q})\cdot(\mathbf{r}_1-\mathbf{r}_2)},\\
    \tilde{V}_\text{COM}(\mathbf{r}_1,\mathbf{r}_2) &=\sum_\mathbf{q}\sum_{\mathbf{G}_1\neq-\mathbf{G}_2}\frac{V(\mathbf{q})}{A}\tilde{w}_{\mathbf{G}_1}\tilde{w}_{\mathbf{G}_2}\times \\
    &\phantom{= \sum_\mathbf{q}\sum_{\mathbf{G}_1\neq-\mathbf{G}_2}}e^{i((\mathbf{G}_1+\mathbf{q})\cdot\mathbf{r}_1+(\mathbf{G}_2-\mathbf{q})\cdot\mathbf{r}_2)}.
\end{split}
\end{equation}
Here we show that the projection of $\tilde{V}_\text{COM}(\mathbf{r}_1,\mathbf{r}_2)$ to LLL is small. The projection of $e^{i((\mathbf{G}_1+\mathbf{q})\cdot\mathbf{r}_1+(\mathbf{G}_2-\mathbf{q})\cdot\mathbf{r}_2)}$ gives rise to two form factors $\lambda_{-\mathbf{q}-\mathbf{G}_1}(\mathbf{k}_1,\mathbf{k}_2)\propto e^{-|\mathbf{q}+\mathbf{G}_1|^2\ell_B^2/4}$ and $\lambda_{\mathbf{q}-\mathbf{G}_2}(\mathbf{k}_3,\mathbf{k}_4)\propto e^{-|\mathbf{q}-\mathbf{G}_2|^2\ell_B^2/4}$. Because of the two Gaussian form factors, it may seem that the summand is small for large $\mathbf{G}$. However, since $\mathbf{q}$ is not restricted to the BZ, it can be large and cancel $\mathbf{G}_i$ in the exponent. Fortunately, since $\mathbf{G}_1\neq -\mathbf{G}_2$, $\mathbf{q}$ can only cancel one of $\mathbf{G}_1$ and $\mathbf{G}_2$, and the Gaussian factor corresponding to the one that is not canceled is going to suppress the summand (see SM~\cite{SM2026} for a more detailed discussion). Thus, $\tilde{V}_\text{COM}(\mathbf{r}_1,\mathbf{r}_2)$ is small and can be neglected. {\it The implication of this is profound: in an ideal band with uniform Berry curvature, the COM dependent part of the interaction can be neglected even if the higher harmonics of the Fourier series of $|h(\mathbf{r})|^2$ are not negligible. Furthermore, negligibility of $\tilde{V}_\text{COM}$ also implies that the continuous magnetic translation symmetry is (approximately) restored, and at filling fraction $\nu=p/q$ with coprime $p,q$, all eigen-energies (not just the ground states) are $q$-fold degenerate~\cite{haldane1985many}}. 

In contrast, projection of $\tilde{V}_\text{rel}(\mathbf{r}_1-\mathbf{r}_2)$ gives rise to $\lambda_{-\mathbf{q}-\mathbf{G}}(\mathbf{k}_1,\mathbf{k}_2)\lambda_{\mathbf{q}+\mathbf{G}}(\mathbf{k}_3,\mathbf{k}_4)\propto e^{-|\mathbf{q}+\mathbf{G}|^2\ell_B^2/2}$, which is not small when $\mathbf{q} = -\mathbf{G}$, hence the effect of $\tilde{w}_{\mathbf{G}\neq 0}$ cannot be neglected. In fact, these are the terms that give rise to the FCI charge-neutral gap enhancement in the $d_s\ll \ell_B$ limit.

\subsection{Appendix C. Gap enhancement in higher-vortexable bands}
\begin{figure}[tbh]
    \centering
\includegraphics[width=0.4\textwidth]{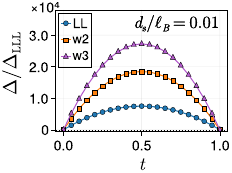}
    \caption{Charge-neutral gap enhancement factor $\Delta/\Delta_{\mathrm{LLL}}$ for higher-vortexable bands with multicomponent wavefunctions for short-range interaction as function of $t$ defined in the main text. The blue circles correspond to hybridized LL bands with $h(\mathbf{r})=1$, while the orange squares and purple triangles correspond to ideal bands with increasing lattice modulation. The details of the lattice modulations of the bands studied here are given in Appendix A.}
    \label{fig:end_ll_mix}
\end{figure}

\clearpage
% \pagebreak
\let\addcontentsline\supplementaddcontentsline
\onecolumngrid
% \vspace{10cm}

%%%%%%%%%%%%%%%%%%%%%%%%%%%%%%%%%%%%%%
%%   Supplementary Information
%%%%%%%%%%%%%%%%%%%%%%%%%%%%%%%%%%%%%%
\makeatletter
\renewcommand \thesection{S-\@arabic\c@section}
\renewcommand\thetable{S\@arabic\c@table}
\renewcommand \thefigure{S\@arabic\c@figure}
\renewcommand \theequation{S\@arabic\c@equation}
\makeatother
\setcounter{equation}{0}  %  this will re-count eq from 1
\setcounter{figure}{0}  %  this will re-count eq from 1
\setcounter{section}{0}  %  this will re-count eq from 1
\counterwithin{figure}{section} 
{
    \center \bf \large 
    Supplemental Material\vspace*{0.1cm}\\ 
    \vspace*{0.0cm}
}
\maketitle
\tableofcontents

\section{ED setup}

\begin{figure}[tbh]
    \centering
    \includegraphics[width=0.75\textwidth]{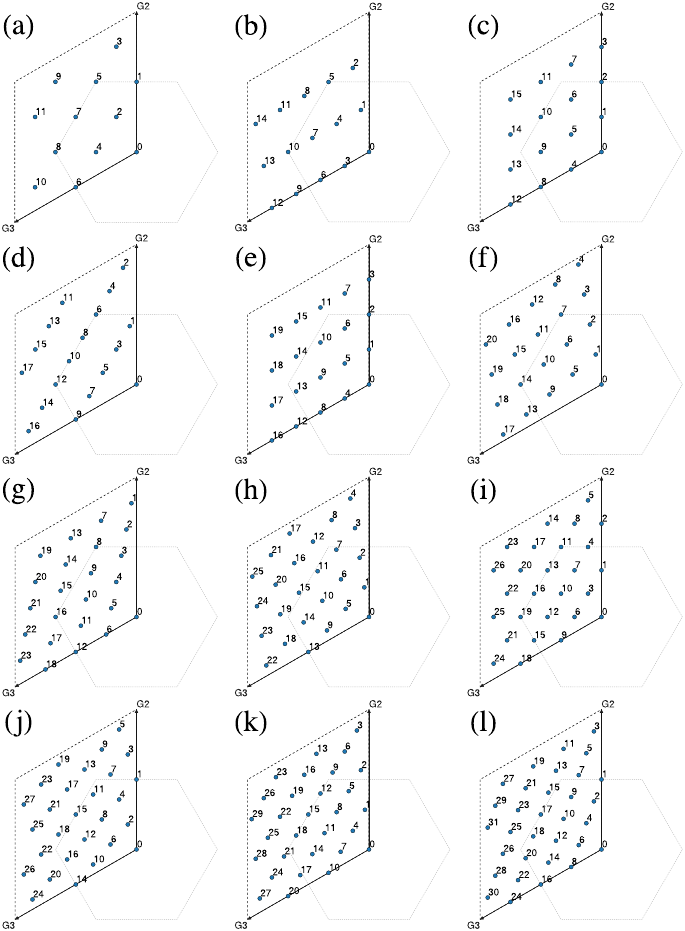}
    \caption{\textbf{Triangular lattice ED momentum clusters.}
    The twelve panels show the finite momentum grids used in our ED calculations, with cluster sizes
    (a) $N_s=12$, (b) $N_s=15$, (c) $N_s=16$, (d) $N_s=18$, (e) $N_s=20$, (f) $N_s=21$, (g) $N_s=24$, (h) $N_s=26$, (i) $N_s=27$, (j) $N_s=28$, (k) $N_s=30$, and (l) $N_s=32$. Blue dots denote the allowed momenta in the finite Brillouin zone, and the integer next to each dot is the momentum label
    $k_n=0,1,\ldots,N_s-1$ used in the ED basis. The light gray hexagon indicates the first Brillouin zone, while the black arrows show the triangular reciprocal lattice directions $\mathbf G_2$ and $\mathbf G_3$.}
    \label{fig:EDclusters}
\end{figure}

Through out this paper, we perform the projected exact diagnolization on finite triangular-lattice momentum clusters, as summarized in Fig.~\ref{fig:EDclusters}.  The reciprocal lattice is generated by two primitive vectors
\begin{equation}\label{eq:Gdefs}
    \mathbf G_1 = |\mathbf G| (0,1),
    \qquad
    \mathbf G_2 = |\mathbf G| \left(-\frac{\sqrt{3}}{2},-\frac{1}{2}\right),
    \qquad
    |\mathbf G|=\frac{2\sqrt{\pi}}{3^{1/4}\ell_B},
\end{equation}
where we set the magnetic length to $\ell_B=1$ unless otherwise stated.  Each finite cluster consists of a discrete set of $N_s$ allowed crystal momenta in the Brillouin zone. The integer label shown next to each momentum point in Fig.~\ref{fig:EDclusters} is the total momentum label $K$ used in the ED spectrum.

\FloatBarrier

\section{Description of vortexable and higher-vortexable bands}

In this section we review the properties of (higher-) vortexable Chern bands following~\cite{wang2021exact}, and derive some of the identities used in the main text. 

% \sss{I think we should just copy paste this part from the main text. I don't know why this called a model, and why the model "starts from" anything. All we need to say here is that any vortexable bands with $|C|=1$ can be written as Eq.~(S2).}
% We first specify the generalized Landau level model used throughout this work.  The model starts from the magnetic Bloch basis
% of a LLL with one flux quantum per unit cell,
% \(A_{\mathrm{uc}}=2\pi\ell_B^2\), and generates an ideal \(|C|=1\) Chern band by
% multiplying each LLL wavefunction by the same \(\mathbf k\)-independent real space modulation factor \(h(\mathbf r)\):
% \begin{equation}
%     \psi_{\mathbf k}(\mathbf r)
%     =
%     \mathcal{N}_{\mathbf k}\,
%     \psi^{\mathrm{LLL}}_{\mathbf k}(\mathbf r)\,
%     h(\mathbf r).
% \end{equation}
% Here \(\mathcal{N}_{\mathbf k}\) is fixed by normalization.Since
% \(\psi_{\mathbf k}(\mathbf r)\) is Bloch periodic while
% \(\psi^{\mathrm{LLL}}_{\mathbf k}(\mathbf r)\) obeys magnetic translation
% symmetry, the modulation factor \(h(\mathbf r)\) must be quasiperiodic so as
% to cancel the magnetic translation phase. Consequently,
% \(|h(\mathbf r)|^2\) is periodic on the lattice. In this sense the model keeps
% the analytic structure of the LLL while allowing the probability density inside
% each unit cell to be redistributed. The uniform LLL is recovered when
% \(h(\mathbf r)\) is constant.

The single particle wavefunctions of any vortexable/ideal band with Chern number $|C|=1$ in momentum space can be written as~\cite{wang2021exact}
\begin{equation}
    \psi_\mathbf{k}(\mathbf{r}) = \mathcal{N}_\mathbf{k}\psi^\text{LLL}_\mathbf{k}(\mathbf{r})h(\mathbf{r}),
\end{equation}
where $\mathcal{N}_\mathbf{k}$ is the normalization factor, $\psi^\text{LLL}_\mathbf{k}(\mathbf{r})$ is the (normalized) LLL wavefunction in a homogeneous magnetic field $B_0$, obeying magnetic
Bloch periodicity: $\psi^\text{LLL}_\mathbf{k}(\mathbf{r}+\mathbf{a}_i) = e^{i\mathbf{a}_i\times \mathbf{r}/2\ell_B^2}e^{i\mathbf{k}\cdot\mathbf{a}_i}\psi^\text{LLL}_\mathbf{k}(\mathbf{r})$, with magnetic length $\ell_B=\sqrt{\hbar/eB_0}$ and lattice vectors $\mathbf{a}_i$ such that every unit cell contain one flux quantum $\Phi_0 = 2\pi\hbar/e$ or equivalently unit cell area $A_\text{uc}$ satisfies $A_\text{uc}=\mathbf{a}_1\times \mathbf{a}_2 = 2\pi\ell_B^2$. The function $h(\mathbf{r})$ is a $\mathbf{k}$-independent modulation that is periodic when describing a Dirac particle in an inhomogeneous periodic field $B(\mathbf{r}) = B_0(1-\ell_B^2\nabla^2\log |h(\mathbf{r})|)$~\cite{aharonov1979ground}, and quasi-periodic, $h(\mathbf{r}+\mathbf{a}_i) = e^{-i\mathbf{a}_i\times \mathbf{r}/2\ell_B^2}h(\mathbf{r})$~\cite{guerci2025layer} for vortexable bands at zero magnetic field. Bands described by these wavefunctions are termed ideal, as they satisfy the ideal trace condition $\mathrm{tr}(g(\mathbf{k})) = |F_{xy}(\mathbf{k})|$ relating the quantum metric $g(\mathbf{k})$ and Berry curvature $F_{xy}(\mathbf{k})$. These bands are termed vortexable because multiplication of the wavefunctions $\psi_\mathbf{k}(\mathbf{r})$ by an arbitrary holomorphic function $f(z)$, with $z=x+iy$, yields a state $f(z)\psi_\mathbf{k}(\mathbf{r})$ that remains fully within the same band. 
% Note that a generalization of Eq.~\eqref{eq:vortexablebandWF} where a unit cell contains multiple flux quanta is also possible~\cite{sarkar2023symmetry,sarkar2025unconventional}, leading to multiple flat bands and unconventional fractional phases. 
Importantly, $h(\mathbf{r})$ encodes the (moir\'e) lattice modulation, and is responsible for Umklapp scattering~\cite{wang2021exact}. Since $h(\mathbf{r})$ is (quasi-) periodic, \(|h(\mathbf r)|^2\) is periodic on the lattice. The uniform LLL is recovered when \(h(\mathbf r)\) is constant.

\subsection{Fourier components}

Since \(|h(\mathbf r)|^2\) is lattice-periodic, we expand it as
\begin{equation}
    |h(\mathbf r)|^2
    =
    \sum_{\mathbf G}
    w_{\mathbf G}e^{i\mathbf G\cdot\mathbf r},
    \qquad
    w_{-\mathbf G}=w_{\mathbf G}^{*},
    \label{eq:h_fourier_supp}
\end{equation}
where \(\mathbf G\) runs over reciprocal lattice vectors.

\begin{figure}[tbh]
    \centering
    \includegraphics[width=0.75\textwidth]{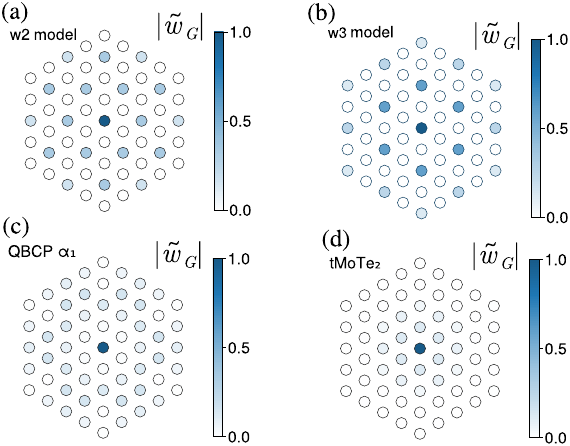}
    \caption{
    \textbf{Fourier components of the vortexable band modulation.}
    Each point is a reciprocal lattice vector \(\mathbf G\), and the color scale denotes the magnitude \(|\tilde w_{\mathbf G}|\).  
    (a) w2 model. 
    (b) w3 model. 
    (c) QBCP model at the magic value \(\alpha_1\). 
    (d) Continuum model \(\mathrm{tMoTe}_2\) top band at twist angle \(3.7^\circ\), where the Fourier coefficients are extracted from the single particle band wave functions.
    }
    \label{fig:sup_fig1}
\end{figure}

To visualize the Fourier expansion in Eq.~\eqref{eq:h_fourier_supp}, 
Fig.~\ref{fig:sup_fig1} shows the reciprocal space picture of the normalized coefficients
\(\tilde w_{\mathbf G}\) used in the main text calculations.  Each circle denotes one reciprocal lattice vector \(\mathbf G\), and the color represents the magnitude of \(\tilde w_{\mathbf G}\).  The first two panels show the w2 and w3 model inputs.  The third panel shows the QBCP model at the magic value \(\alpha_1\).  The fourth panel shows the coefficients extracted from the single particle wave functions of the top band of the \(3.7^\circ\) continuum \(\mathrm{tMoTe}_2\) model~\cite{wu2019topological,dong2023composite}.  Only \(|\tilde w_{\mathbf G}|\) is shown in the figure; the signs and phases of \(\tilde w_{\mathbf G}\) are kept in the projected-interaction calculation.

For reproducibility, the symmetry-independent input coefficients used to generate the four panels are, rounded to two significant digits,
% \begin{equation}
% \begin{array}{c|cccccccc}
% \text{model}
% & \tilde w_1 & \tilde w_2 & \tilde w_3 & \tilde w_4
% & \tilde w_5 & \tilde w_6 & \tilde w_7 & \tilde w_8
% \\ \hline
% \text{ w2 model}
% & 0 & -1/3 & 0 & 0 & 1/3 & -1/6 & 0 & 0
% \\
% \text{ w3 model}
% & 0 & 0 & 3/5 & 0 & 0 & 1/4 & 0 & 1/8
% \\
% \text{QBCP } \alpha_1
% & 0.017 & 0.076 & -0.12 & -0.15 & 0.0039 & 0.089 & 0.058 & 0.0073
% \\
% \mathrm{tMoTe}_2\ (3.7^\circ)
% & -0.12 & 0.074 & -0.043 & -0.0032 & -0.0032 & -0.0013 & -0.0035 & -0.0029
% \end{array}
% \label{eq:wG_model_inputs}
% \end{equation}
\begingroup
\begin{equation}
\setlength{\arraycolsep}{2pt}
\begin{array}{@{}c|cccccccc|cc@{}}
\text{model}
& \tilde w_1 & \tilde w_2 & \tilde w_3 & \tilde w_4
& \tilde w_5 & \tilde w_6 & \tilde w_7 & \tilde w_8
& M_2 & M_3
\\ \hline
\text{ w2 model}
& 0 & -1/3 & 0 & 0 & 1/3 & -1/6 & 0 & 0
& 2.50 & 7.78
\\
\text{ w3 model}
& 0 & 0 & 3/5 & 0 & 0 & 1/4 & 0 & 1/8
& 3.63 & 17.09
\\
\text{QBCP } \alpha_1
& 0.017 & 0.076 & -0.12 & -0.15 & 0.0039 & 0.089 & 0.058 & 0.0073
& 1.48 & 2.57
\\
\mathrm{tMoTe}_2\ (3.7^\circ)
& -0.12 & 0.074 & -0.043 & -0.0032 & -0.0032 & -0.0013 & -0.0035 & -0.0029
& 1.13 & 1.43
\end{array}
\label{eq:wG_model_inputs}
\end{equation}
\endgroup
where the convention for \(\tilde w_i\), \(i=1,\ldots,8\), is defined in the End Matter. Note that all the bands considered here have $C_{6z}$ symmetry (in reality, tMoTe$_2$ only has $C_{3z}$ symmetry per valley, but due to the first harmonic approximation to the moir\'e potential/tunneling, the continuum Hamiltonian per valley has an extra pseudo-inversion/spinless $C_{2z}$ symmetry~\cite{yu2024fractional}).

\subsection{Form factor and normalization}
First we start from the LLL form factor:
\begin{equation}
    \lambda^{\mathrm{LLL}}_{\mathbf q}(\mathbf k,\mathbf k')
    =
    \left\langle
    \psi^{\mathrm{LLL}}_{\mathbf k}
    \left|
    e^{-i\mathbf q\cdot\mathbf r}
    \right|
    \psi^{\mathrm{LLL}}_{\mathbf k'}
    \right\rangle .
    \label{eq:LLL_lambda_q_def_supp}
\end{equation}
Due to Bloch periodicity, \(\lambda^{\mathrm{LLL}}_{\mathbf q}(\mathbf k,\mathbf k')\) is nonzero only when
\begin{equation}
    \mathbf q=\mathbf k'-\mathbf k+\mathbf G .
\end{equation}
We therefore introduce another equivalent notation:
\begin{equation}
    \lambda_{\mathbf G}^{\mathrm{LLL};\mathbf k\mathbf k'}
    \equiv
    \lambda^{\mathrm{LLL}}_{\mathbf k'-\mathbf k+\mathbf G}
    (\mathbf k,\mathbf k')
    =
    \left\langle
    \psi^{\mathrm{LLL}}_{\mathbf k}
    \left|
    e^{-i(\mathbf k'-\mathbf k+\mathbf G)\cdot\mathbf r}
    \right|
    \psi^{\mathrm{LLL}}_{\mathbf k'}
    \right\rangle .
    \label{eq:LLL_lambda_G_def_supp}
\end{equation}
For a given allowed momentum transfer
\(\mathbf q=\mathbf k'-\mathbf k+\mathbf G\),
\begin{equation}
    \lambda^{\mathrm{LLL}}_{\mathbf q}(\mathbf k,\mathbf k')
    =
    \lambda_{\mathbf G}^{\mathrm{LLL};\mathbf k\mathbf k'},
    \qquad
    \mathbf G=\mathbf q-\mathbf k'+\mathbf k .
    \label{eq:lambda_q_to_G_supp}
\end{equation}
This \(\lambda_{\mathbf q}^{\mathrm{LLL}}\) is the physical density form factor
labeled by the total momentum transfer \(\mathbf q\), while
\(\lambda_{\mathbf G}^{\mathrm{LLL};\mathbf k\mathbf k'}\) is the same object
after resolving the reciprocal lattice Umklapp channel \(\mathbf G\). 

In the magnetic Bloch gauge choice used in this paper, the explicit LLL form factor is~\cite{wang2021exact}
\begin{equation}
    \lambda_{\mathbf G}^{\mathrm{LLL};\mathbf k\mathbf k'}
    =
    \eta_{\mathbf G}\,
    \exp\left[
        -\frac{\ell_B^2}{4}
        |\mathbf k-\mathbf k'-\mathbf G|^2
        +\frac{i\ell_B^2}{2}\,
        \mathbf k\times\mathbf k'
        +\frac{i\ell_B^2}{2}\,
        (\mathbf k+\mathbf k')\times\mathbf G
    \right].
    \label{eq:LLL_formfactor_G_supp}
\end{equation}
The normalization condition \(\langle \psi_{\mathbf k}|\psi_{\mathbf k}\rangle=1\)
then gives
\begin{equation}
    \mathcal{N}_{\mathbf k}^{-2}
=\left\langle
    \psi^{\mathrm{LLL}}_{\mathbf k}
    \left|
    |h(\mathbf r)|^2
    \right|
    \psi^{\mathrm{LLL}}_{\mathbf k}
    \right\rangle
    \nonumber=
    \sum_{\mathbf G}
    w_{\mathbf G}
    \left\langle
    \psi^{\mathrm{LLL}}_{\mathbf k}
    \left|
    e^{i\mathbf G\cdot\mathbf r}
    \right|
    \psi^{\mathrm{LLL}}_{\mathbf k}
    \right\rangle
    \nonumber=
    \sum_{\mathbf G}
    w_{\mathbf G}\,
    \lambda^{\mathrm{LLL}}_{-\mathbf G}(\mathbf k,\mathbf k).
    \label{eq:Nk_wG_formfactor_supp}
\end{equation}
In the magnetic Bloch phase convention used in the main text,
\begin{equation}
    \lambda^{\mathrm{LLL}}_{-\mathbf G}(\mathbf k,\mathbf k)
    =
    \eta_{\mathbf G}\,
    \exp\left[
        i\ell_B^2\,\mathbf G\times\mathbf k
        -
        \frac{\ell_B^2|\mathbf G|^2}{4}
    \right],
    \label{eq:LLL_diag_lambda_supp}
\end{equation}
where \(\eta_{\mathbf G}\) is a convention-dependent magnetic-translation phase.
Thus
\begin{equation}
    \mathcal{N}_{\mathbf k}^{-2}
    =
    \sum_{\mathbf G}
    \eta_{\mathbf G}w_{\mathbf G}
    \exp\left[
        i\ell_B^2\,\mathbf G\times\mathbf k
        -
        \frac{\ell_B^2|\mathbf G|^2}{4}
    \right].
    \label{eq:Nk_explicit_supp}
\end{equation}
This is the normalization formula used in the main text. The Gaussian factor
\(\exp[-\ell_B^2|\mathbf G|^2/4]\) strongly suppresses large \(|\mathbf G|\)
harmonics in \(\mathcal{N}_{\mathbf k}\). Therefore high Fourier components of
\(|h(\mathbf r)|^2\) can appreciably modify interaction matrix elements while
having only a weak effect on the Berry-curvature variation.

We now derive the form factor for a vortexable Chern band used in the projected interaction:
\begin{equation}
    \lambda_{\mathbf q}(\mathbf k,\mathbf k')
    =
    \left\langle
    \psi_{\mathbf k}
    \left|
    e^{-i\mathbf q\cdot\mathbf r}
    \right|
    \psi_{\mathbf k'}
    \right\rangle .
    \label{eq:lambda_def_supp}
\end{equation}
where \(\psi_\mathbf{k}(\mathbf{r}) = \mathcal{N}_\mathbf{k}\psi^\text{LLL}_\mathbf{k}(\mathbf{r})h(\mathbf{r})\). Using the Fourier expansion
of \(|h(\mathbf r)|^2\), we find
\begin{align}
    \lambda_{\mathbf q}(\mathbf k,\mathbf k')
    &=
    \mathcal{N}_{\mathbf k}\mathcal{N}_{\mathbf k'}
    \left\langle
    \psi^{\mathrm{LLL}}_{\mathbf k}
    \left|
    |h(\mathbf r)|^2
    e^{-i\mathbf q\cdot\mathbf r}
    \right|
    \psi^{\mathrm{LLL}}_{\mathbf k'}
    \right\rangle
    \nonumber\\
    &=
    \mathcal{N}_{\mathbf k}\mathcal{N}_{\mathbf k'}
    \sum_{\mathbf G}
    w_{\mathbf G}
    \left\langle
    \psi^{\mathrm{LLL}}_{\mathbf k}
    \left|
    e^{-i(\mathbf q-\mathbf G)\cdot\mathbf r}
    \right|
    \psi^{\mathrm{LLL}}_{\mathbf k'}
    \right\rangle
    \nonumber\\
    &=
    \mathcal{N}_{\mathbf k}\mathcal{N}_{\mathbf k'}
    \sum_{\mathbf G}
    w_{\mathbf G}\,
    \lambda^{\mathrm{LLL}}_{\mathbf q-\mathbf G}
    (\mathbf k,\mathbf k').
    \label{eq:lambda_convolution_supp}
\end{align}
Equation~\eqref{eq:lambda_convolution_supp} is a convolution formula:
each Fourier component \(w_{\mathbf G}\) shifts the LLL momentum transfer from \(\mathbf q\) to \(\mathbf q-\mathbf G\). These shifted terms are the Umklapp form factor channels entering the projected interaction.

\subsection{LL hybridization}\label{sec:llhybridization}
We now generalize this vortexable band construction to higher Landau levels following~\cite{liu2025theory}.
Let
\(\{|\psi_{\mathbf k}^{\mathrm{LL},n}\rangle\}\) be the conventional magnetic
Bloch basis of the \(n\)th Landau level.  We first multiply all the LLs by the same envelope \(h(\mathbf r)\) to get the modulated LL basis \( |e_{n\mathbf k}\rangle\),
\begin{equation}
    |e_{n\mathbf k}\rangle
    =
    h(\mathbf r)
    |\psi_{\mathbf k}^{\mathrm{LL},n}\rangle ,
    \qquad
    |h(\mathbf r)|^2
    =
    \sum_{\mathbf G}
    w_{\mathbf G}e^{i\mathbf G\cdot\mathbf r}.
    \label{eq:raw_modulated_LL_basis_supp}
\end{equation}
For \(h(\mathbf r)=1\), it reduces to the ordinary LL basis, which is orthogonal in the LL index.  For nontrivial \(h\), the states are no longer orthogonal.

The raw modulated density matrix element is defined directly as
\begin{align}
    \lambda_{\mathbf G}^{\mathrm{raw},mn;\mathbf k\mathbf k'}
    &\equiv
    \left\langle
    e_{m\mathbf k}
    \left|
    e^{-i(\mathbf k'-\mathbf k+\mathbf G)\cdot\mathbf r}
    \right|
    e_{n\mathbf k'}
    \right\rangle .
    \label{eq:raw_lambda_def_supp}
\end{align}
Substituting \(|e_{n\mathbf k}\rangle=h(\mathbf r)|\psi_{\mathbf k}^{\mathrm{LL},n}\rangle\) and using
\(|h(\mathbf r)|^2=\sum_{\mathbf G'}w_{\mathbf G'}e^{i\mathbf G'\cdot\mathbf r}\), this becomes
\begin{equation}
    \lambda_{\mathbf G}^{\mathrm{raw},mn;\mathbf k\mathbf k'}
    =
    \sum_{\mathbf G'}
    w_{\mathbf G'}
    \left\langle
    \psi_{\mathbf k}^{\mathrm{LL},m}
    \left|
    e^{-i[\mathbf k'-\mathbf k+(\mathbf G-\mathbf G')]\cdot\mathbf r}
    \right|
    \psi_{\mathbf k'}^{\mathrm{LL},n}
    \right\rangle .
\end{equation}
Recall the conventional LL form factor between any state \(m\) and \(n\),
\begin{equation}
    \lambda_{\mathbf G}^{\mathrm{LL},mn;\mathbf k\mathbf k'}
    =
    \left\langle
    \psi_{\mathbf k}^{\mathrm{LL},m}
    \left|
    e^{-i(\mathbf k'-\mathbf k+\mathbf G)\cdot\mathbf r}
    \right|
    \psi_{\mathbf k'}^{\mathrm{LL},n}
    \right\rangle ,
    \label{eq:LL_lambda_mn_orth_supp}
\end{equation}
Hence the raw modulated form factor is the convolution
\begin{equation}
    \lambda_{\mathbf G}^{\mathrm{raw},mn;\mathbf k\mathbf k'}
    =
    \sum_{\mathbf G'}
    w_{\mathbf G'}\,
    \lambda_{\mathbf G-\mathbf G'}^{\mathrm{LL},mn;\mathbf k\mathbf k'} .
    \label{eq:raw_lambda_convolution_supp}
\end{equation}
The overlap matrix of the raw basis is the special case
\(\mathbf k'=\mathbf k\) and \(\mathbf G=\mathbf 0\):
\begin{equation}
    S_{mn}(\mathbf k)
    =
    \langle e_{m\mathbf k}|e_{n\mathbf k}\rangle
    =
    \lambda_{\mathbf 0}^{\mathrm{raw},mn;\mathbf k\mathbf k}
    =
    \sum_{\mathbf G'}
    w_{\mathbf G'}\,
    \lambda_{-\mathbf G'}^{\mathrm{LL},mn;\mathbf k\mathbf k}.
    \label{eq:S_matrix_orth_supp}
\end{equation}
For nontrivial \(h(\mathbf r)\), \(S_{mn}(\mathbf k)\) is generally not
diagonal in the LL indices.  Therefore a scalar normalization is not sufficient
once more than one LL is included.

We orthonormalize the raw states at each \(\mathbf k\) by defining the
orthonormalized basis \(|\psi_{n\mathbf k}\rangle\) through the Gram--Schmidt procedure 
\begin{equation}
    |\psi_{n\mathbf k}\rangle
    =
    \sum_m
    |e_{m\mathbf k}\rangle
    U_{mn}(\mathbf k),
    \qquad
    U^\dagger(\mathbf k)S(\mathbf k)U(\mathbf k)
    =
    \mathbf 1 .
    \label{eq:orthonormalized_basis_supp}
\end{equation}

We denote the form factor matrix in the orthonormalized basis by
\(\lambda^{\mathrm{orth}}\), from~\ref{eq:raw_lambda_convolution_supp},the matrix elements are
\begin{align}
    \lambda_{\mathbf G}^{\mathrm{orth},mn;\mathbf k\mathbf k'}
    &\equiv
    \left\langle
    \psi_{m\mathbf k}
    \left|
    e^{-i(\mathbf k'-\mathbf k+\mathbf G)\cdot\mathbf r}
    \right|
    \psi_{n\mathbf k'}
    \right\rangle
    =
    \sum_{a,b}
    U^*_{am}(\mathbf k)
    \left\langle
    e_{a\mathbf k}
    \left|
    e^{-i(\mathbf k'-\mathbf k+\mathbf G)\cdot\mathbf r}
    \right|
    e_{b\mathbf k'}
    \right\rangle
    U_{bn}(\mathbf k')
    \nonumber\\
    &=
    \left\{
    U^\dagger(\mathbf k)
    \left[
    \sum_{\mathbf G'}
    w_{\mathbf G'}\,
    \lambda_{\mathbf G-\mathbf G'}^{\mathrm{LL};\mathbf k\mathbf k'}
    \right]
    U(\mathbf k')
    \right\}_{mn}.
    \label{eq:orth_lambda_final_supp}
\end{align}
Here \(\lambda_{\mathbf G}^{\mathrm{LL};\mathbf k\mathbf k'}\) is a matrix in
ordinary LL indices, with \(\left[
    \lambda_{\mathbf G}^{\mathrm{LL};\mathbf k\mathbf k'}
    \right]_{mn}
    =
    \lambda_{\mathbf G}^{\mathrm{LL},mn;\mathbf k\mathbf k'} .\)

The projected form factor of the \(n\)-th orthonormalized modulated LL is the
diagonal matrix element
\begin{equation}
    \lambda_{n,\mathbf G}^{\mathrm{orth};\mathbf k\mathbf k'}
    \equiv
    \lambda_{\mathbf G}^{\mathrm{orth},nn;\mathbf k\mathbf k'} .
    \label{eq:projected_orth_lambda_n_supp}
\end{equation}
This is the form factor used in the projected many-body Hamiltonian for the LL\(n\) (higher) vortexable state.

The quantum geometry of the \(n\)-th orthonormalized modulated LL can be
obtained from the small-\(\delta\mathbf k\) overlap
\begin{equation}
    \left\langle
    \widetilde u_{n,\mathbf k+\delta\mathbf k}
    \middle|
    \widetilde u_{n,\mathbf k}
    \right\rangle
    =
    \lambda_{n,\mathbf 0}^{\mathrm{orth};
    \mathbf k+\delta\mathbf k,\mathbf k}.
    \label{eq:orth_overlap_geometry_supp}
\end{equation}
For \(n\ge 1\), the pointwise ideal relation
\(\operatorname{tr}g_n(\mathbf k)=(2n+1)|F_{xy}^{(n)}(\mathbf k)|\) is generally not
preserved.  Instead, the Brillouin-zone averaged trace obeys~\cite{liu2025theory}
\begin{equation}
    \frac{
    \langle \operatorname{tr}g_n\rangle_{\mathrm{BZ}}
    }{
    |\langle F_{xy}^{(n)}\rangle_{\mathrm{BZ}}|
    }
    =
    2n+1,
    \label{eq:orth_integrated_trace_relation_supp}
\end{equation}
where
\begin{equation}
    \langle \operatorname{tr}g_n\rangle_{\mathrm{BZ}}
    =
    \frac{1}{A_{\mathrm{BZ}}}
    \int_{\mathrm{BZ}}d^2k\,
    \operatorname{tr}g_n(\mathbf k),
    \qquad
    \langle F_{xy}^{(n)}\rangle_{\mathrm{BZ}}
    =
    \frac{2\pi C_n}{A_{\mathrm{BZ}}}.
\end{equation}
For a Chern-one band with one flux quantum per unit cell,
\(A_{\mathrm{BZ}}=2\pi/\ell_B^2\), so
\begin{equation}
    \langle \operatorname{tr}g_n\rangle_{\mathrm{BZ}}
    =
    (2n+1)\ell_B^2 .
    \label{eq:orth_avg_metric_C1_supp}
\end{equation}

We also used a multicomponent LL-hybridized model in the main text.  For a set of LL weights
\(\{t_n\}_{n\in\mathcal I}\), with
\begin{equation}
    t_n\ge 0,
    \qquad
    \sum_{n\in\mathcal I}t_n=1,
\end{equation}
we define the orthonormal multicomponent modulated LL-hybridized state to be
\begin{equation}
    \psi_{\mathbf k}^{\mathrm{MC}}(\mathbf r)
    =
    \left\{
    \sqrt{t_{i_1}}\,\psi_{i_1\mathbf k}(\mathbf r),
    \sqrt{t_{i_2}}\,\psi_{i_2\mathbf k}(\mathbf r),
       \ldots,
    \sqrt{t_{i_n}}\,\psi_{i_n\mathbf k}(\mathbf r)
    \right\}, \qquad i_1, i_2, \ldots, i_n \in \mathcal I.
\label{eq:MC_LL_hybrid_wavefunction_supp}
\end{equation}
where the different LL components are treated as orthogonal internal components.  Therefore the projected form factor is simply
\begin{equation}
    \lambda_{\mathbf G}^{\mathrm{MC};\mathbf k\mathbf k'}
    =
    \sum_{n\in\mathcal I}
    t_n\,
    \lambda_{n,\mathbf G}^{\mathrm{orth};\mathbf k\mathbf k'} .
    \label{eq:MC_LL_formfactor_supp}
\end{equation}
For example, for an LL01 model with weights \(t_0=1-t\) and \(t_1=t\),
\begin{equation}
    \lambda_{\mathbf G}^{\mathrm{MC};\mathbf k\mathbf k'}
    =
    (1-t)\,
    \lambda_{0,\mathbf G}^{\mathrm{orth};\mathbf k\mathbf k'}
    +
    t\,
    \lambda_{1,\mathbf G}^{\mathrm{orth};\mathbf k\mathbf k'} .
    \label{eq:LL01_formfactor_supp}
\end{equation}

It is useful to discuss the quantum geometry of the multicomponent state. We calculate the quantum metric directly from
the wavefunction overlap at small momentum separation.  Let
\(|u_{n,\mathbf k}\rangle\) denote the cell-periodic part of the \(n\)-th
orthonormalized modulated LL state \(|\psi_{n\mathbf k}\rangle\), and let the
cell-periodic multicomponent state be
\begin{equation*}
    |u_{\mathbf k}^{\mathrm{MC}}\rangle
    =
    \left\{
    \sqrt{t_{i_1}}\,|u_{i_1,\mathbf k}\rangle,
    \sqrt{t_{i_2}}\,|u_{i_2,\mathbf k}\rangle,
    \ldots,
    \sqrt{t_{i_n}}\,|u_{i_n,\mathbf k}\rangle
    \right\},
    \qquad
    \langle u_{m,\mathbf k}|u_{n,\mathbf k}\rangle
    =
    \delta_{mn}.
\end{equation*}
Define
\begin{equation}
    S_{\mathrm{MC}}(\mathbf k,\delta\mathbf k)
    \equiv\left\langle
    u_{\mathbf k+\delta\mathbf k}^{\mathrm{MC}}
    \middle|
    u_{\mathbf k}^{\mathrm{MC}}
    \right\rangle=
    \lambda_{\mathbf 0}^{\mathrm{MC};
    \mathbf k+\delta\mathbf k,\mathbf k}
    =
    \sum_{n\in\mathcal I}
    t_n S_n(\mathbf k,\delta\mathbf k),
    \label{eq:SMC_small_momentum_supp}
\end{equation}
where
\(
    S_n(\mathbf k,\delta\mathbf k)
    \equiv
    \left\langle
    u_{n,\mathbf k+\delta\mathbf k}
    \middle|
    u_{n,\mathbf k}
    \right\rangle
    =
    \lambda_{n,\mathbf 0}^{\mathrm{orth};
    \mathbf k+\delta\mathbf k,\mathbf k}.\) The quantum metric of the multicomponent state is extracted from
\begin{equation}
    1-|S_{\mathrm{MC}}(\mathbf k,\delta\mathbf k)|^2
    =
    g_{\mu\nu}^{\mathrm{MC}}(\mathbf k)
    \delta k_\mu \delta k_\nu
    +O(\delta k^3).
    \label{eq:MC_metric_from_overlap_supp}
\end{equation}
Expanding the left side using the identity \(\sum t_n=1\) we get
\begin{equation}
    1-\left|\sum_n t_n S_n\right|^2
    =
    \sum_n t_n(1-|S_n|^2)
    +
    \frac12\sum_{m,n}t_m t_n |S_m-S_n|^2 ,
    \label{eq:MC_overlap_variance_identity_supp}
\end{equation}
which separates the weighted average of the individual metric contributions
from the pairwise mismatch between different LL components.

For the unmodulated hybridized LL, \(h(\mathbf r)=1\), the ordinary LL overlaps
follow from the \(\mathbf G=\mathbf 0\) pure-LL form factor
\(e^{\frac{i\ell_B^2}{2}\mathbf k\times\delta\mathbf k}L_n(x)e^{-x/2}\),
with \(x=\ell_B^2|\delta\mathbf k|^2/2\).  Since
\(L_n(x)e^{-x/2}=1-(n+\frac12)x+O(x^2)\), this gives
\begin{equation}
    S_n^{\mathrm{LL}}(\mathbf k,\delta\mathbf k)
    =
    e^{\frac{i\ell_B^2}{2}\mathbf k\times\delta\mathbf k}
    \left[
    1-\frac{2n+1}{4}\ell_B^2|\delta\mathbf k|^2
    +O(\delta k^4)
    \right].
    \label{eq:pure_LL_overlap_small_dk_supp}
\end{equation}
Therefore
\(S_m^{\mathrm{LL}}-S_n^{\mathrm{LL}}=O(\delta k^2)\), so the mismatch term in
Eq.~\eqref{eq:MC_overlap_variance_identity_supp} is only \(O(\delta k^4)\) and
does not contribute to the metric.  Hence the trace metric of the unmodulated
hybridized LL is exactly the weighted sum of the trace metrics of the component
LLs:
\begin{equation}
    \left\langle \operatorname{tr}g^{\mathrm{LLMC}}\right\rangle_{\mathrm{BZ}}
    =
    \ell_B^2\sum_{n\in\mathcal I}t_n(2n+1).
    \label{eq:MC_LL_trace_weighted_rule_supp}
\end{equation}
For example, the LL02 model with \(t_0=1-t\) and \(t_2=t\) gives
\(\langle \operatorname{tr}g^{\mathrm{MC}}\rangle_{\mathrm{BZ}}
=\ell_B^2(1+4t)\).

For the modulated hybridized LL, the same identity
Eq.~\eqref{eq:MC_overlap_variance_identity_supp} still applies, but the second
term generally contributes to the quantum metric.  The reason is that the
orthonormalized modulated LL overlaps have the expansion
\begin{equation}
    S_n(\mathbf k,\delta\mathbf k)
    =
    1+iA_\mu^{(n)}(\mathbf k)\delta k_\mu+O(\delta k^2),
    \label{eq:modulated_LL_overlap_linear_supp}
\end{equation}
where \(A_\mu^{(n)}\) is the Berry connection of the \(n\)-th orthonormalized
modulated LL.  Unlike the unmodulated LL case, these Berry connections are
generally different for different \(n\).  Thus
\begin{equation}
    S_m(\mathbf k,\delta\mathbf k)-S_n(\mathbf k,\delta\mathbf k)
    =
    i\left[A_\mu^{(m)}(\mathbf k)-A_\mu^{(n)}(\mathbf k)\right]\delta k_\mu
    +O(\delta k^2),
\end{equation}
and the mismatch term in Eq.~\eqref{eq:MC_overlap_variance_identity_supp}
contributes to \(1-|S_{\mathrm{MC}}|^2\) at order \(\delta k^2\).  In the fixed
component convention used in Eq.~\eqref{eq:MC_LL_hybrid_wavefunction_supp}, this
gives
\begin{equation}
    g_{\mu\nu}^{\mathrm{MC}}(\mathbf k)
    =
    \sum_n t_n g_{\mu\nu}^{(n)}(\mathbf k)
    +
    \frac12\sum_{m,n}t_m t_n
    \left[
    A_\mu^{(m)}(\mathbf k)-A_\mu^{(n)}(\mathbf k)
    \right]
    \left[
    A_\nu^{(m)}(\mathbf k)-A_\nu^{(n)}(\mathbf k)
    \right].
    \label{eq:MC_modulated_metric_tensor_supp}
\end{equation}
Taking the trace and averaging over the Brillouin zone gives
\begin{equation}
    \left\langle \operatorname{tr}g^{\mathrm{MC}}\right\rangle_{\mathrm{BZ}}
    =
    \ell_B^2\sum_n t_n(2n+1)
    +
    \frac12\sum_{m,n}t_m t_n
    \left\langle
    \left|
    \mathbf A^{(m)}-\mathbf A^{(n)}
    \right|^2
    \right\rangle_{\mathrm{BZ}}.
    \label{eq:MC_modulated_trace_metric_average_supp}
\end{equation}
Here we used the averaged trace relation for each individual orthonormalized
modulated LL, \(    \left\langle \operatorname{tr}g_n\right\rangle_{\mathrm{BZ}}
    =
    (2n+1)\ell_B^2 .\)The second term in
Eq.~\eqref{eq:MC_modulated_trace_metric_average_supp} is expected to be small in
the small Berry curvature fluctuation regime.  In the complex-coordinate notation of 
Ref.~\cite{liu2025theory} Eq.~(34) gives the diagonal Berry connection as
\(A^{(n)}(\mathbf k)=-i\bar z_{\mathbf k}/2+i\alpha_{n\mathbf k}\), where \(\alpha_{n\mathbf k}=\sum_{m=0}^{n}\partial\log \mathcal N_{m\mathbf k}\), and
\(\mathcal N_{m\mathbf k}\) is defined in Ref.~\cite{liu2025theory} Eq.~(34)
%is the normalization coefficient of the orthonormalized modulated \(m\)-th LL. 
The term $-i\bar z_{\mathbf k}/2$ cancels in \(\mathbf A^{(m)}-\mathbf A^{(n)}\).  
The remaining \(n\)-dependence is then fixed by $\mathcal{N}_{m\mathbf k}$.
Similar to
Eq.~\eqref{eq:Nk_explicit_supp}, $\mathcal{N}_{m\mathbf k}$ vary only through
Gaussian-suppressed Fourier components of the modulation.  Thus, when the
first-shell coefficient \(\tilde w_1\) is small, the leading Berry-curvature
fluctuation and \(\partial\log \mathcal N_{m\mathbf k}\) are small.  Since the
second term in Eq.~\eqref{eq:MC_modulated_trace_metric_average_supp} is
quadratic in these Berry-connection differences, it gives only a very small
correction.

To summarize, a single modulated state obeys the same averaged trace
relation as an ordinary LL.  However, for a multicomponent modulated LL, the quantum metric generally 
does not follow a linear weighted sum over components. This linear relation holds exactly only for an unmodulated multicomponent LL.

\section{Real space density distribution of the FCI ground states}
For a many-body state projected to a single band, the real-space density is
\begin{equation}
   \rho (\mathbf r)
    =
    \langle \Psi|\hat\psi^\dagger(\mathbf r)\hat\psi(\mathbf r)|\Psi\rangle
    =
    \sum_{\mathbf k,\mathbf k'}
    \psi_{\mathbf k}^{*}(\mathbf r)
    \psi_{\mathbf k'}(\mathbf r)
    \langle \Psi|
    c_{\mathbf k}^{\dagger}c_{\mathbf k'}
    |\Psi\rangle ,
    \label{eq:density_general}
\end{equation}
where the projected field operator is
\begin{equation}
    \hat\psi(\mathbf r)
    =
    \sum_{\mathbf k}\psi_{\mathbf k}(\mathbf r)c_{\mathbf k}.
\end{equation}
For an ED eigenstate \(|\Psi_{\mathbf K}\rangle\) with total
momentum \(\mathbf K\), translation symmetry makes the one-body density matrix
diagonal in single-particle momentum, 
\begin{equation}
    \langle \Psi_{\mathbf K}|
    c_{\mathbf k}^{\dagger}c_{\mathbf k'}
    |\Psi_{\mathbf K}\rangle
    =
    \rho_{\mathbf k}^{(\mathbf K)}
    \delta_{\mathbf k,\mathbf k'},
    \qquad
    \rho_{\mathbf k}^{(\mathbf K)}
    =
    \langle \Psi_{\mathbf K}|
    c_{\mathbf k}^{\dagger}c_{\mathbf k}
    |\Psi_{\mathbf K}\rangle .
    \label{eq:one_body_diag}
\end{equation}
Therefore
\begin{equation}
   \rho^{(\mathbf K)} (\mathbf r)
    =
    \sum_{\mathbf k}
    \rho_{\mathbf k}^{(\mathbf K)}
    |\psi_{\mathbf k}(\mathbf r)|^2 .
    \label{eq:density_diag}
\end{equation}
In the main-text calculation, we use the occupation averaged over the three FCI ground states at total momenta \(\mathbf K_\alpha\), \(\rho_{\mathbf k}=\frac{1}{3}\sum_{\alpha=1}^{3}\rho_{\mathbf k}^{(\mathbf K_\alpha)}\), and denote this averaged quantity by \(\rho_{\mathbf k}\) below.

Since \(|\psi_{\mathbf k}(\mathbf r)|^2\) is unit-cell periodic, the density is periodic over the moir\'e unit cell due to Eq.~\eqref{eq:density_diag}. Hence, we write it as a Fourier series 
\begin{equation}
   \rho (\mathbf r)
    =
    \frac{1}{A_{\mathrm{uc}}}
    \sum_{\mathbf G}
    \rho_{\mathbf G}e^{i\mathbf G\cdot\mathbf r},
    \qquad
    \rho_{\mathbf G}
    =
    \int_{\mathrm{uc}} d^2\mathbf r\,
    e^{-i\mathbf G\cdot\mathbf r}
   \rho (\mathbf r).
    \label{eq:density_fourier}
\end{equation}
Substituting Eq.~\eqref{eq:density_diag}, we obtain
\begin{equation}
    \rho_{\mathbf G}
    =
    \sum_{\mathbf k}
    \rho_{\mathbf k}
    \int_{\mathrm{uc}} d^2\mathbf r\,
    e^{-i\mathbf G\cdot\mathbf r}
    |\psi_{\mathbf k}(\mathbf r)|^2 .
    \label{eq:nG_intermediate}
\end{equation}

Now let \(N_{\mathrm{uc}}\) be the number of unit cells in the finite system. The single-particle states are normalized over the full system area
\( A=N_{\mathrm{uc}}A_{\mathrm{uc}}\). 
\begin{equation}
    \lambda_{\mathbf G}^{\mathbf k\mathbf k}
    =
    \left\langle
    \psi_{\mathbf k}
    \left|
    e^{-i\mathbf G\cdot\mathbf r}
    \right|
    \psi_{\mathbf k}
    \right\rangle
    =
    N_{\mathrm{uc}}
    \int_{\mathrm{uc}} d^2\mathbf r\,
    e^{-i\mathbf G\cdot\mathbf r}
    |\psi_{\mathbf k}(\mathbf r)|^2 .
    \label{eq:diag_lambda_density}
\end{equation}
Therefore
\begin{equation}
    \rho_{\mathbf G}
    =
    \frac{1}{N_{\mathrm{uc}}}
    \sum_{\mathbf k}
    \rho_{\mathbf k}
    \lambda_{\mathbf G}^{\mathbf k\mathbf k}.
    \label{eq:nG_diaglambda}
\end{equation}
Fourier transform from~\ref{eq:density_fourier},
\begin{equation}
   \rho (\mathbf r)
    =
    \frac{1}{N_{\mathrm{uc}}A_{\mathrm{uc}}}
    \sum_{\mathbf G}
    e^{i\mathbf G\cdot\mathbf r}
    \sum_{\mathbf k}
    \rho_{\mathbf k}
    \lambda_{\mathbf G}^{\mathbf k\mathbf k},
    \label{eq:density_from_diaglambda}
\end{equation}
where the vortexable-band form factor can be calculated by
\begin{equation}
    \lambda_{\mathbf G}^{\mathbf k\mathbf k}
    =
    \mathcal{N}_{\mathbf k}^2
    \sum_{\mathbf G'}
    w_{\mathbf G'}\,
    \lambda_{\mathbf G-\mathbf G'}^{\mathrm{LLL};\mathbf k\mathbf k}.
    \label{eq:diag_lambda_convolution_density}
\end{equation}
Thus the real-space density is determined by the occupation numbers
\(\rho_{\mathbf k}\), the coefficients \(w_{\mathbf G}\), and the ordinary LLL form factors.

\subsection{For FCI ground states in ideal bands with nearly uniform Berry curvature $A_\text{uc}\rho(\mathbf{r}) \approx \nu  |\tilde{h}(\mathbf{r})|^2$}
\label{sec:density_modulation_approx}
For a vortexable band
\begin{equation}
    \psi_{\mathbf k}(\mathbf r)
    =
    \mathcal{N}_{\mathbf k}
    \psi_{\mathbf k}^{\mathrm{LLL}}(\mathbf r)h(\mathbf r),
    \label{eq:psi_factorized_density}
\end{equation}
Eq.~\eqref{eq:density_diag} becomes
\begin{equation}
   \rho(\mathbf r)
    =
    |h(\mathbf r)|^2
    \sum_{\mathbf k}
    \rho_{\mathbf k}
    \mathcal{N}_{\mathbf k}^2
    |\psi_{\mathbf k}^{\mathrm{LLL}}(\mathbf r)|^2 .
    \label{eq:density_product}
\end{equation}
If the Berry curvature is nearly flat, then
\(\mathcal{N}_{\mathbf k}\approx w_{\mathbf 0}^{-1/2}\), and hence
\begin{equation}
   \rho(\mathbf r)
    \simeq
    |\tilde h(\mathbf r)|^2
    \sum_{\mathbf k}
    \rho_{\mathbf k}
    |\psi_{\mathbf k}^{\mathrm{LLL}}(\mathbf r)|^2,
    \qquad
    |\tilde h(\mathbf r)|^2
    =
    \frac{|h(\mathbf r)|^2}
    {\langle |h(\mathbf r)|^2\rangle_{\mathrm{uc}}}
    =
    \frac{|h(\mathbf r)|^2}{w_{\mathbf 0}} .
    \label{eq:density_flat_geometry}
\end{equation}
% This gives a simple physical estimate of the density. 
As we showed in the main text, the interacting problem projected to a vortexable band with nearly uniform Berry curvature is (to a good approximation) equivalent to the same interaction projected to LLL up to a renormalization factor. Furthermore, in LLL, magnetic translation symmetry implies that $\rho_\mathbf{k}$ is constant.  So, we can take
\(\rho_{\mathbf k}\approx \overline{\rho}\), with
\begin{equation}
    \overline{\rho}
    =
    \frac{1}{N_{\mathrm{uc}}}
    \sum_{\mathbf k}\rho_{\mathbf k}
    =
    \frac{N_e}{N_{\mathrm{uc}}}
    =
    \nu .
\end{equation}
More over, since a filled LLL has uniform density distribution, we \(\sum_{\mathbf k}
    |\psi_{\mathbf k}^{\mathrm{LLL}}(\mathbf r)|^2
    =
    1/A_{\mathrm{uc}},\).
Plugging all these into Eq.~\eqref{eq:density_flat_geometry}, we obtain
\begin{equation}
    A_{\mathrm{uc}}\rho(\mathbf r)
    \approx
    \nu |\tilde h(\mathbf r)|^2
    =
    \nu\frac{|h(\mathbf r)|^2}{w_{\mathbf 0}} .
    \label{eq:density_simple_estimate}
\end{equation}
% In the convention \(w_{\mathbf 0}=1\), this reduces to
% \(A_{\mathrm{uc}}\rho(\mathbf r)\simeq \nu |h(\mathbf r)|^2\).
% In the actual calculation below, we do not rely on this estimate; instead, we keep the full \(\rho_{\mathbf k}\) and evaluate the density using the projected form factors.

\section{Anyon spectrum}
We also examine whether the energy enhancement persists away from the \(\nu=1/3\) Laughlin state.  For \(N_e=9\), the system with \(N_k=27\) realizes the $1/3$ Laughlin state, whereas
\begin{equation}
    N_k=3N_e-1=26,
    \qquad
    N_k=3N_e+1=28
\end{equation}
correspond to adding one quasielectron and one quasihole, respectively. The \(N_k=28\) spectrum probes the quasihole anyon with all the ground states at zero energy, which has a clean interpretation that quasiholes are zero modes of the Laughlin pseudopotential Hamiltonian.  The \(N_k=26\) spectrum probes the quasielectron anyon, which is more sensitive to microscopic details.  
\begin{figure*}[t]
    \centering
    \includegraphics[width=0.92\textwidth]{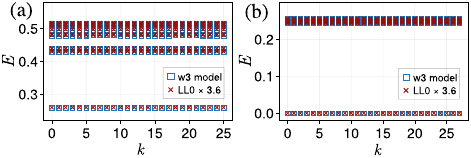}
    \caption{
    Anyon excitation spectrum $E=(E_{N_k,N_e}-E_{\text{FCI}})/\tilde{d}_s^3$ at $\tilde{d}_s=0.01$ near the \(\nu=1/3\) Laughlin state for
    \(N_e=9\). The blue squares denote the spectrum of the B3 model,
    while red crosses denote the LL0 spectrum rescaled by
    \(M_2=\sum_{\mathbf G}|\tilde w_{\mathbf G}|^2=3.6\).  Panel (a) shows
    \(N_k=26=3N_e-1\), corresponding to adding one quasielectron.  Panel (b) shows
    \(N_k=28=3N_e+1\), corresponding to adding one quasihole.  The collapse of the
    two spectra demonstrates that the same scaling effect controls both the
    neutral Laughlin spectrum and the charged anyon spectrum.
    }
    \label{fig:anyon_spectrum_scaling}
\end{figure*}

Figure~\ref{fig:anyon_spectrum_scaling} compares the charged anyon spectrum of the B3 model with the corresponding LL0 spectrum.  The LL0 energies are multiplied by the same scale used for the \(\nu=1/3\) Laughlin spectrum in Fig. 1 in the main text.
\begin{equation}
    E_{\rm model}    \simeq    M_2 E_{\rm LL0},
    \qquad
    M_2=\sum_{\mathbf G}|\tilde w_{\mathbf G}|^2 .
\end{equation}
After this single overall scaling, the agreement holds for all momentum sectors and all energy levels between the two spectra. This indicates that the enhancement acts at the level of the projected interaction itself.

\section{2-particle spectrum and pseudopotentials}\label{sec:pseudopotentials}

\begin{figure}[tbh]
    \centering
    \includegraphics[width=0.95\textwidth]{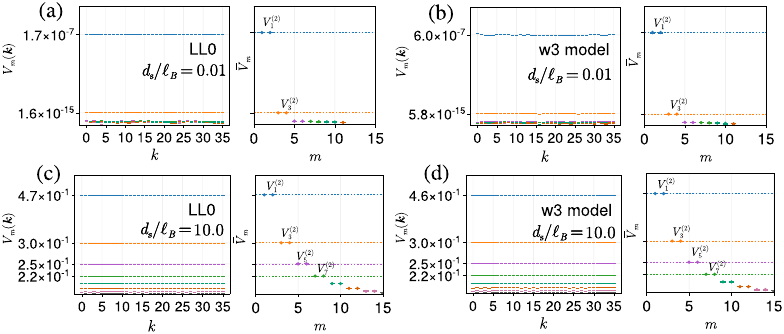}
    \caption{
    \textbf{Two-particle spectrum  diagnosis of pseudopotentials.}
    Two-particle ED spectrum for an \(N_k=6\times6=36\) cluster with \(N_e=2\).
    In each panel, the left subpanel shows the momentum-resolved ED spectrum, while the right subpanel shows the
    \(k\)-averaged energies \(\overline V_m\), representing the pseudopotentials~\cite{lauchli2013hierarchy}.  The small
    dispersion with \(k\) in the left subpanels shows that the center-of-mass splitting remains small in the
    modulated models.  Panels (a,b) use a
    short screening length \(d_s/\ell_B=0.01\), for (a) the LLL and (b) the
    w3 model.  In this limit the leading two-body pseudopotential
    \(V_1^{(2)}\) is the only dominant channel, and its renormalization is
    controlled by the \(M_2=\sum_{\mathbf G}|\tilde w_{\mathbf G}|^2\)
    factor discussed in the main text.  Panels (c,d) use
    \(d_s/\ell_B=10.0\), where several pseudopotential channels become
    comparable and the enhancement of \(V_1^{(2)}\) is
    reduced, approaching 1.}
    \label{fig:sup_fig4}
\end{figure}

To compute the Haldane pseudopotentials, we first isolate the relative-coordinate interaction from the full two-body interaction, starting from the \( \tilde V_{\mathrm{rel}}\) expression derived in the main text,
\begin{equation}
    \tilde V_{\mathrm{rel}}(\mathbf r_1-\mathbf r_2)
    =
    \frac{1}{A}
    \sum_{\mathbf q}
    V_{\mathrm{eff}}(\mathbf q)
    e^{i\mathbf q\cdot(\mathbf r_1-\mathbf r_2)},
    \qquad
    V_{\mathrm{eff}}(\mathbf q)
    =
    \sum_{\mathbf G}
    |\tilde w_{\mathbf G}|^2
    V(\mathbf q-\mathbf G).
    \label{eq:SM_compact_Veff}
\end{equation}
Thus the vortexable modulation enters the relative interaction through a reciprocal space convolution.  We now study its effect on the 
Haldane pseudopotentials in the short screening distance regime.

\subsection{Pure LL0 case}

For the screened Coulomb interaction
\begin{equation}
    V(\mathbf q)
    =
    \frac{2\pi e^2}{\epsilon |\mathbf q|}
    \tanh(d_s|\mathbf q|),
    \label{eq:SM_compact_screened_Coulomb}
\end{equation}
define dimensionless variables
\begin{equation}
    \mathbf p=\ell_B\mathbf q, \quad p=|\mathbf p|,
    \qquad
    \mathbf g=\ell_B\mathbf G,\quad g=|\mathbf g|,
    \qquad
    d=\frac{d_s}{\ell_B},
    \label{eq:SM_compact_dimensionless}
\end{equation}
and the Coulomb interaction scale \(E_C=e^2/\epsilon\ell_B.\) Then the interaction can be writen as
\begin{equation}
    V(\mathbf q)
    =
    2\pi E_C\ell_B^2\,\mathcal V(p),
    \qquad
    \mathcal V(p)
    =
    \frac{\tanh(dp)}{p}.
    \label{eq:SM_compact_dimensionless_V}
\end{equation}
Here \(\mathcal V(p)\) is the dimensionless screened interaction.  Then from Eq.~\eqref{eq:SM_compact_Veff}, the dimensionless effective interaction is
\begin{equation}
    \mathcal V_{\mathrm{eff}}(\mathbf p)
    =
    \sum_{\mathbf G}
    |\tilde w_{\mathbf G}|^2
    \frac{\tanh(d|\mathbf p-\mathbf g|)}
    {|\mathbf p-\mathbf g|}.
    \label{eq:SM_compact_dimensionless_Veff}
\end{equation}
The Haldane pseudopotentials depend only on the rotationally invariant part of
the interaction, so we first angular-average over the relative angle between
\(\mathbf p\) and each \(\mathbf g\):
\begin{equation}
    \overline{|\mathbf p-\mathbf g|^{2j}}
    =
    \sum_{\ell=0}^{j}
    \binom{j}{\ell}^{2}
    p^{2\ell}|\mathbf g|^{2(j-\ell)} .
    \label{eq:SM_compact_angular_average}
\end{equation}
We define normalized modulation moments
\begin{equation}
    M_{2}^{(r)} = \ell_B^{2r}\frac{\langle  |\partial^r h(\mathbf{r})|^2\rangle_\text{uc}}{\langle  |h(\mathbf{r})|^2\rangle_\text{uc}}
    =
    \sum_{\mathbf G}
    |\tilde w_{\mathbf G}|^2
    |\ell_B\mathbf G|^{2r},
    \label{eq:SM_compact_M2r_def}
\end{equation}
where 
\begin{equation}
    |\partial^r h(\mathbf{r})|^2 = \sum_{r_1,\dots,r_n\in \{x,y\}}|\left(\partial_{r_1}\dots\partial_{r_n}h(\mathbf{r})\right)|^2
\end{equation}
The familiar enhancement factor is \(M_2\equiv M_{2}^{(0)}\).  Using the small-\(y\) expansion
\begin{equation}
    \frac{\tanh y}{y}
    =
    1-\frac{y^2}{3}
    +\frac{2y^4}{15}
    -\frac{17y^6}{315}
    +O(y^8),
    \label{eq:SM_compact_tanh_series}
\end{equation}
we obtain
\begin{equation}
\begin{split}
    \overline{\mathcal V_{\mathrm{eff}}}( p)
    &=
    dM_{2}^{(0)}
    -
    \frac{d^3}{3}
    \left(
        M_{2}^{(0)} p^2+M_{2}^{(1)}
    \right)
    +
    \frac{2d^5}{15}
    \left(
        M_{2}^{(0)} p^4+4M_{2}^{(1)} p^2+M_{2}^{(2)}
    \right)
    \\
    &\quad
    -
    \frac{17d^7}{315}
    \left(
        M_{2}^{(0)} p^6+9M_{2}^{(1)} p^4+9M_{2}^{(2)} p^2+M_{2}^{(3)}
    \right)
    +O(d^9),
\end{split}
\label{eq:SM_compact_Veff_average_LL0}
\end{equation}

For LL0, the pseudopotential \(V_m\) is
\begin{equation}
    V_m^{\mathrm{LL0}}
    =
    E_C
    \int_0^\infty
     p\,d p\,
    \overline{\mathcal V_{\mathrm{eff}}}( p)\,
    L_m( p^2)
    e^{- p^2}.
    \label{eq:SM_compact_Vm_LL0_def}
\end{equation}
This gives the first few LL0 pseudopotentials
\begin{align}
    V_0^{\mathrm{LL0}}
    &=
    E_C
    \left[
        \frac{M_{2}^{(0)}}{2}d
        -
        \frac{M_{2}^{(0)}+M_{2}^{(1)}}{6}d^3
        +
        \frac{2M_{2}^{(0)}+4M_{2}^{(1)}+M_{2}^{(2)}}{15}d^5
    \right]
    +O(d^7),
    \label{eq:SM_compact_V0_LL0}
    \\
    V_1^{\mathrm{LL0}}
    &=
    E_C
    \left[
        \frac{M_{2}^{(0)}}{6}d^3
        -
        \frac{4(M_{2}^{(0)}+M_{2}^{(1)})}{15}d^5
    \right]
    +O(d^7),
    \label{eq:SM_compact_V1_LL0}
    \\
    V_2^{\mathrm{LL0}}
    &=
    E_C
    \left[
        \frac{2M_{2}^{(0)}}{15}d^5
    \right]
    +O(d^7),
    \label{eq:SM_compact_V2_LL0}
    \\
    V_3^{\mathrm{LL0}}
    &=
    E_C
    \left[
        \frac{17M_{2}^{(0)}}{105}d^7
    \right]
    +O(d^9).
    \label{eq:SM_compact_V3_LL0}
\end{align}
The leading fermionic Laughlin channel in Eq.~\eqref{eq:SM_compact_V1_LL0}
is the LL0 result used in the main text: the vortexable modulation
renormalizes the leading short-range \(V_1\) pseudopotential by the factor
\(M_2\equiv M_{2}^{(0)}\).  Since the \(\nu=1/3\) Laughlin state is controlled by the
fermionic \(V_1\) channel, its short-screening energy scale in LL0 inherits
this cubic scaling \(V_1\propto d_s^3\).  The next \(d^5\) correction in
Eq.~\eqref{eq:SM_compact_V1_LL0} has a negative sign, which explains why the
enhancement ratio decreases as \(d\) becomes larger.
\subsection{Multicomponent LL form factors}
\label{subsec:SM_multicomponent_LL_formfactor_compact}

We next generalize the LL0 calculation to a multicomponent LL-hybridized
form factor.  We use the orthogonal component form factor from
Eq.~\eqref{eq:MC_LL_formfactor_supp}, with weights \(t_n\ge0\) and
\(\sum_{n\in\mathcal I}t_n=1\). The LL dependence is contained only in the Laguerre factors \(L_n\!\left(
        \ell_B^2q^2/2\right)\), where \(q=|\mathbf q|\). Define
\begin{equation}
    \Lambda_t(x)
    =
    \sum_{n\in\mathcal I}
    t_nL_n\!\left(\frac{x}{2}\right),
    \qquad
    x=q^2\ell_B^2 .
    \label{eq:SM_compact_Lambda_t}
\end{equation}
Since a two-body interaction contains two density form factors, the LL factor
multiplying the effective interaction is
\begin{equation}
    F_t(x)
    =
    \Lambda_t(x)^2
    =
    \left[
        \sum_{n\in\mathcal I}
        t_nL_n\!\left(\frac{x}{2}\right)
    \right]^2 .
    \label{eq:SM_compact_Ft}
\end{equation}
The projected interaction is therefore
\begin{equation}
  V_{\mathrm{eff}}^{(t)}(\mathbf q)
    =
    F_t(q^2\ell_B^2)\,
    V_{\mathrm{eff}}(\mathbf q).
    \label{eq:SM_compact_projected_interaction_t}
\end{equation}

The corresponding pseudopotential is
\begin{equation}
    V_m^{(t)}
    =
    E_C
    \int_0^\infty
    p\,dp\,
    \overline{\mathcal V^{(t)}_{\mathrm{eff}}}(p)\,
    F_t(p^2)
    L_m(p^2)
    e^{-p^2}.
    \label{eq:SM_compact_Vm_t_def}
\end{equation}
which can be written as,
\begin{equation}
    V_m^{(t)}
    =
    \frac{E_C}{2}
    \sum_{j=0}^{\infty}
    \tau_j d^{2j+1}
    \sum_{\ell=0}^{j}
    \binom{j}{\ell}^{2}
    M_{2}^{(j-\ell)}
    \mathcal I_m^{(\ell)},
    \label{eq:SM_compact_Vm_all_order}
\end{equation}
where
\begin{equation}
    \frac{\tanh y}{y}
    =
    \sum_{j=0}^{\infty}\tau_jy^{2j},
    \qquad
    \mathcal I_m^{(\ell)}
    =
    \int_0^\infty dx\,
    e^{-x}
    x^\ell
    F_t(x)L_m(x).
    \label{eq:SM_compact_Mmell}
\end{equation}
This formula generates the hybridized-LL pseudopotentials without repeating
the LL0 calculation.

\subsection{Example results for \(V_1\) and \(V_3\)}
\label{subsec:SM_example_V1_V3}

We list the two odd channels most relevant for the discussion below.  All
results use \(M_2\equiv M_{2}^{(0)}\) and \(d=d_s\).

\textbf{LL1}

For pure LL1,
\begin{align}
    V_1
    &=
    E_C
    \left[
        \frac{M_{2}^{(0)}}{4}d^3
        -
        \frac{2(5M_{2}^{(0)}+3M_{2}^{(1)})}{15}d^5
    \right]
    +O(d^7),
    \label{eq:SM_compact_V1_LL1}
    \\
    V_3
    &=
    E_C
    \left[
        \frac{M_{2}^{(0)}}{4}d^3
        -
        \frac{2(3M_{2}^{(0)}+M_{2}^{(1)})}{5}d^5
    \right]
    +O(d^7).
    \label{eq:SM_compact_V3_LL1}
\end{align}

\textbf{Equal LL01 hybridization}

For the equal LL01 hybridization case, \(t_0=t_1=1/2,\) the corresponding odd pseudopotentials are
\begin{align}
    V_1
    &=
    E_C
    \left[
        \frac{M_{2}^{(0)}}{8}d
        +
        \left(
            \frac{M_{2}^{(0)}}{48}
            -
            \frac{M_{2}^{(1)}}{24}
        \right)d^3
        -
        \left(
            \frac{M_{2}^{(0)}}{15}
            +
            \frac{M_{2}^{(1)}}{30}
            -
            \frac{M_{2}^{(2)}}{60}
        \right)d^5
    \right]
    +O(d^7),
    \label{eq:SM_compact_V1_LL01_half}
    \\
    V_3
    &=
    E_C
    \left[
        \frac{M_{2}^{(0)}}{16}d^3
        -
        \left(
            \frac{M_{2}^{(0)}}{5}
            +
            \frac{M_{2}^{(1)}}{10}
        \right)d^5
    \right]
    +O(d^7).
    \label{eq:SM_compact_V3_LL01_half}
\end{align}

\textbf{Equal LL12 hybridization}

For \(t_1=t_2=1/2,\) the two odd channels are
\begin{align}
    V_1
    &=
    E_C
    \left[
        \frac{M_{2}^{(0)}}{32}d
        +
        \frac{23M_{2}^{(0)}-2M_{2}^{(1)}}{192}d^3
        -
        \frac{58M_{2}^{(0)}+46M_{2}^{(1)}-M_{2}^{(2)}}{240}d^5
    \right]
    +O(d^7),
    \label{eq:SM_compact_V1_LL12}
    \\
    V_3
    &=
    E_C
    \left[
        \frac{3M_{2}^{(0)}}{32}d
        -
        \frac{M_{2}^{(0)}+M_{2}^{(1)}}{32}d^3
        +
        \frac{-4M_{2}^{(0)}+4M_{2}^{(1)}+M_{2}^{(2)}}{80}d^5
    \right]
    +O(d^7).
    \label{eq:SM_compact_V3_LL12}
\end{align}

\begin{figure}[tbh]
    \centering
    \includegraphics[width=0.78\textwidth]{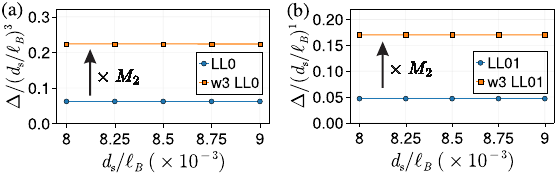}
    \caption{
    \textbf{Small-\(d_{\mathrm{s}}\) scaling of the \(\nu=1/3\) Laughlin gap.}
    The ED gap \(\Delta\) is divided by the expected leading power of
    \(d=d_{\mathrm{s}}/\ell_B\).  A nearly constant curve therefore directly confirms the
    small-\(d_{\mathrm{s}}\) power law.  (a) For the unmodulated LL0 and the
    w3 model-modulated LL0, we plot \(\Delta/d^3\).  The flat behavior is
    consistent with the leading pseudopotential
    \(V_1^{\mathrm{LL0}}\propto M_{2} d^3\) in
    Eq.~\eqref{eq:SM_compact_V1_LL0}.  (b) For the equal LL01 hybridized model (with \(\psi_{\mathbf k}^{\mathrm{MC}}(\mathbf r)
    =
    \left\{
    \sqrt{0.5}\,\psi_{0\mathbf k}(\mathbf r),
    \sqrt{0.5}\,\psi_{1\mathbf k}(\mathbf r)
    \right\}\))
    and its w3 model-modulated counterpart, \(\Delta/d\) follows the linear leading term
    \(V_1^{\mathrm{LL01}}\propto M_{2} d\) in
    Eq.~\eqref{eq:SM_compact_V1_LL01_half}.  In both panels the w3
    modulation preserves the small-\(d_{\mathrm{s}}\) exponent but increases the
    prefactor, as indicated by the \(M_2\) enhancement.
    }
    \label{fig:sup_fig5}
\end{figure}

Figure~\ref{fig:sup_fig5} presents a gap--\(d_{\mathrm{s}}\) scaling analysis as a test for the pseudopotentials in small-\(d_{\mathrm{s}}\) regime. We divide out the expected leading power of the pseudopotentials the ED gap.  For the
\(\nu=1/3\) Laughlin state, the relevant pseudopotential is the
\(V_1\) pseudopotential.  In the LL0 case,
Eq.~\eqref{eq:SM_compact_V1_LL0} gives
\(V_1^{\mathrm{LL0}}\sim M_{2} d^3\), so \(\Delta/d^3\) is approximately
constant in panel (a).  For the equal LL01 hybridized band,
Eq.~\eqref{eq:SM_compact_V1_LL01_half} gives
\(V_1^{\mathrm{LL01}}\sim M_{2} d\), so \(\Delta/d\) is approximately constant
in panel (b).  The w3 model-modulated data follow the same power scaling, but with enhancement of the \(M_2\) factor, which is the renormalization factor of the dominant short-range pseudopotential.

\begin{figure}[tbh]
    \centering
    \includegraphics[width=0.95\textwidth]{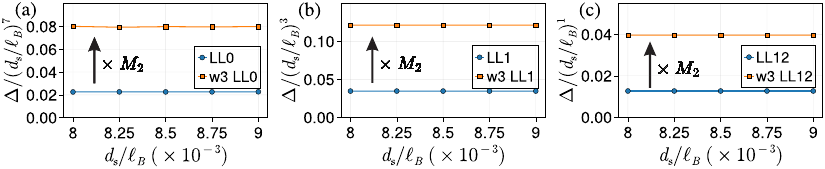}
    \caption{
    \textbf{Small-\(d_{\mathrm{s}}\) scaling of the \(\nu=1/5\) Laughlin gap.}
    The ED gap \(\Delta\) is divided by the leading power of
    \(d=d_{\mathrm{s}}/\ell_B\) expected from the leading pseudopotential.
    (a) For LL0 and w3 model-modulated LL0, we plot \(\Delta/d^7\), consistent
    with the leading behavior
    \(V_3^{\mathrm{LL0}}\propto M_{2} d^7\) in
    Eq.~\eqref{eq:SM_compact_V3_LL0}.  (b) For LL1 and w3 model-modulated LL1,
    we plot \(\Delta/d^3\), consistent with
    \(V_3^{\mathrm{LL1}}\propto M_{2} d^3\) in
    Eq.~\eqref{eq:SM_compact_V3_LL1}.  (c) For the equal LL12 hybridized model (with \(\psi_{\mathbf k}^{\mathrm{MC}}(\mathbf r)
    =
    \left\{
    \sqrt{0.5}\,\psi_{1\mathbf k}(\mathbf r),
    \sqrt{0.5}\,\psi_{2\mathbf k}(\mathbf r)
    \right\}\))
    and its w3 model-modulated counterpart, we plot \(\Delta/d\), consistent
    with the linear leading term
    \(V_3^{\mathrm{LL12}}\propto M_{2} d\) in
    Eq.~\eqref{eq:SM_compact_V3_LL12}.  The w3 modulation gives gap enhancement factor \(M_2\).
    }
    \label{fig:sup_fig6}
\end{figure}

Figure~\ref{fig:sup_fig6} shows the corresponding gap-\(d_\mathrm{s}\) scaling test for
the \(\nu=1/5\) Laughlin state.  In this case the leading relevant
pseudopotential is
\(V_3\).  For LL0,
Eq.~\eqref{eq:SM_compact_V3_LL0} gives
\(V_3^{\mathrm{LL0}}\sim M_{2} d^7\), so panel (a) shows a nearly constant
\(\Delta/d^7\).  For pure LL1,
Eq.~\eqref{eq:SM_compact_V3_LL1} has leading power
\(V_3^{\mathrm{LL1}}\sim M_{2} d^3\), giving the flat \(\Delta/d^3\) behavior
in panel (b).  For equal LL12 hybridization,
Eq.~\eqref{eq:SM_compact_V3_LL12} gives
\(V_3^{\mathrm{LL12}}\sim M_{2} d\), so panel (c) shows an approximately
constant \(\Delta/d\).  The w3 model modulation preserves these powers while
enhancing the gap by \(M_2\).

\section{Effect of LL hybridization and screening length}

In this section, we test the robustness of the \(\nu=1/3\) FCI gap enhancement by varying both the screening length \(d_s\) and the amount of Landau level hybridization.  For the multicomponent vortexable bands, we use
\begin{equation}
\psi^\text{MC}_\mathbf{k} = \{\sqrt{t}\mathcal{N}_{1\mathbf{k}}\psi_\mathbf{k}^\text{LL1}(\mathbf{r}),\sqrt{1-t}\mathcal{N}_{0\mathbf{k}}\psi_\mathbf{k}^\text{LLL}(\mathbf{r})\}h(\mathbf{r}),
\label{eq:MC_LL01_wavefunction}
\end{equation}
where the parameter \(t\) tunes between a LLL-like vortexable band at \(t=0\), an equal LL0/LL1 mixture at \(t=1/2\), denoted as LL01 below, and a LL1-like higher vortexable band at \(t=1\).

\begin{figure}[tbh]
    \centering
\includegraphics[width=0.8\textwidth]{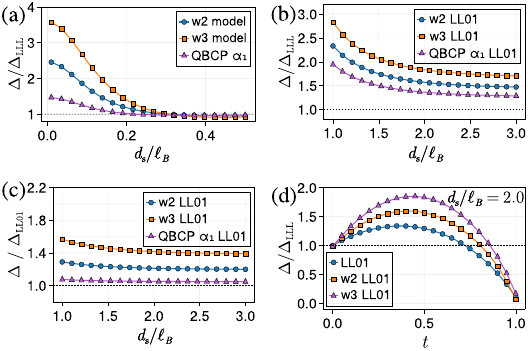}
\caption{
    \textbf{Screening length and LL-hybridization dependence of the \(\nu=1/3\) FCI gap enhancement.}
    (a) Gap ratio \(\Delta/\Delta_{\mathrm{LLL}}\) for vortexable bands as a function of the screening length \(d_s/\ell_B\).  The dashed line denotes the LLL reference.  In the short screening length limit, the enhancement approaches the factor
    \(M_2=\sum_{\mathbf G}|\tilde w_{\mathbf G}|^2\), while for large \(d_s\) the ratio approaches unity.
    (b) The same ratio for multicomponent LL01 bands, corresponding to \(t=1/2\) in Eq.~\eqref{eq:MC_LL01_wavefunction}.  LL01 hybridization makes the enhancement persist for much larger \(d_s/\ell_B\).
    (c) The LL01 data normalized by the unmodulated LL01 gap, \(\Delta_{\mathrm{LL01}}\), which isolates the additional enhancement due to the vortexable modulation \(h(\mathbf r)\) beyond the baseline enhancement from LL mixing.
    (d) Gap ratio at fixed \(d_s/\ell_B=2.0\) as a function of the LL0/LL1 hybridization weight \(t\).  The enhancement is largest near intermediate LL01 hybridization and is further amplified by the w2 and w3 model modulations.}
\label{fig:sup_fig7}
\end{figure}

Throughout Fig.~\ref{fig:sup_fig7}, the many-body ground state remains in the \(\nu=1/3\) FCI phase, so the plotted quantity directly measures the enhancement of the FCI gap under changes of screening length and LL0/LL1 hybridization.  Panel (a) shows that, for vortexable bands without LL01 hybridization, the gap enhancement is strongest in the short screening length limit.  In this regime the interaction is dominated by the $V_1$ pseudopotential and the effect of the modulation is to rescale the pseudopotential by the factor
\begin{equation}
M_2
=
\sum_{\mathbf G}|\tilde w_{\mathbf G}|^2
\label{eq:M2_enhancement}
\end{equation}
Accordingly, \(\Delta/\Delta_{\mathrm{LLL}}\) approaches \(M_2\) as \(d_s/\ell_B\to 0\).  For larger \(d_s/\ell_B\), the interaction has ~$1/|\mathbf q-\mathbf G|$ suppression and becomes less sensitive to the modulation encoded by the nonzero \(\mathbf G\) components of \(h(\mathbf r)\), and the gap ratio gradually returns to the LLL value.  Panels (b) and (c) demonstrate that LL01 hybridization changes this behavior qualitatively: the hybridizated LL0/LL1 form factor reshapes the |\(V_1\) pseudopotential depnedence on $d_s$, allowing the enhancement to persist to much larger \(d_s/\ell_B\).  After normalizing by the unmodulated LL01 gap in panel (c), the w2, w3 model, and QBCP data remain above unity, showing that the modulation provides an additional enhancement on top of the baseline LL01-hybridization effect.  Finally, panel (d) shows that the gap enhancement is optimized near intermediate LL0/LL1 hybridization weight \(t\) \cite{zhang2025beyond}. The larger enhancement of the w2 and w3 models compared with the unmodulated LL01 case supports the physical picture that more robust FCI gaps can be engineered by combining real-space modulation with controlled higher vortexability.

\FloatBarrier

\section{No upper bound for the enhancement factor $M_n$}
\label{sec:M2_no_upper_bound}

In this section, we show that the two-body interaction enhancement $M_2$ has no
cutoff-independent upper bound. More precisely, we construct admissible Fourier
coefficients whose enhancement grows quadratically with the shell cutoff $S$.
A similar conclusion can be generalized to the $n$-body interaction case.

Recall
\begin{equation}
|\tilde{h}(\mathbf r)|^2
=
\sum_{\mathbf G}\tilde w_{\mathbf G} e^{i\mathbf G\cdot \mathbf r},
\qquad
\tilde w_{\mathbf 0}=1,
\qquad
|\tilde{h}(\mathbf r)|^2\ge 0,
\end{equation}
and the two-body enhancement factor
\begin{equation}
M_2=\sum_{\mathbf G}|\tilde w_{\mathbf G}|^2 .
\end{equation}
For the explicit construction below, the Fourier coefficients are real, so this
reduces to \(M_2=\sum_{\mathbf G}\tilde w_{\mathbf G}^2\).

For
\begin{equation}
\mathbf G=n\mathbf G_1+m\mathbf G_2,
\end{equation}
we define the shell index
\begin{equation}
q(\mathbf G)\equiv q(n,m)=n^2+m^2-nm .
\end{equation}
This measures the squared distance of \(\mathbf G\) from the origin in the
triangular reciprocal lattice. We impose the finite-shell cutoff
\begin{equation}
\tilde w_{\mathbf G}=0
\qquad \text{for} \qquad
q(\mathbf G)>S^2,
\end{equation}
and the first-shell constraint
\begin{equation}
\tilde w_{\mathbf G}=0
\qquad \text{for} \qquad
q(\mathbf G)=1.
\end{equation}
In the setting considered here, this first-shell condition is the condition for
the flat-normalization limit \(\mathcal{N}_{\mathbf k}=1\).

We now want an explicit positive modulation \(|\tilde h(\mathbf r)|^2\)
whose Fourier weight is spread over many reciprocal-lattice vectors, while the
zeroth Fourier coefficient remains fixed and the first shell is absent.  A
natural way to build such a function is to start from a nonnegative Fourier polynomial with unit average.  This leads us to the Fej\'er kernel with a cutoff $N$. 
\begin{equation}
F_N(\theta)
=
\sum_{k=-(N-1)}^{N-1}
\left(1-\frac{|k|}{N}\right)e^{ik\theta}
=
\frac{1}{N}
\left|
\sum_{j=0}^{N-1}e^{ij\theta}
\right|^2 .
\end{equation}
It is nonnegative for all \(\theta\) and has unit zeroth Fourier coefficient.
Let
\begin{equation}
u=\mathbf G_1\cdot \mathbf r,
\qquad
v=\mathbf G_2\cdot \mathbf r.
\end{equation}
We choose the trial function $f$ as
\begin{equation}
f_N(\mathbf r)
=
f_N(u,v)
=
F_N(3u)F_N(3v) \geq 0.
\label{eq:trialfN}
\end{equation}
Moreover,
\(f_N\) can be written explicitly as a modulus square,
\begin{equation}
f_N(u,v)
=
\left|
\frac{1}{N}
\sum_{j,\ell=0}^{N-1}
e^{3i(ju+\ell v)}
\right|^2,
\end{equation}
so it is an admissible choice of \(|h(\mathbf r)|^2\).

Expanding Eq.~\eqref{eq:trialfN}, we get
\begin{equation}
f_N(u,v)
=
\sum_{a,b=-(N-1)}^{N-1}
\left(1-\frac{|a|}{N}\right)
\left(1-\frac{|b|}{N}\right)
e^{i(3au+3bv)}.
\end{equation}
Therefore the only nonzero Fourier coefficients occur at
\begin{equation}
(n,m)=(3a,3b),
\end{equation}
with
\begin{equation}
\tilde w^{(N)}_{3a,3b}
=
\left(1-\frac{|a|}{N}\right)
\left(1-\frac{|b|}{N}\right),
\qquad
|a|,|b|\le N-1,
\end{equation}
and all other coefficients vanish. In particular,
\begin{equation}
\tilde w^{(N)}_{\mathbf 0}=1, \qquad
\left.\tilde w^{(N)}_{\mathbf G}\right|_{q(\mathbf G)=1}=0.
\end{equation}
All the constraints are automatically satisfied.

We now relate the cutoff parameter \(N\) to the shell cutoff \(S\). For the
nonzero coefficients,
\begin{equation}
q(n,m)
=
q(3a,3b)
=
9(a^2+b^2-ab).
\end{equation}
Since \(|a|,|b|\le N-1\),
\begin{equation}
a^2+b^2-ab\le 3(N-1)^2.
\end{equation}
Thus
\begin{equation}
q(n,m)\le 27(N-1)^2.
\end{equation}
Hence all nonzero Fourier coefficients lie inside the cutoff \(q\le S^2\)
provided
\begin{equation}
27(N-1)^2\le S^2.
\end{equation}
A convenient admissible choice is
\begin{equation}
N
=
\left\lfloor
\frac{S}{3\sqrt{3}}
\right\rfloor
+1.
\label{eq:NchoiceSM}
\end{equation}

For this construction, the achieved value of \(M_2\) is
\begin{equation}
M_2[f_N]
=
\sum_{a,b=-(N-1)}^{N-1}
\left(1-\frac{|a|}{N}\right)^2
\left(1-\frac{|b|}{N}\right)^2
\nonumber
=
\left(\frac{2N^2+1}{3N}\right)^2.
\end{equation}
% The one-dimensional sum is
% \begin{align}
% \sum_{a=-(N-1)}^{N-1}
% \left(1-\frac{|a|}{N}\right)^2
% &=
% 1+2\sum_{a=1}^{N-1}
% \left(1-\frac{a}{N}\right)^2
% \nonumber\\
% &=
% 1+\frac{2}{N^2}\sum_{\ell=1}^{N-1}\ell^2
% \nonumber\\
% &=
% \frac{2N^2+1}{3N}.
% \end{align}
% Therefore
% \begin{equation}
% M_2[f_N]
% =
% \left(\frac{2N^2+1}{3N}\right)^2.
% \label{eq:exacttrialvalue}
% \end{equation}
At large \(N\),
\begin{equation}
M_2[f_N]
\sim
\frac{4}{9}N^2.
\end{equation}
Using Eq.~\eqref{eq:NchoiceSM}, this gives
\begin{equation}
M_2[f_N]
=
\frac{4}{243}S^2+O(S).
\label{eq:lowerboundS}
\end{equation}
Thus the enhancement can grow at least quadratically with the shell cutoff.

If exact \(C_3\) or \(C_6\) symmetry is required, one may average \(f_N\) over \(C_3\) or \(C_6\) rotations; this preserves all constraints and does not change the \(\Omega(S^2)\) scaling.

Finally, we show that the enhancement cannot grow faster than quadratically.
For any admissible \(f\ge 0\) with \(\tilde w_{\mathbf 0}=1\),
\begin{equation}
|\tilde w_{n,m}|
=
\left|
\frac{1}{(2\pi)^2}
\int_0^{2\pi}\!\!\int_0^{2\pi}
f(u,v)e^{-i(nu+mv)}\,du\,dv
\right|
\nonumber \le
\frac{1}{(2\pi)^2}
\int_0^{2\pi}\!\!\int_0^{2\pi}
f(u,v)\,du\,dv
=
\tilde w_{0,0}
=
1.
\end{equation}
Hence \(|\tilde w_{n,m}|^2\le 1\). So \(M_2\le \#\{(n,m)\in\mathbb Z^2:q(n,m)\le S^2\}=O(S^2)\).

Consequently, the two-body enhancement factor $M_2$ is bounded for every fixed finite cutoff
\(S\), but it has no universal upper bound independent of \(S\).

\section{$\text{tMoTe}_2$ discussion}

In this section, we first provide details of single particle wave functions and ideal band approximation of the top valence band of tMoTe$_2$ at twist angle $3.7^\circ$. Then we show supplement the many-body spectra results of the main text by providing the same in the limit $d_s\ll a$. We conclude the section by discussing the difference between two hole filling fractions $\nu_h =2/3$ and $\nu_h=1/3$.

\subsection{Ideal band approximation}

\begin{figure}[tbh]
    \centering
    \includegraphics[width=0.85\textwidth]{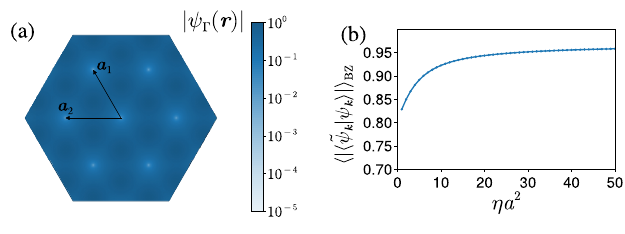}
    \caption{\textbf{Ideal band approximation for the top valence band of \(\mathrm{tMoTe}_2\) at
    \(\theta=3.7^\circ\).}
    (a) Real space magnitude \(|\psi_\Gamma(\mathbf r)|\) of the continuum model Bloch wave function, showing a near zero at \(\mathbf r=0\).  This motivates introducing the rotationally symmetric
    regulator \(f_\eta(\mathbf r)=1-e^{-\eta|\mathbf r|^2}\) before extracting the ideal-band modulation
    \(h(\mathbf r)\).  (b) Brillouin zone averaged overlap
    \(\langle |\langle \tilde{\psi}_{\mathbf{k}}|\psi_{\mathbf{k}}\rangle| \rangle_{\mathrm{BZ}}\) between the
    tMoTe$_2$ continuum wave functions and the ideal band approximation as a function of \(\eta a^2\).  The high overlap
    at large \(\eta\) shows that the approximation preserves the tMoTe$_2$ wave functions well.
    }
    \label{fig:sup_tmote2_a}
\end{figure}

In the main text, using the continuum moir\'e Hamiltonian and parameters from~\cite{wang2024fractional},  we reported that the the top valence band of tMoTe$_2$ at twist angle $3.7^\circ$ is nearly ideal with  $\langle\text{tr}(g)\rangle_\text{BZ}/\langle F_{xy}\rangle_\text{BZ}\approx 1.15$. Here we describe the procedure we employed to approximate the band to an ideal band and extract the Fourier coefficients $\tilde{w}_\mathbf{G}$. 

We mentioned before that the Bloch wavefunctions of any ideal band with Chern number $|C|=1$ can be written as 
\begin{equation}
    \psi_\mathbf{k}(\mathbf{r}) = \mathcal{N}_\mathbf{k} \psi^\text{LLL}_\mathbf{k}(\mathbf{r})h(\mathbf{r}),
\end{equation}
where $\psi^\text{LLL}_\mathbf{k}(\mathbf{r})$ are the LLL Bloch wavefunction on a torus, and $\mathcal{N}_\mathbf{k}$ is the normalization factor. For an ideal band, if we know the Bloch functions $\psi_\mathbf{k}(\mathbf{r})$ numerically, we can extract $h(\mathbf{r})$ by evaluating $\psi_\mathbf{k}(\mathbf{r})/\psi_\mathbf{k}^\text{LLL}(\mathbf{r})$. To write an explicit expression of $\psi^\text{LLL}_\mathbf{k}(\mathbf{r})$, we choose Landau gauge for the vector potential $\mathbf{A}(\mathbf{r}) = B_0 (\hat{G}_2\cdot (\mathbf{r}-\mathbf{r}_0))(\hat{z}\times\hat{G}_2)$, where $\mathbf{r}_0$ is an arbitrary constant vector, and $\hat{G}_2 = \mathbf{G}_2/|\mathbf{G}_2|$ is a unit vector along the reciprocal lattice vector $\mathbf{G}_2$ ($\mathbf{G}_i$ are defined in Eq.~\eqref{eq:Gdefs}). In this gauge, $\psi^\text{LLL}_\mathbf{k}(\mathbf{r})$ can be written as
\begin{equation}
    \psi^\text{LLL}_\mathbf{k}(\mathbf{r}) = e^{i(\mathbf{k}\cdot\mathbf{a}_1)z/a_1}\vartheta_1\left(\left.\frac{z-z_0+ik\ell_B^2}{a_1}\right|\tau\right) e^{-\frac{1}{2\ell_B^2}(\hat{G}_2\cdot (\mathbf{r}-\mathbf{r}_0))^2},
\end{equation}
where $\mathbf{a}_i$ are the lattice vectors corresponding to reciprocal lattice vectors $\mathbf{G}_i$ in Eq.~\eqref{eq:Gdefs}, $a_i = (\mathbf{a}_i)_x+i(\mathbf{a}_i)_y$ are the complexified primitive moir\'e lattice vectors, $\tau=a_2/a_1$, $k = k_x+ik_y$, and $\theta_1$ is the Jacobi theta function of the first type $\vartheta(z|\tau) = \sum_{n=-\infty}^\infty e^{\pi i \tau(n+1/2)^2+\pi i(2n+1)z}$~\cite{ledwith2020fractional,sarkar2025unconventional}. Importantly, $\psi^\text{LLL}_\mathbf{k}(\mathbf{r})$ has a zero at position $n_1\mathbf{a}_1+n_2\mathbf{a}_2+\mathbf{r}_0 - \hat{z}\times \mathbf{k}\ell_B^2$ in every unit cell. Hence, any ideal band wavefunction must also have a zero in every unit cell. Conversely, for an ideal band, by numerically evaluating $\psi_\mathbf{k}(\mathbf{r})$, if we find zeros at $n_1\mathbf{a}_1+n_2\mathbf{a}_2+\bar{\mathbf{r}}_0(\mathbf{k})$, we can choose $\mathbf{r}_0 = \bar{\mathbf{r}}_0(\mathbf{k})+\hat{z}\times \mathbf{k}\ell_B^2$ such that $\psi_\mathbf{k}^\text{LLL}(\mathbf{r})$ also has zeros at $n_1\mathbf{a}_1+n_2\mathbf{a}_2+\bar{\mathbf{r}}_0(\mathbf{k})$, and we can evaluate $h(\mathbf{r})$ (without blowing up somewhere).

Since the top valence band of tMoTe$_2$ at twist angle $3.7^\circ$ is not exactly ideal, its Bloch wavefunctions generically do not have exact zeros in the unit cell. However, since the band is nearly vortexable, the wavefunctions have near zeros. For example, as shown in Fig.~\ref{fig:sup_tmote2_a}(a), we show $|\psi_\Gamma(\mathbf{r})|$ which is very close to zero at the center of the unit cell $\mathbf{r} = 0$ (which is the $\mathcal{R}^M_M$ stacking position~\cite{wu2019topological}). Hence, we can multiply it by a regulator $f_\eta(\mathbf{r})$ which vanishes at $\mathbf{r} = 0$ but is close to 1 everywhere else. Also requiring rotational symmetry, $f_\eta(\mathbf{r})$ can only depend on $|\mathbf{r}|$. One such choice is $f_\eta(\mathbf{r}) = (1-e^{-\eta|\mathbf{r}|^2)}$ with some constant $\eta$. Approximating $\psi_\Gamma(\mathbf{r})$ to $\tilde{\psi}_\Gamma(\mathbf{r}) =\psi_\Gamma(\mathbf{r})f_\eta(\mathbf{r})$, we get $h(\mathbf{r}) = \tilde{\psi}_\Gamma(\mathbf{r})/\tilde{\psi}_\Gamma^\text{LLL}(\mathbf{r})$ (with $\mathbf{r}_0 = 0$). Then the ideal band approximated wavefunctions would be
\begin{equation}
    \tilde{\psi}_\mathbf{k}(\mathbf{r}) = \mathcal{N}_\mathbf{k} \psi^\text{LLL}_\mathbf{k}(\mathbf{r})h(\mathbf{r}) = \mathcal{N}_\mathbf{k}e^{i(\mathbf{k}\cdot\mathbf{a}_1)z/a_1}\vartheta_1\left(\left.\frac{z+ik\ell_B^2}{a_1}\right|\tau\right) \frac{\tilde{\psi}_\Gamma(\mathbf{r})}{\vartheta_1\left(\left.z/a_1\right|\tau\right)}.
\end{equation}
Note that this ideal band approximation scheme was originally proposed for graphene monolayer under spatially inhomogeneous strain field~\cite{gao2023untwisting}. Next, to examine how close the approximated wavefunctions are to the original wavefunction, we evaluate absolute value of their overlap averaged over the BZ for a range of values of $\eta$ as shown in Fig.~\ref{fig:sup_tmote2_a}(b). We find that $\eta a^2>10$ the average overlap is larger than $0.90$, which indicates that this a good approximation. Finally, $\tilde{w}_\mathbf{G}$'s reported in Eq.~\eqref{eq:wG_model_inputs} were extracted expanding $|h(\mathbf{r})|^2 = |\tilde{\psi}_\Gamma(\mathbf{r})/\tilde{\psi}_\Gamma^\text{LLL}(\mathbf{r})|^2$ in Fourier series for $\eta = 500$, when $\langle |\langle \tilde{\psi}_{\mathbf{k}}|\psi_{\mathbf{k}}\rangle| \rangle_{\mathrm{BZ}}=0.967$.

\subsection{Many-body spectrum for very short range interaction}

\begin{figure}[h]
    \centering
    \includegraphics[width=0.85\textwidth]{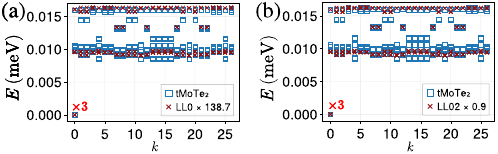}
    \caption{
    \textbf{ED spectrum of \(\mathrm{tMoTe}_2\) in the short-range interaction regime.}
    Energy spectrum at \(\nu_h=1/3\) hole filling for the continuum
    \(\mathrm{tMoTe}_2\) band with \(d_s/a=0.01\) and dielectric constant
    \(\epsilon=15\).  The single-particle dispersion is not included, so the
    spectrum reflects only the projected interaction.  (a) Comparison between
    the \(\mathrm{tMoTe}_2\) spectrum and the pure LL0 spectrum.  The LL0
    spectrum is multiplied by a factor of 138.7, demonstrating the large enhancement of the interaction scale in
    \(\mathrm{tMoTe}_2\) relative to pure LL0 at short screening length.
    (b) Comparison with a hybridized LL02 model with  \(
    \psi_{\mathbf k}^{\mathrm{MC}}(\mathbf r)
    =
    \left\{
    \sqrt{0.96175}\,\psi_{0\mathbf k}(\mathbf r),
    \sqrt{0.03825}\,\psi_{2\mathbf k}(\mathbf r)
    \right\}
    \),
    chosen such that
    \(\langle \mathrm{tr}\,g\rangle_{\mathrm{tMoTe}_2}
    =
    \langle \mathrm{tr}\,g\rangle_{\mathrm{LL02}}\).
    The close agreement between the \(\mathrm{tMoTe}_2\) and LL02 spectra
    indicates that the short-range enhancement is well captured by the
    LL-hybridization effect, which strongly increases the relevant
    pseudopotentials in the short-range interaction regime.
    }
    \label{fig:sup_tmote2_b}
\end{figure}

Here we discuss the many-body spectra at $\nu_h=1/3$ filling in short range interaction limit $d_s\ll a$ ($a$ is the moir\'e lattice constant). To isolate the effect of band wavefunctions (via the formfactors) on stability of the FCI ground state, here we do not include the band dispersion. In Fig.~\ref{fig:sup_tmote2_b}(a), we compare the many-body spectra of tMoTe$_2$ at $\nu_h=1/3$ with that of LLL at $\nu=1/3$ at $d_s/a = 0.01$. {\it Remarkably, we find that the charge neutral gap of the FCI ground state in tMoTe$_2$ is approximately $138$ times larger than that in the LLL.} Since $M_2\approx 1.13$ for this band (Eq.~\eqref{eq:h_fourier_supp}), $M_2$ does not explain the this large gap enhancement. To understand the origin of this large gap enhancement over LLL, we examine the scaling of the charge neutral gap in tMoTe$_2$ as a function of $d_s$ in $d_s\ll a$ limit, and find that the charge neutral gap is $\propto d_s$ unlike LLL where the charge neutral gap is $\propto d_s^3$. This clearly indicates that the contact potential $d_s \delta(\mathbf{r})$ part of the short range Coulomb interaction $\text{tanh}(d_sq)/q$ is the main contributor to the charge neutral gap in tMoTe$_2$. As we learned from Sec.~\ref{sec:pseudopotentials}, the contact potential $\delta(\mathbf{r})$ part contributes to the effective $V_1$ pseudopotenital in multicomponent higher vortexable bands. Also, recall that for the top valence band of tMoTe$_2$ at twist angle $3.7^\circ$,  $\langle\text{tr}(g)\rangle_\text{BZ}/\langle F_{xy}\rangle_\text{BZ}\approx 1.15$, which is consistent with multicomponent hybridized LL bands that also have $\langle\text{tr}(g)\rangle_\text{BZ}/\langle F_{xy}\rangle_\text{BZ}>1$ (as described in Sec.~\ref{sec:llhybridization}). However, there are inifintely many choices of hybridized LL bands that can have the same $\langle\text{tr}(g)\rangle_\text{BZ}/\langle F_{xy}\rangle_\text{BZ}$. Hence, we wish to know which LL hybridized band the tMoTe$_2$ top valence band resemble. As shown in Fig.~\ref{fig:sup_tmote2_b}(b), we numerically find that the many-body spectrum of a hybridized LL with zeroth and second LL closely resemble the spectrum of tMoTe$_2$ nearly identical gap enhancement over LLL.

\subsection{Filling fractions $\nu_h = 2/3$ and $\nu_h=1/3$}

\begin{figure}[tbh]
    \centering
    \includegraphics[width=0.98\textwidth]{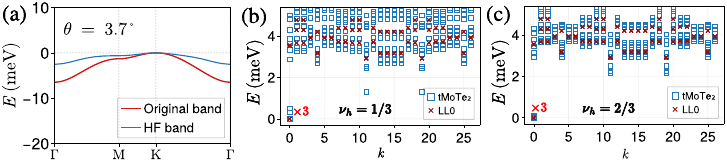}
    \caption{
    \textbf{Particle-hole/Hartree-Fock effective band and projected ED spectra.}
    (a) Single-particle dispersion of the top continuum model band of twisted MoTe\(_2\) at
    \(\theta=3.7^\circ\).  The red curve is the original electron band \(E(\mathbf k)\), which enters the
    hole-basis ED Hamiltonian as \(-E(\mathbf k)\).  The blue curve is the PH/HF effective band
    \(E(\mathbf{k})-\epsilon_H(\mathbf{k})+\epsilon_F(\mathbf{k})\), computed at \(d_s=30\,\mathrm{nm}\) and
    \(\epsilon=15.0\), and shifted by an irrelevant constant.  This blue band is not used directly in ED;
    it is shown to diagnose the PH-transformed effective band.  (b) ED spectrum at \(\nu_h=1/3\), computed
    with \(H_h[-E]\).  (c) ED spectrum at \(\nu_h=2/3\), also computed with \(H_h[-E]\).  Blue squares denote
    the tMoTe\(_2\) spectrum, and red crosses denote the LL0 reference spectrum.  The \(\nu_h=2/3\) spectrum
    shows better agreement with LL0 and a larger many-body gap in comparison to \(\nu_h=1/3\) spectrum.
    }
    \label{fig:sup_tmote2_c}
\end{figure}

In experiments on twisted MoTe\(_2\), the FCI is most prominent near \(\nu_h=2/3\), while the FCI is weaker at 
\(\nu_h=1/3\)~\cite{pan2026optical}. In qualitative agreement with this, we also find that the FCI charge neutral gap is smaller at $\nu_h =1/3$ than at $\nu_h=2/3$ in ED as shown in Figs.~\ref{fig:sup_tmote2_c}(b,c) for experimentally relevant parameters. On the other hand, in our theoretical analysis of the interaction Hamiltonian in the main text, we found that as long as the Berry curvature of the band is reasonably uniform and the band is close to ideal, the many-body problem is approximately equivalent to the many-body problem in LLL at $d_s \gtrsim a$ regardless of the filling fraction. Since LLL is particle-hole symmetric, one would then expect \(\nu_h=2/3\) and \(\nu_h=1/3\) to have similar stability, which is in contradiction with experiments. In this section, we show that the key to explaining this apparent contradiction is the band dispersion (which was not taken into account in the theoretical analysis) and the small (but not negligible with respect to the band width and interaction scale) particle-hole asymmetry of the tMoTe$_2$ top valence band as explained below.

The standard projected hole Hamiltonian is
\begin{equation}
    H_h[E,\tilde\gamma]
    =
    -\sum_{\mathbf k}
    E(\mathbf k)\,
    \tilde\gamma_{\mathbf k}^{\dagger}\tilde\gamma_{\mathbf k}
    + \frac{1}{2A} \sum_{\mathbf{k}_1\mathbf{k}_2\mathbf{k}_3\mathbf{k}_4}^\text{BZ} \tilde\gamma^\dagger_{\mathbf{k}_1}\tilde\gamma^\dagger_{\mathbf{k}_3}\tilde\gamma_{\mathbf{k}_4}\tilde\gamma_{\mathbf{k}_2}\sum_\mathbf{q}V(\mathbf{q})\,\lambda_\mathbf{q}(\mathbf{k}_1,\mathbf{k}_2)\lambda_{-\mathbf{q}}(\mathbf{k}_3,\mathbf{k}_4),
    \label{eq:Hh_minus_E_one_band}
\end{equation}
where $\tilde\gamma_{\mathbf k}^\dagger$ creates a hole in the top valence band of tMoTe$_2$, $E(\mathbf{k})$ is the dispersion of the top valence band as shown in red in Fig.~\ref{fig:sup_tmote2_c}(a), $\lambda_\mathbf{q}(\mathbf{k}_1,\mathbf{k}_2)$ is the form factor, and $A$ is the system area. Note that the negative sign in front of the single particle term signifies that we are considering hole band~\cite{yu2024fractional}. 

We wish to understand the difference between the filling fraction $\nu_h=1/3$ and $\nu_h=2/3$. One way to compare the these two filling fractions is to do a particle-hole transformation of $H_h[E,\tilde\gamma]\to \mathcal{C}H_h[E,\tilde\gamma]\mathcal{C}^{-1}$ for the $\nu_h=2/3$ case such that 
\begin{equation}
    \mathcal C \tilde\gamma_{\mathbf k}^{\dagger}\mathcal C^{-1}
    =
    \tilde c_{\mathbf k},
    \qquad
    \mathcal C \tilde\gamma_{\mathbf k}\mathcal C^{-1}
    =
    \tilde c_{\mathbf k}^{\dagger}.
\end{equation}
Note that the spectrum of $\mathcal{C}H_h\mathcal{C}^{-1}$ at $\nu=1/3$ is exactly the same as the spectrum of $H_h$ at $\nu_h=2/3$. Performing this transform explicitly, we obtain
\begin{equation}
\begin{split}
    &\mathcal{C}H_h[E,\tilde\gamma]\mathcal{C}^{-1}\\
    =& -\sum_{\mathbf k}
    E(\mathbf k)\,
    \tilde c_{\mathbf k}\tilde c_{\mathbf k}^{\dagger}
    + \frac{1}{2A} \sum_{\mathbf{k}_1\mathbf{k}_2\mathbf{k}_3\mathbf{k}_4}^\text{BZ} \tilde c_{\mathbf{k}_1}\tilde c_{\mathbf{k}_3}\tilde c_{\mathbf{k}_4}^\dagger\tilde c_{\mathbf{k}_2}^\dagger\sum_\mathbf{q}V(\mathbf{q})\,\lambda_\mathbf{q}(\mathbf{k}_1,\mathbf{k}_2)\lambda_{-\mathbf{q}}(\mathbf{k}_3,\mathbf{k}_4)\\
    =& \sum_\mathbf{k} -(-E(\mathbf{k})+\epsilon_H(\mathbf{k})-\epsilon_F(\mathbf{k}))\tilde c^\dagger_\mathbf{k}\tilde c_\mathbf{k} + \frac{1}{2A} \sum_{\mathbf{k}_1\mathbf{k}_2\mathbf{k}_3\mathbf{k}_4}^\text{BZ}\tilde c_{\mathbf{k}_2}^\dagger\tilde c_{\mathbf{k}_4}^\dagger\tilde c_{\mathbf{k}_3} \tilde c_{\mathbf{k}_1}\sum_\mathbf{q}V(\mathbf{q})\,\lambda_\mathbf{q}(\mathbf{k}_1,\mathbf{k}_2)\lambda_{-\mathbf{q}}(\mathbf{k}_3,\mathbf{k}_4)+\text{const.}\\
    =& \sum_\mathbf{k} -(-E(\mathbf{k})+\epsilon_H(\mathbf{k})-\epsilon_F(\mathbf{k}))\tilde c^\dagger_\mathbf{k}\tilde c_\mathbf{k} + \frac{1}{2A} \sum_{\mathbf{k}_1\mathbf{k}_2\mathbf{k}_3\mathbf{k}_4}^\text{BZ}\tilde c_{\mathbf{k}_2}^\dagger\tilde c_{\mathbf{k}_4}^\dagger\tilde c_{\mathbf{k}_3} \tilde c_{\mathbf{k}_1}\sum_{\mathbf{q}}V(-\mathbf{q})\,\lambda_{-\mathbf{q}}(\mathbf{k}_1,\mathbf{k}_2)\lambda_\mathbf{q}(\mathbf{k}_3,\mathbf{k}_4)+\text{const.}\\
    =& \sum_\mathbf{k} -(-E(\mathbf{k})+\epsilon_H(\mathbf{k})-\epsilon_F(\mathbf{k}))\tilde c^\dagger_\mathbf{k}\tilde c_\mathbf{k} + \frac{1}{2A} \sum_{\mathbf{k}_1\mathbf{k}_2\mathbf{k}_3\mathbf{k}_4}^\text{BZ}\tilde c_{\mathbf{k}_2}^\dagger\tilde c_{\mathbf{k}_4}^\dagger\tilde c_{\mathbf{k}_3} \tilde c_{\mathbf{k}_1}\sum_{\mathbf{q}}V(\mathbf{q})\,\left(\lambda_{\mathbf{q}}(\mathbf{k}_2,\mathbf{k}_1)\lambda_{-\mathbf{q}}(\mathbf{k}_4,\mathbf{k}_3)\right)^*+\text{const.}\\
    =&(H_h[-E+\epsilon_H-\epsilon_F,\tilde c])^*+\text{const.},
\end{split}
\end{equation}
where
\begin{equation}
\begin{split}
    \epsilon_H(\mathbf{k}) &= \frac{1}{A}\sum_\mathbf{G} V(\mathbf{G}) \lambda_\mathbf{G}(\mathbf{k},\mathbf{k})\sum_{\mathbf{k}'}^\text{BZ} \lambda_{-\mathbf{G}}(\mathbf{k}',\mathbf{k}'),\\
    \epsilon_F(\mathbf{k}) &= \frac{1}{A}\sum_\mathbf{G}\sum_{\mathbf{k}'}^\text{BZ} V(\mathbf{k}-\mathbf{k}'+\mathbf{G}) \lambda_\mathbf{G}(\mathbf{k}',\mathbf{k})\lambda_{-\mathbf{G}}(\mathbf{k},\mathbf{k}')
\end{split}
\end{equation}
are just the Hartree and Fock potentials of the hole filled band, $^*$ stands for complex conjugation, and we used $V(\mathbf{q})=V(-\mathbf{q})$. 

This gives us a simple way to compare $\nu_h=1/3$ and $\nu_h=2/3$ filling fractions, namely, we can just compare $H_h[E,\tilde\gamma]$ and $(H_h[-E+\epsilon_H-\epsilon_F,\tilde c])^*$ at filling fraction $\nu=1/3$. Furthermore, since the spectrum is real, the complex conjugation does not affect it; hence we can compare $H_h[E,\tilde\gamma]$ and $H_h[-E+\epsilon_H-\epsilon_F,\tilde c]$. Note that interaction part of $H_h[E,\tilde\gamma]$ and $H_h[-E+\epsilon_H-\epsilon_F,\tilde c]$ are the same, it is only the kinetic dispersion that is different. We plotted $E(\mathbf{k})$ and $E(\mathbf{k})-\epsilon_H(\mathbf{k})+\epsilon_F(\mathbf{k})$ in Fig.~\ref{fig:sup_tmote2_c}(a) in red and blue, respectively. Clearly, the Hartree-Fock corrected $E(\mathbf{k})-\epsilon_H(\mathbf{k})+\epsilon_F(\mathbf{k})$ has smaller bandwidth ($\sim 2.5$ meV) with respect to the bare band $E(\mathbf{k})$ ($\sim 6.7$ meV) for $d_s = 30\text{ nm}$ and $\epsilon=15.0$ (this bandwidth reduction is consistent with previous Hartree Fock results, see for example~\cite{dong2023composite}). Comparing these bandwidths to the Coulomb energy scale $e^2/\epsilon a \sim 17.6 \text{ meV}$ for these parameters, we conclude that the bandwidth reduction is significant in $E(\mathbf{k})-\epsilon_H(\mathbf{k})+\epsilon_F(\mathbf{k})$. Thus the larger many-body gap at \(\nu_h=2/3\) can be understood as a consequence of the interaction-generated HF correction, which improves the effective flatness of the dispersion.

\section{Corrections to the $M_2$ scaling and the role of low harmonics}
\label{sec:M2_corrections}

In the main text, we showed that in the short-screening-length limit the interaction projected to a vortexable band is approximately the LLL interaction multiplied by the factor
\(
M_2=\sum_{\mathbf G}|\tilde w_{\mathbf G}|^2
\),
provided that the low harmonics of the lattice modulation are small. Here we start from the full projected interaction and isolate the two corrections to this scaling: (i) the center-of-mass (COM) dependent interaction and (ii) the momentum dependence of the normalization $\mathcal N_{\mathbf k}$. Both are controlled primarily by the lowest reciprocal-lattice harmonics.

\subsection{Full projected interaction and approximations}

The form factor of a vortexable band is given by Eq.~\eqref{eq:lambda_convolution_supp},
\begin{equation}
    \lambda_{\mathbf q}(\mathbf k,\mathbf k')
    =
    \mathcal N_{\mathbf k}\mathcal N_{\mathbf k'}
    \sum_{\mathbf G}
    w_{\mathbf G}
    \lambda^{\mathrm{LLL}}_{\mathbf q-\mathbf G}
    (\mathbf k,\mathbf k').
    \label{eq:M2_lambda_full}
\end{equation}
Below we use the normalized Fourier coefficients
\(
\tilde w_{\mathbf G}=w_{\mathbf G}/w_{\mathbf 0}
\), with $\tilde w_{\mathbf 0}=1$.

Substituting Eq.~\eqref{eq:M2_lambda_full} into the projected interaction Hamiltonian gives the exact projected interaction
\begin{equation}
H_{\mathrm{int}}
=
\frac{1}{2A}
\sum_{\mathbf k_1\mathbf k_2\mathbf k_3\mathbf k_4}^{\mathrm{BZ}}
c_{\mathbf k_1}^{\dagger}
c_{\mathbf k_3}^{\dagger}
c_{\mathbf k_4}
c_{\mathbf k_2}
\sum_{\mathbf q}V(\mathbf q)
\prod_{i=1}^{4}\mathcal N_{\mathbf k_i}
\times
\sum_{\mathbf G_1,\mathbf G_2}
w_{\mathbf G_1}w_{\mathbf G_2}
\lambda^{\mathrm{LLL}}_{\mathbf q-\mathbf G_1}
(\mathbf k_1,\mathbf k_2)
\lambda^{\mathrm{LLL}}_{-\mathbf q-\mathbf G_2}
(\mathbf k_3,\mathbf k_4).
\label{eq:M2_full_projected_H}
\end{equation}
Equation~\eqref{eq:M2_full_projected_H} is the starting point for the following discussion. The simple $M_2$ result follows after two approximations.
Because the product $\prod_{i=1}^{4}\mathcal N_{\mathbf k_i}$ multiplies the entire reciprocal-lattice sum in Eq.~\eqref{eq:M2_full_projected_H}, the normalization error is multiplicative: it dresses both the $\mathbf K=0$ relative interaction $V_{\mathrm{rel}}$ and the $\mathbf K\neq0$ COM interaction $V_{\mathrm{COM}}$.
The first approximation is to neglect the momentum dependence of the normalization,
\begin{equation}
    \mathcal N_{\mathbf k}\simeq w_{\mathbf 0}^{-1/2}.
    \label{eq:M2_flat_N_approx}
\end{equation}
This is because we focus on bands with nearly uniform Berry curvature, for which the momentum dependence of $\mathcal N_{\mathbf k}$ is small, as discussed in the main text.
Using $w_{\mathbf G}=w_{\mathbf 0}\tilde w_{\mathbf G}$, the four normalization factors in Eq.~\eqref{eq:M2_full_projected_H} cancel the two factors of $w_{\mathbf 0}$:
\begin{equation}
    \left(\prod_{i=1}^{4}\mathcal N_{\mathbf k_i}\right)
    w_{\mathbf G_1}w_{\mathbf G_2}
    \simeq
    \tilde w_{\mathbf G_1}\tilde w_{\mathbf G_2}.
    \label{eq:M2_weight_cancellation}
\end{equation}
Because the $\lambda^{\mathrm{LLL}}$ factors in Eq.~\eqref{eq:M2_full_projected_H} are plane-wave matrix elements within the LLL, the Hamiltonian can be read as the LLL projection of a real-space interaction $\tilde V$. We use this representation to separate the translation-invariant part that depends only on the relative coordinate from the lattice-dependent. part that depends on the center-of-mass coordinate. Only the former can be compared directly with the LLL interaction and produce the simple $M_2$ rescaling, while the latter gives a correction. Using Eq.~\eqref{eq:M2_weight_cancellation}, relabeling $\mathbf q\to-\mathbf q$, and noting that $V(-\mathbf q)=V(\mathbf q)$ for the screened Coulomb interaction, we obtain
\begin{equation}
\begin{split}
    \tilde V(\mathbf r_1,\mathbf r_2)
    =
    \frac{1}{A}
    \sum_{\mathbf q}
    \sum_{\mathbf G_1,\mathbf G_2}
    V(\mathbf q)
    \tilde w_{\mathbf G_1}
    \tilde w_{\mathbf G_2}
    e^{i(\mathbf q+\mathbf G_1)\cdot\mathbf r_1}
    e^{i(-\mathbf q+\mathbf G_2)\cdot\mathbf r_2}.
\end{split}
\label{eq:M2_full_interaction_supp}
\end{equation}
%Using $|\tilde h(\mathbf r)|^2=\sum_{\mathbf G}\tilde w_{\mathbf G}e^{i\mathbf G\cdot\mathbf r}$, Eq.~\eqref{eq:M2_full_interaction_supp} can equivalently be written as $\tilde V(\mathbf r_1,\mathbf r_2)=V(\mathbf r_1-\mathbf r_2)|\tilde h(\mathbf r_1)|^2|\tilde h(\mathbf r_2)|^2$.

Introducing the relative and center-of-mass (COM) coordinates
$\mathbf r=\mathbf r_1-\mathbf r_2$ and
$\mathbf R=(\mathbf r_1+\mathbf r_2)/2$, we define
$\mathbf K=\mathbf G_1+\mathbf G_2$.  This is the total
reciprocal-lattice momentum carried by the two modulation harmonics and hence
labels the spatial harmonics of the COM dependence.  It therefore provides a
natural way to separate the translation-invariant sector, $\mathbf K=0$, from
the lattice-dependent COM corrections, $\mathbf K\neq0$.  In these variables,
the phase in Eq.~\eqref{eq:M2_full_interaction_supp} becomes
$e^{i\mathbf K\cdot\mathbf R}
e^{i[\mathbf q+(\mathbf G_1-\mathbf G_2)/2]\cdot\mathbf r}$.
For $\mathbf K=0$, we have $\mathbf G_2=-\mathbf G_1$, so the COM phase
drops out and the interaction depends only on $\mathbf r$.  After shifting the
summation momentum, this relative-coordinate contribution is
\begin{equation}
    \tilde V_{\mathrm{rel}}(\mathbf r)
    =
    \frac{1}{A}
    \sum_{\mathbf q,\mathbf G}
    |\tilde w_{\mathbf G}|^2
    V(\mathbf q-\mathbf G)
    e^{i\mathbf q\cdot\mathbf r}.
    \label{eq:M2_Vrel_K0}
\end{equation}
This is the relative interaction appearing in Eq.~\eqref{eq:SM_compact_Veff}.

The $\mathbf K\neq0$ terms depend on $\mathbf R$ and form the COM-dependent
interaction.  We first retain the form used in Eq.~(A2) of the main text,
\begin{equation}
    \tilde V_{\mathrm{COM}}(\mathbf r_1,\mathbf r_2)
    ={}\frac{1}{A}
    \sum_{\mathbf q}
    \sum_{\mathbf G_1\neq-\mathbf G_2}
    V(\mathbf q)\,
    \tilde w_{\mathbf G_1}
    \tilde w_{\mathbf G_2}\,
    e^{i[(\mathbf q+\mathbf G_1)\cdot\mathbf r_1
    +(-\mathbf q+\mathbf G_2)\cdot\mathbf r_2]}.
\label{eq:M2_VCOM_maintext_form}
\end{equation}
To make the different COM-momentum channels explicit, we then set
$\mathbf G_1=\mathbf G$ and $\mathbf G_2=\mathbf K-\mathbf G$ and group the
terms by $\mathbf K$:
\begin{equation}
\begin{split}
    \tilde V_{\mathrm{COM}}(\mathbf r,\mathbf R)
    =
    \frac{1}{A}
    \sum_{\mathbf K\neq0}
    e^{i\mathbf K\cdot\mathbf R}
    \sum_{\mathbf q,\mathbf G}
    V(\mathbf q)\,
    \tilde w_{\mathbf G}
    \tilde w_{\mathbf K-\mathbf G}
    e^{i(\mathbf q+\mathbf G-\mathbf K/2)\cdot\mathbf r}.
\end{split}
\label{eq:M2_VCOM_K}
\end{equation}
Our second approximation is to neglect Eq.~\eqref{eq:M2_VCOM_K} and retain only the relative-coordinate interaction. Under the short-range condition of Sec.~\ref{sec:pseudopotentials}, it is enhanced by $M_2=\sum_{\mathbf G}|\tilde w_{\mathbf G}|^2=\langle|\tilde h(\mathbf r)|^4\rangle_{\mathrm{uc}}$, so that $H_{\mathrm{rel}}\simeq M_2H_{\mathrm{LLL}}$ and $\Delta/\Delta_{\mathrm{LLL}}\simeq M_2$, as used in the main text. The omitted $\mathbf K\neq0$ channels and the momentum dependence of $\mathcal N_{\mathbf k}$ give the two corrections discussed below.

\subsection{Correction from the momentum-dependent normalization}

Factoring out $w_{\mathbf 0}$ from Eq.~\eqref{eq:Nk_explicit_supp}, we write
\begin{equation}
    \mathcal N_{\mathbf k}^{-2}
    =w_{\mathbf 0}(1+\delta n_{\mathbf k}),
    \qquad
    \delta n_{\mathbf k}
    =\sum_{\mathbf G\ne0}\eta_{\mathbf G}\tilde w_{\mathbf G}
    e^{i\ell_B^2\mathbf G\times\mathbf k}
    e^{-|\mathbf G|^2\ell_B^2/4}.
    \label{eq:M2_delta_nk}
\end{equation}
For $|\delta n_{\mathbf k}|\ll1$,
\begin{equation}
    \mathcal N_{\mathbf k}
    =w_{\mathbf 0}^{-1/2}
    \left(1-\frac{1}{2}\delta n_{\mathbf k}
    +\frac{3}{8}\delta n_{\mathbf k}^2+\cdots\right).
    \label{eq:M2_Nk_expansion}
\end{equation}
Consequently, the four normalization factors in Eq.~\eqref{eq:M2_full_projected_H} give
\begin{equation}
    \prod_{i=1}^{4}\mathcal N_{\mathbf k_i}
    =w_{\mathbf 0}^{-2}
    \left[1-\frac{1}{2}\sum_{i=1}^{4}\delta n_{\mathbf k_i}
    +O(\!\delta n^2)\right].
    \label{eq:M2_Nk_product}
\end{equation}
The leading normalization correction is therefore linear in $\tilde w_{\mathbf G}$, but each harmonic carries the Gaussian suppression factor $e^{-|\mathbf G|^2\ell_B^2/4}$.  (By contrast, the relative-coordinate enhancement $M_2$ begins quadratically in $\tilde w_{\mathbf G}$ but has no such Gaussian suppression.)  For the first two triangular-lattice shells,
\begin{equation}
    e^{-|\mathbf G^{(1)}|^2\ell_B^2/4}
    =e^{-\pi/\sqrt{3}}\simeq0.16,
    \qquad
    e^{-|\mathbf G^{(2)}|^2\ell_B^2/4}
    =e^{-\sqrt{3}\pi}\simeq0.0043.
    \label{eq:M2_Nk_shell_factors}
\end{equation}
Thus the second shell correction is much more strongly suppressed. We then just need to consider the correction from the first harmonic. Below, we give a rough estimation of the magnitude of this correction for the $C_6$-symmetric models considered in the main text, where all six vectors in the first reciprocal-lattice shell have the same real coefficient $\tilde w_1$.  In our magnetic Bloch convention, $\eta_{\mathbf G}=-1$ for all first-shell vectors, so their contribution to the normalization can be written as
\begin{equation}
    \delta n_{\mathbf k}^{(1)}
    =
    e^{-\pi/\sqrt{3}}\tilde w_1 f_1(\mathbf k),
    \qquad
    f_1(\mathbf k)
    =
    \sum_{\mathbf G\in S_1}
    \eta_{\mathbf G}
    e^{i\ell_B^2\mathbf G\times\mathbf k}.
\end{equation}
The six first-shell terms sum up giving $    -6\le f_1(\mathbf k)\le3.$ Consequently, $    |\delta n_{\mathbf k}^{(1)}|\le 6e^{-\pi/\sqrt3}|\tilde w_1|\simeq0.98|\tilde w_1|.$
Thus, although each first-shell component is suppressed by the Gaussian factor
$e^{-\pi/\sqrt3}\simeq0.16$, the sixfold shell sum can partially compensate this suppression.  The first harmonic can therefore produce an appreciable momentum dependence of $\mathcal N_{\mathbf k}$ even when $|\tilde w_1|$ is small.

Expanding the four normalization factors entering a two-body matrix element gives
\begin{equation}
    \prod_{i=1}^{4}\mathcal N_{\mathbf k_i}
    =
    w_{\mathbf 0}^{-2}
    \left[
    1-
    \frac{e^{-\pi/\sqrt3}\tilde w_1}{2}
    \sum_{i=1}^{4}f_1(\mathbf k_i)
    +O(\tilde w_1^2)
    \right].
\end{equation}
A conservative bound on the resulting linear change of an individual interaction matrix element is therefore of order
\begin{equation}
    \left|
    w_{\mathbf 0}^{2}
    \prod_{i=1}^{4}\mathcal N_{\mathbf k_i}-1
    \right|
    \lesssim
    2|\tilde w_1|.
    \label{eq:M2_first_shell_product_bound}
\end{equation}
This bound corresponds to all four momenta sampling the extrema of the
first-shell modulation simultaneously, and therefore substantially
overestimates a typical matrix element. A more representative scale is obtained from the Brillouin-zone fluctuations.
Since
\(
\langle f_1(\mathbf k)\rangle_{\mathrm{BZ}}=0
\)
and
\(
\langle f_1^2(\mathbf k)\rangle_{\mathrm{BZ}}=6
\),
the root-mean-square single-particle correction is
\begin{equation}
    \sqrt{
    \left\langle
    |\delta n_{\mathbf k}^{(1)}|^2
    \right\rangle_{\mathrm{BZ}}
    }
    =
    \sqrt6\,e^{-\pi/\sqrt3}|\tilde w_1|
    \simeq0.4|\tilde w_1|.
    \label{eq:M2_first_shell_rms}
\end{equation}
For a two-body scattering process the momenta satisfy
$\mathbf k_1+\mathbf k_3=\mathbf k_2+\mathbf k_4$ modulo a reciprocal lattice
vector.  Nevertheless, under a uniform average over all momentum-conserving
quadruples, every pair of external momenta is still uniformly distributed over
the Brillouin zone.  Since
$\langle\delta n_{\mathbf k}^{(1)}\rangle_{\mathrm{BZ}}=0$, the cross
correlations vanish, and
\begin{equation}
    \sqrt{\left\langle
    \left|
    w_{\mathbf 0}^2\prod_{i=1}^{4}\mathcal N_{\mathbf k_i}-1
    \right|^2
    \right\rangle}\simeq
    \frac{1}{2}
    \sqrt{\left\langle
    \left|\sum_{i=1}^{4}\delta n_{\mathbf k_i}^{(1)}\right|^2
    \right\rangle}=
    \sqrt{\left\langle
    |\delta n_{\mathbf k}^{(1)}|^2
    \right\rangle_{\mathrm{BZ}}}
    \simeq0.4|\tilde w_1|.
    \label{eq:M2_first_shell_product_rms}
\end{equation}
Thus $0.4|\tilde w_1|$ gives the typical normalization-induced variation of
a two-body matrix element.  A physically weighted average, which includes the
interaction and form factors, need not have exactly vanishing correlations, so
this value should be regarded as a characteristic scale rather than a direct
prediction for the many-body gap shift.

This multiplicative factor acts on the full projected interaction, so its expansion generates corrections to both $V_{\mathrm{rel}}$ and $V_{\mathrm{COM}}$, including coupled COM-normalization contributions.
Physically, this correction does not act as a uniform rescaling of the
interaction because its Brillouin-zone average vanishes.  Instead, it introduces
a momentum-dependent perturbation that breaks the approximate magnetic
translation symmetry of the LLL problem.  This is closely related to the
mechanism discussed in Ref.~\cite{shi2025effects}: using the single-mode
approximation, it was shown that a first-harmonic modulation splits the
symmetry-related magnetoroton states linearly in the modulation strength. However, note that the normalization correction
alone does not fix the sign of the shift of the spectrum and
therefore does not by itself imply a smaller many-body gap. The ED results below find a net
reduction of gap in the sampling space.  In our formulation, the momentum
dependence of $\mathcal N_{\mathbf k}$ provides one microscopic contribution
to this first-harmonic perturbation.

\subsection{Correction from the COM-dependent interaction}

The second correction comes from the $\mathbf K\neq0$ terms in
Eq.~\eqref{eq:M2_VCOM_K}.  After projection to the LLL, the two density form
factors contribute Gaussian factors whose product can be written as
\begin{equation}
    e^{-\ell_B^2|\mathbf q+\mathbf G_1|^2/4}
    e^{-\ell_B^2|\mathbf q-\mathbf G_2|^2/4}=
    e^{-\frac{\ell_B^2}{2}
    \left|\mathbf q+\frac{\mathbf G_1-\mathbf G_2}{2}\right|^2}
    e^{-|\mathbf K|^2\ell_B^2/8},
    \qquad
    \mathbf K=\mathbf G_1+\mathbf G_2 .
\label{eq:M2_COM_Gaussian_supp}
\end{equation}
Although the unrestricted $\mathbf q$ sum can compensate the first
exponential, it cannot remove the second.  Thus every nonzero COM harmonic
retains the suppression factor
\begin{equation}
    e^{-|\mathbf K|^2\ell_B^2/8}.
    \label{eq:M2_COM_suppression_factor}
\end{equation}
This makes the correction controlled primarily by the smallest nonzero COM
momenta.

As Eq.~\eqref{eq:M2_VCOM_K} shows, each COM channel $\mathbf K$ contains a
pair of harmonics $\mathbf G$ and $\mathbf K-\mathbf G$.  We estimate
the channel weight at each $\mathbf K$ by defining
\begin{equation}
    C_{\mathbf K}
    =
    \sum_{\mathbf G}
    \left|
    \tilde w_{\mathbf G}
    \tilde w_{\mathbf K-\mathbf G}
    \right|.
    \label{eq:M2_CK_def}
\end{equation}
Here $C_{\mathbf0}=M_2$, whereas $C_{\mathbf K\neq0}$ measures the total
Fourier-pair weight of processes that transfer momentum $\mathbf K$ to the
pair COM.
Including the
Gaussian projection factor, the relative correction is estimated as
\begin{equation}
    \epsilon_{\mathrm{COM}}(\mathbf K)
    \sim
    \frac{C_{\mathbf K}}{M_2}
    e^{-|\mathbf K|^2\ell_B^2/8}.
    \label{eq:M2_COM_error_estimate}
\end{equation}
This is a conservative characteristic scale rather than a strict operator-norm bound,
because the full matrix element also contains relative-momentum dependence.
It bounds the magnitude but does not determine the sign of the resulting
many-body energy correction.
Nevertheless, it directly shows that the relevant quantity is the
low-$|\mathbf K|$ absolute convolution $C_{\mathbf K}$.

For the triangular lattice in Eq.~\eqref{eq:Gdefs}, the first two COM shells
have suppression factors
\begin{equation}
    e^{-|\mathbf G^{(1)}|^2\ell_B^2/8}
    \simeq0.40,
    \qquad
    e^{-|\mathbf G^{(2)}|^2\ell_B^2/8}
    \simeq0.066.
    \label{eq:M2_COM_numerical_suppression}
\end{equation}
The first COM shell therefore gives the dominant correction.

For the $C_6$-symmetric two-shell model, for any
$\mathbf K$ in the first reciprocal-lattice shell,
\begin{equation}
    C_{\mathbf K}
    =
    2|\tilde w_1|
    +2|\tilde w_1|^2
    +4|\tilde w_1\tilde w_2|.
    \label{eq:M2_CK_first_shell}
\end{equation}
Consequently, the leading first-shell COM correction per $\mathbf K$ has the
characteristic scale
\begin{equation}
    \epsilon_{\mathrm{COM}}^{(1)}
    \sim
    \frac{2e^{-\pi/(2\sqrt3)}}{M_2}
    |\tilde w_1|
    \simeq
    \frac{0.8}{M_2}|\tilde w_1|.
    \label{eq:M2_COM_first_shell_scale}
\end{equation}
Thus the COM correction, like the normalization correction discussed above,
appears already at linear order in $\tilde w_1$. When
$\tilde w_1=0$, its leading COM contribution lies in the second reciprocal
lattice shell,
\begin{equation}
    C_{\mathbf K\in S_2}
    =
    2|\tilde w_2|+O(\tilde w^2),
\end{equation}
with Gaussian factor only $0.066$.  Its bare scale is therefore
$2e^{-3\pi/(2\sqrt3)}|\tilde w_2|\simeq0.13|\tilde w_2|$.  By contrast, the
same harmonic increases the $\tilde V_{\mathrm{rel}}$ through the unsuppressed term
$6|\tilde w_2|^2$ in $M_2$.  Thus a pure $\tilde w_2$ modulation produces a
weak linear COM correction but a quadratic relative-interaction enhancement.
As $|\tilde w_2|$ grows, the latter can dominate, which is the case discussed in the main text.

The $\mathbf K\neq0$ COM interaction is also qualitatively different from the
$\mathbf K=0$ relative interaction.  Rather than rescaling all LLL energies uniformly, it
breaks magnetic-translation symmetry and disperses the
symmetry-related magnetoroton branches~\cite{shi2025effects}. 

Finally, by Cauchy--Schwarz,
\begin{equation}
    C_{\mathbf K}
    \le
    \sum_{\mathbf G}|\tilde w_{\mathbf G}|^2
    =
    M_2,
\end{equation}
together with Eq.~\eqref{eq:M2_COM_error_estimate}, this bound shows that the $M_2$ rescaling remains
controlled when $C_{\mathbf K}$ is small at the first few harmonics.

\subsection{The two-shell model and the Gaussian hierarchy of low-harmonic corrections}
\label{subsec:M2_finite_shell_corrections}

\begin{figure}[tbh]
    \centering
    \includegraphics[width=\textwidth]{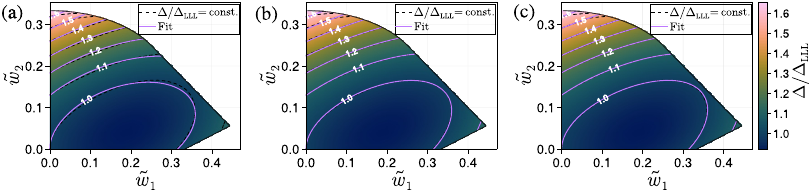}
    \caption{
    \textbf{Two-shell gap landscape and the hierarchy of low-harmonic fits.}
    The color map shows the ED gap ratio $\Delta/\Delta_{\mathrm{LLL}}$ for the
    $C_6$-symmetric two-shell model with $N_k=27$ and $N_e=9$ on the sampled
    branch $\tilde w_1,\tilde w_2\geq0$; all higher-shell coefficients vanish.
    Black dashed curves are constant-gap contours of the ED data.  Purple
    curves show (a) Fit~1, (b) Fit~2, and (c) Fit~3, defined by
    Eqs.~\eqref{eq:M2_fit_zeta}, \eqref{eq:M2_fit_zeta2}, and
    \eqref{eq:M2_fit_zeta3}, respectively, using the $N_k=27$ coefficients in
    Table~\ref{tab:M2_fit_hierarchy}.  The solid black curve marks the boundary
    imposed by $|\tilde h(\mathbf r)|^2\geq0$. Exact-diagonalization data are evaluated 
    on a uniform $21\times19$ grid over $\tilde{w}_1\in[0,4/9]$ and $\tilde{w}_2\in[0,1/3]$, 
    with spacings $\Delta\tilde{w}_1=1/45$ and $\Delta\tilde{w}_2=1/54$; only density-positive 
    points are retained. The heat map and contours are bilinearly interpolated onto 
    a $10\times$-refined $201\times181$ display grid for visualization only, with no 
    additional ED calculations performed on the refined grid.
    }
    \label{fig:M2_two_shell_ED}
\end{figure}

To test the $M_2$ rescaling approximation and quantify its leading corrections,
we numerically study the gap enhancement ratio in the $C_6$-symmetric
two-shell model
\begin{equation}
    |\tilde h(\mathbf r)|^2
    =
    1+\tilde w_1F_1(\mathbf r)+\tilde w_2F_2(\mathbf r),
    \qquad
    F_s(\mathbf r)
    =
    \sum_{\mathbf G\in S_s}e^{i\mathbf G\cdot\mathbf r},
    \label{eq:M2_two_shell_modulation}
\end{equation}
where $S_1$ and $S_2$ are the first two reciprocal-lattice shells, each
containing six vectors, and all higher-shell coefficients are set to zero.
Figure~\ref{fig:M2_two_shell_ED} compares the ED gap ratio with the three nested
fitting forms introduced below.  In the numerical scan,
$\tilde w_1,\tilde w_2\geq0$.  All fitted expressions below refer to this
nonnegative branch.  They should not be directly extrapolated to negative
harmonics, because the linear corrections can be sign asymmetric and the
identity of the lowest neutral excitation can change.

It is useful to define the first-COM-shell Gaussian factor
\begin{equation}
    \zeta
    \equiv
    e^{-|\mathbf G^{(1)}|^2\ell_B^2/8}
    =
    e^{-\pi/(2\sqrt3)}
    \simeq0.404.
    \label{eq:M2_zeta_def}
\end{equation}
The powers of $\zeta$ organize the corrections arising from different COM
shells and from the momentum-dependent normalization.  Although
$\zeta\simeq0.4$ is not an asymptotically small parameter, the successive
scales
\(
\zeta,\zeta^2,\zeta^3,\ldots
\)
provide a useful hierarchy.

\begin{table}[tbh]
    \centering
    \caption{
    \textbf{Gaussian hierarchy of corrections in the two-shell model.}
    The polynomial column lists the structures generated through cubic order
    in $\tilde w_1$ and $\tilde w_2$.  ``New structures'' denotes monomials that
    are not already present at a lower Gaussian order.
    }
    \label{tab:M2_gaussian_hierarchy}
    \setlength{\tabcolsep}{5pt}
    \renewcommand{\arraystretch}{1.1}
    \begin{tabular}{l c c l l}
        \hline\hline
        Source
        & Scale
        & Value
        & Polynomial content
        \\
        \hline
        Relative interaction
        & $\zeta^0$
        & $1$
        & $\tilde w_1^2,\tilde w_2^2$
        \\
        First COM shell
        & $\zeta^1$
        & $0.404$
        & $\tilde w_1,\tilde w_1^2,
        \tilde w_1\tilde w_2$
        \\
        First-shell normalization and
        second-order first-COM response
        & $\zeta^2$
        & $0.163$
        & $\tilde w_1,\tilde w_1^2,\tilde w_1^3,
        \tilde w_1^2\tilde w_2,\tilde w_1\tilde w_2^2$
        \\
        Second COM shell, COM--normalization interference,
        and third-order first-COM response
        & $\zeta^3$
        & $0.0658$
        & $\tilde w_2,\tilde w_2^2,\tilde w_1^2,
        \tilde w_1^3,\tilde w_1^2\tilde w_2$
        \\
        Third COM shell
        & $\zeta^4$
        & $0.0266$
        & higher-order corrections
        \\
        Second-shell normalization
        & $\zeta^6$
        & $0.0043$
        & $\tilde w_2,\cdots$
        \\
        \hline\hline
    \end{tabular}
\end{table}

The relative-coordinate contribution is not Gaussian suppressed and is fixed
by
\begin{equation}
    M_2
    =
    1+6\tilde w_1^2+6\tilde w_2^2.
    \label{eq:M2_two_shell_exact}
\end{equation}
For a COM momentum transfer $\mathbf K$ in the first reciprocal-lattice shell,
the two-shell convolution is
\begin{equation}
    C_{\mathbf K\in S_1}
    =
    2\tilde w_1
    +2\tilde w_1^2
    +4\tilde w_1\tilde w_2.
    \label{eq:M2_C1_two_shell}
\end{equation}
Therefore, at order $\zeta$, the COM correction generates the three structures
\(
\tilde w_1
\),
\(
\tilde w_1^2
\),
and
\(
\tilde w_1\tilde w_2
\).
The Gaussian and shell-counting analysis determines which monomials appear,
but it does not determine the signs or magnitudes of their contributions to the
many-body gap.

At order $\zeta^2$, two additional effects enter.  First, the first-shell
normalization correction is proportional to
\(
\zeta^2\tilde w_1
\).
When it multiplies the $M_2$-enhanced relative interaction, its polynomial
content through cubic order is
\begin{equation}
    \tilde w_1M_2
    =
    \tilde w_1
    +6\tilde w_1^3
    +6\tilde w_1\tilde w_2^2.
    \label{eq:M2_normalization_polynomial_content}
\end{equation}
Second, the many-body spectrum has a second-order response to the first-COM
perturbation.  The corresponding polynomial content follows from
\begin{equation}
    C_{\mathbf K\in S_1}^2=
    \left(
    2\tilde w_1+2\tilde w_1^2
    +4\tilde w_1\tilde w_2
    \right)^2=
    4\tilde w_1^2
    +8\tilde w_1^3
    +16\tilde w_1^2\tilde w_2
    +O(\tilde w^4).
\label{eq:M2_C1_squared}
\end{equation}
Thus the new monomials first appearing at order $\zeta^2$ are
\begin{equation}
    \tilde w_1^3,
    \qquad
    \tilde w_1^2\tilde w_2,
    \qquad
    \tilde w_1\tilde w_2^2.
\end{equation}

At order $\zeta^3$, the second COM shell enters directly.  For
$\mathbf K\in S_2$,
\begin{equation}
    C_{\mathbf K\in S_2}
    =
    2\tilde w_2
    +2\tilde w_1^2
    +2\tilde w_2^2.
    \label{eq:M2_C2_two_shell}
\end{equation}
This produces the first pure linear-$\tilde w_2$ correction, together with a
pure $\tilde w_2^2$ correction.  At the same Gaussian order, interference
between the first-shell normalization and the first COM shell gives
\begin{equation}
\begin{split}
    \tilde w_1 C_{\mathbf K\in S_1}
    &=
    2\tilde w_1^2
    +2\tilde w_1^3
    +4\tilde w_1^2\tilde w_2,
\end{split}
\label{eq:M2_normalization_COM_interference}
\end{equation}
These terms renormalize polynomial structures already present at lower order.
The genuinely new monomials at order $\zeta^3$ are therefore
\(
\tilde w_2
\)
and
\(
\tilde w_2^2
\).

Collecting these results, the gap ratio can be organized through order
$\zeta^3$ and cubic order in the harmonics as
\begin{equation}
\begin{split}
    \frac{\Delta}{\Delta_{\mathrm{LLL}}}
    ={}&
    \overbrace{\zeta^0
    \left(
        1+6\tilde w_1^2+6\tilde w_2^2
    \right)}^{M_2}
    \\
    &+
    \zeta
    \left(
        a_{10}^{(1)}\tilde w_1
        +a_{20}^{(1)}\tilde w_1^2
        +a_{11}^{(1)}\tilde w_1\tilde w_2
    \right)
    \\
    &+
    \zeta^2
    \left(
        a_{10}^{(2)}\tilde w_1
        +a_{20}^{(2)}\tilde w_1^2
        +a_{30}^{(2)}\tilde w_1^3
        +a_{21}^{(2)}\tilde w_1^2\tilde w_2
        +a_{12}^{(2)}\tilde w_1\tilde w_2^2
    \right)
    \\
    &+
    \zeta^3
    \left(
        a_{01}^{(3)}\tilde w_2
        +a_{02}^{(3)}\tilde w_2^2
        +a_{20}^{(3)}\tilde w_1^2
        +a_{30}^{(3)}\tilde w_1^3
        +a_{21}^{(3)}\tilde w_1^2\tilde w_2
    \right)
    \\
    &+
    O(\zeta^4,\tilde w^4).
\end{split}
\label{eq:M2_zeta3_systematic_expansion}
\end{equation}
The coefficients in Eq.~\eqref{eq:M2_zeta3_systematic_expansion} are
state-dependent many-body response coefficients.  They contain the
relative-momentum dependence of the COM interaction, normalization effects,
and virtual transitions to other many-body states.

Because $\zeta$ is fixed by the triangular-lattice geometry, a numerical fit
cannot separately resolve contributions with the same polynomial dependence
but different powers of $\zeta$.  We therefore use three nested effective
fitting forms.

The leading fitting form retains the monomials generated by the first COM
shell:
\begin{equation}
\begin{split}
    R_{\zeta}
    \equiv
    \frac{\Delta}{\Delta_{\mathrm{LLL}}}
    \simeq{}&
    M_2
    +A_{10}\tilde w_1
    +A_{20}\tilde w_1^2
    +A_{11}\tilde w_1\tilde w_2.
\end{split}
\label{eq:M2_fit_zeta}
\end{equation}
For $N_k=27$ and $N_e=9$, the current leading fit is
\begin{equation}
    R_{\zeta}
    \simeq
    1+2.65\tilde w_1^2+6\tilde w_2^2
    -0.83\tilde w_1
    -2.79\tilde w_1\tilde w_2.
    \label{eq:M2_N27_fit}
\end{equation}

The next fit includes the new cubic structures generated by the first-shell
normalization and by the second-order response to the first COM shell:
\begin{equation}
    R_{\zeta^2}
    \simeq{}
    M_2
    +A_{10}\tilde w_1
    +A_{20}\tilde w_1^2
    +A_{11}\tilde w_1\tilde w_2+
    B_{30}\tilde w_1^3
    +B_{21}\tilde w_1^2\tilde w_2
    +B_{12}\tilde w_1\tilde w_2^2.
\label{eq:M2_fit_zeta2}
\end{equation}
The third fit further includes the pure second-harmonic structures generated
by the second COM shell:
\begin{equation}
    R_{\zeta^3}
    \simeq{}
    M_2
    +A_{10}\tilde w_1
    +A_{20}\tilde w_1^2
    +A_{11}\tilde w_1\tilde w_2+
    B_{30}\tilde w_1^3
    +B_{21}\tilde w_1^2\tilde w_2
    +B_{12}\tilde w_1\tilde w_2^2+
    C_{01}\tilde w_2
    +C_{02}\tilde w_2^2.
\label{eq:M2_fit_zeta3}
\end{equation}
The coefficients in each of
Eqs.~\eqref{eq:M2_fit_zeta}--\eqref{eq:M2_fit_zeta3} should be refitted
independently.  In particular, adding higher-order terms can change the
effective values of $A_{10}$, $A_{20}$, and $A_{11}$.

\begin{table}[tbh]
    \centering
    \caption{
    \textbf{Finite-size comparison of Fits 1--3.}
    Fit~1, Fit~2, and Fit~3 denote
    Eqs.~\eqref{eq:M2_fit_zeta},
    \eqref{eq:M2_fit_zeta2}, and
    \eqref{eq:M2_fit_zeta3}, respectively.
    $R^2$ is the coefficient of determination and RMSE is the
    root-mean-square error of the fitted gap ratio.
    All entries are obtained from independent fits at each system size.
    }
    \label{tab:M2_fit_hierarchy}
    \resizebox{0.7\textwidth}{!}{
    \begin{tabular}{
        c c c |
        c c c |
        c c c |
        c c |
        c c
    }
        \hline\hline
        $N_k$
        & $N_e$
        & Fit
        & $A_{10}$
        & $A_{20}$
        & $A_{11}$
        & $B_{30}$
        & $B_{21}$
        & $B_{12}$
        & $C_{01}$
        & $C_{02}$
        & $R^2$
        & RMSE
        \\
        \hline

        \multirow{3}{*}{12}
        & \multirow{3}{*}{4}
        & Fit~1
        & -1.384 & -2.995 & -2.715
        & -- & -- & -- & -- & --
        & 0.99865 & 0.00648
        \\
        & & Fit~2
        & -1.380 & -3.410 & -1.626
        & 1.183 & -2.690 & -1.916
        & -- & --
        & 0.99916 & 0.00510
        \\
        & & Fit~3
        & -1.363 & -3.648 & -1.068
        & 1.783 & -4.608 & -1.216
        & -0.012 & -0.173
        & 0.99967 & 0.00319
        \\
        \hline

        \multirow{3}{*}{15}
        & \multirow{3}{*}{5}
        & Fit~1
        & -1.143 & -3.287 & -2.866
        & -- & -- & -- & -- & --
        & 0.99877 & 0.00588
        \\
        & & Fit~2
        & -1.123 & -3.718 & -2.269
        & 1.229 & -2.397 & -0.194
        & -- & --
        & 0.99936 & 0.00422
        \\
        & & Fit~3
        & -1.108 & -3.952 & -1.719
        & 1.831 & -4.367 & 0.616
        & -0.008 & -0.193
        & 0.99997 & 0.00096
        \\
        \hline

        \multirow{3}{*}{18}
        & \multirow{3}{*}{6}
        & Fit~1
        & -0.890 & -3.332 & -2.822
        & -- & -- & -- & -- & --
        & 0.99693 & 0.00872
        \\
        & & Fit~2
        & -0.730 & -4.981 & -2.139
        & 3.548 & -2.605 & -0.323
        & -- & --
        & 0.99953 & 0.00340
        \\
        & & Fit~3
        & -0.757 & -4.790 & -2.527
        & 3.250 & -2.178 & 0.621
        & 0.053 & -0.155
        & 0.99959 & 0.00319
        \\
        \hline

        \multirow{3}{*}{21}
        & \multirow{3}{*}{7}
        & Fit~1
        & -0.846 & -3.343 & -2.811
        & -- & -- & -- & -- & --
        & 0.99848 & 0.00596
        \\
        & & Fit~2
        & -0.800 & -4.037 & -2.136
        & 1.813 & -2.872 & -0.066
        & -- & --
        & 0.99977 & 0.00233
        \\
        & & Fit~3
        & -0.796 & -4.132 & -1.900
        & 2.091 & -3.872 & 0.528
        & 0.004 & -0.130
        & 0.99996 & 0.00094
        \\
        \hline

        \multirow{3}{*}{27}
        & \multirow{3}{*}{9}
        & Fit~1
        & -0.834 & -3.355 & -2.788
        & -- & -- & -- & -- & --
        & 0.99868 & 0.00554
        \\
        & & Fit~2
        & -0.778 & -4.116 & -2.127
        & 1.870 & -2.535 & -0.319
        & -- & --
        & 0.99983 & 0.00200
        \\
        & & Fit~3
        & -0.775 & -4.195 & -1.931
        & 2.106 & -3.402 & 0.231
        & 0.005 & -0.119
        & 0.99998 & 0.00072
        \\
        \hline\hline
    \end{tabular}
    }
\end{table}

An increase in $R^2$ together with a reduction in RMSE indicates that the
additional terms capture systematic residual structure.  The improvement
should be compared across system sizes: a higher-order term is physically
meaningful only if its coefficient and the reduction in fitting error remain
stable as $N_k$ increases.

The fitted coefficients in Table~\ref{tab:M2_fit_hierarchy} follow the expected hierarchy of corrections.  The $A$ coefficients describe the leading first-shell response at order $\zeta$; they show clear convergence with increasing system size and therefore provide the most robust quantitative characterization of the low-harmonic correction.  The $B$ coefficients enter at order $\zeta^2$ and receive contributions from several sources, including the momentum-dependent normalization, its multiplicative coupling to the $M_2$-enhanced interaction, and the second-order many-body response to the first COM harmonic.  Because the corresponding cubic monomials are strongly correlated over the allowed parameter region, the individual $B$ coefficients are more sensitive to the fitting basis and should be interpreted collectively through their contribution to the fitted gap rather than as independent microscopic response constants.  Finally, the $C$ coefficients describe the new structures appearing at order $\zeta^3$, most directly from the second COM shell.  Their small fitted magnitudes are consistent with the substantially stronger Gaussian suppression of this shell, supporting the hierarchy inferred from the analytic expansion.

Using the leading fit in Eq.~\eqref{eq:M2_N27_fit}, the boundary
$\Delta=\Delta_{\mathrm{LLL}}$ is
\begin{equation}
    2.65\tilde w_1^2
    -
    (0.83+2.79\tilde w_2)\tilde w_1
    +
    6\tilde w_2^2
    =
    0.
    \label{eq:M2_enhancement_boundary}
\end{equation}
On its small-$\tilde w_1$ branch, the critical values are
$\tilde w_{1,c}\simeq0.016$, $0.064$, and $0.168$ for
$\tilde w_2=0.05$, $0.10$, and $0.15$, respectively.  The suppression region
closes at $\tilde w_2\simeq0.16$ within Fit~1.  Thus there is no universal
upper bound on $|\tilde w_1|$ alone: increasing the beneficial second-harmonic
weight allows a larger first harmonic before the gap falls below its LLL
value.  Fits~2 and~3 provide systematic refinements of this boundary if their
additional coefficients remain stable and significantly improve the fitting
error.

\section{Physical meaning and practical limit of large $M_2$}
\label{sec:M2_physical_limit}

As shown in Supplemental Sec.~\ref{sec:M2_no_upper_bound}, within the ideal
projected single-band theory, $M_2$ can grow without bound as the
reciprocal lattice cutoff increases.  This does not mean that the FCI gap in a real
material can grow without bound.  To see this, define the
normalized lattice-modulation factor
$f(\mathbf r)=|\tilde h(\mathbf r)|^2$, with
$\langle f\rangle_{\rm uc}=1$.  Since
$M_2=\langle f^2\rangle_{\rm uc}$, $M_2=1$ for a uniform modulation, while a
larger $M_2$ means that the modulation weight is concentrated into a smaller
part of the unit cell.  As discussed in Supplemental
Sec.~\ref{sec:density_modulation_approx}, for nearly uniform Berry curvature
and approximately uniform momentum occupations for the FCI state, the physical density is approximately
$\rho(\mathbf r)\simeq(\nu/A_{\rm uc})f(\mathbf r)$.  Thus the physical density
is concentrated in the same regions as the modulation.  Equivalently, the effective area occupied by the
modulation is
$A_{\rm eff}=(\int_{\rm uc}f)^2/\int_{\rm uc}f^2=A_{\rm uc}/M_2$.  The Gaussian-peak $|\tilde h(\mathbf r)|^2$ construction in
Supplemental Sec.~\ref{sec:gaussian_wG} illustrates this picture directly:
narrowing the modulation peaks in real space concentrates the modulation weight into a progressively smaller
area and thereby increases $M_2$.

In a real material, however, the modulation cannot vary on arbitrarily short
length scales: its ultraviolet cutoff is set by atomic scales. Only reciprocal vectors below the corresponding atomic-scale
cutoff can contribute, so the attainable $M_2$ remains finite. Also, we need to consider the remote bands mixing effects. In particular,
let $E_0$ denote the characteristic interaction energy before the $M_2$
enhancement, such as the dominant pseudopotential, and let
$\Delta_{\rm remote}$ be the separation to other bands.  The fixed-band
projection requires $M_2E_0\ll\Delta_{\rm remote}$.  

Thus the formal unbounded $M_2$ result should be viewed as a design principle: concentrating
the intra-unit-cell modulation and shifting Fourier weight away from the lowest
harmonics can enhance the FCI gap while preserving an LLL-like spectrum.
In a real material, however, the enhancement is expected to saturate
once the isolated-band approximation is no longer valid.

\section{Explicit construction of a $|\tilde{h}(\mathbf{r})|^2$ with large $M_2$ and corresponding gap enhancement}
\label{sec:gaussian_wG}

In Sec.~\ref{sec:M2_no_upper_bound}, we showed analytically that $M_2$ does not have any upper bound. However, there we did not provide any concrete example of $|\tilde{h}(\mathbf{r})|^2$ along with corresponding ED results showing that even for very large $M_2$, $\Delta/\Delta_\text{LLL}$ can be approximately $M_2$. To give a concrete example (and as a corollary to show how localization of lattice modulation enhances many-body gaps), here we construct a simple positive modulation profile whose localization can be
tuned continuously.  The identity
\begin{equation}
    M_2
    =
    \sum_{\mathbf G}|\tilde w_{\mathbf G}|^2
    =
    \left\langle |\tilde h(\mathbf r)|^4\right\rangle_{\rm uc}
\end{equation}
shows that $M_2$ measures the mean-square variation of the normalized
modulation $|\tilde h(\mathbf r)|^2$.  Therefore, concentrating a fixed total modulation weight
into a smaller part of the unit cell increases $M_2$.  A natural tunable
realization of this idea is a periodic array of Gaussian peaks,
\begin{equation}
    |\tilde h(\mathbf r)|^2
    =
    \frac{A_{\rm uc}}{4}
    \sum_{\alpha=1}^{4}
    \sum_{\mathbf R}
    g_\sigma(\mathbf r-\boldsymbol\tau_\alpha-\mathbf R),
    \qquad
    g_\sigma(\mathbf r)
    =
    \frac{1}{2\pi\sigma^2}
    e^{-|\mathbf r|^2/(2\sigma^2)},
    \label{eq:gaussian_density_model}
\end{equation}
with equal-weight peaks at the four fractional coordinates inside the unit cell
\begin{equation}
    \boldsymbol\tau_\alpha
    =
    (0,0),\quad
    \left(\frac12,0\right),\quad
    \left(0,\frac12\right),\quad
    \left(\frac12,\frac12\right).
\end{equation}
The prefactor is chosen such that
$\langle|\tilde h|^2\rangle_{\rm uc}=1$, and hence $w_{\mathbf 0}=1$. The corresponding Fourier coefficients are
\begin{equation}
    \tilde w_{\mathbf G}
    =
    e^{-\sigma^2|\mathbf G|^2/2}
    S_{\mathbf G},
    \qquad
    S_{\mathbf G}
    =
    \frac14\sum_{\alpha=1}^{4}
    e^{-i\mathbf G\cdot\boldsymbol\tau_\alpha}.
    \label{eq:gaussian_wG}
\end{equation}
For a reciproal lattice vector $\mathbf G_{mn}=m\mathbf b_1+n\mathbf b_2$, the four-site motif gives
\begin{equation}
    S_{mn}
    =
    \frac14
    \left[1+(-1)^m\right]
    \left[1+(-1)^n\right],
    \label{eq:gaussian_structure_factor}
\end{equation}
so only the reciprocal vectors with both $m$ and $n$ even are retained. {\it This expression also clarifies why We chose four Gaussian peaks per unit cell with centers at $\boldsymbol{\tau}_\alpha$ in Eq.~\eqref{eq:gaussian_wG}: this choice automatically makes sure that first and second harmonic $\tilde w_\mathbf{G} = 0$, the lowest harmonic (after the zeroth) with nonzero weight is the third harmonic.}

The Gaussian peak width $\sigma$ then controls the extent of the surviving
reciprocal-space harmonics.  Decreasing $\sigma$ sharpens the real-space peaks
and broadens the Gaussian envelope in reciprocal space, allowing progressively
larger $|\mathbf G|$ components to contribute.  Consequently,
\begin{equation}
    M_2(\sigma)
    =
    \sum_{\mathbf G}
    e^{-\sigma^2|\mathbf G|^2}
    |S_{\mathbf G}|^2.
\end{equation}
For the four-peak motif in Eq.~\eqref{eq:gaussian_structure_factor},
$S_{\mathbf G}=1$ only when both reciprocal-lattice indices are even and
$S_{\mathbf G}=0$ otherwise.  Thus the surviving reciprocal vectors form a sublattice
with one quarter of the density of the original reciprocal lattice.  In the
$\sigma\rightarrow0$ limit, where many reciprocal vectors contribute, the sum
can be approximated by an integral,
\begin{equation}
    M_2(\sigma)\simeq
    \frac{A_{\rm uc}}{4(2\pi)^2}
    \int d^2\mathbf G\,e^{-\sigma^2G^2}=
    \frac{A_{\rm uc}}{16\pi\sigma^2},
\label{eq:gaussian_M2_asymptotic}
\end{equation}
Therefore $M_2\propto\sigma^{-2}$ and formally diverges as $\sigma\rightarrow0$ within
the projected single-band construction. Thus, within an isolated projected single-band
description, this construction provides a direct realization of arbitrarily
large $M_2$ by concentrating the real-space modulation weight into increasingly narrow
regions.

\begin{figure}[tbh]
    \centering
    \includegraphics[width=\textwidth]{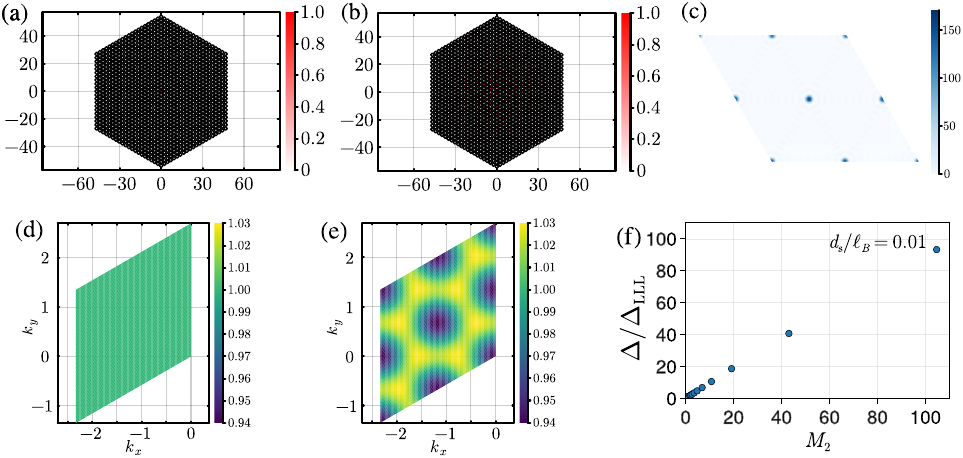}
    \caption{
    \textbf{The Gaussian-peak $|\tilde h(\mathbf r)|^2$ construction for enhancing $M_2$.}
    (a,b) Reciprocal-space Fourier coefficients $\tilde w_{\mathbf G}$ for
    (a) the LLL and (b) the Gaussian peak model with $\sigma=0.0125$.
    The LLL contains only the uniform component, whereas decreasing the
    Gaussian peak width generates substantial weight over increasingly many
    reciprocal-lattice shells.
    (c) Real-space modulation profile $|\tilde h(\mathbf r)|^2$ for the
    $\sigma=0.0125$ Gaussian peak model, showing the strong concentration of the
    modulation weight around the chosen Gaussian peak centers.
    (d,e) Momentum-resolved trace of the quantum metric,
    $\mathrm{tr}\,g(\mathbf k)$, for (d) the LLL and
    (e) the $\sigma=0.0125$ Gaussian peak model.
    (f) Exact-diagonalization gap enhancement ratio
    $\Delta/\Delta_{\mathrm{LLL}}$ versus the theoretical $M_2$ for Gaussian peak
    models with different widths $\sigma$, at $d_s/\ell_B=0.01$.
    The data points lie only slightly below the $y=x$ line, confirming the
    theoretical prediction.
    }
    \label{fig:gaussian_wG}
\end{figure}

Figure~\ref{fig:gaussian_wG} illustrates this construction, the resulting quantum geometry,
and the enhancement of the many-body gap. Exact diagonalization in the
short-screening regime shows that the resulting gap enhancement follows the
increase of $M_2$, providing a direct numerical test of the $M_2$ scaling.

\FloatBarrier

\section{Finite-size analysis of the spectral scaling}
\label{sec:finite_size_scaling}

We now examine whether the spectral scaling observed in the main text is
stable as the exact-diagonalization cluster is enlarged.  Here $N_k$ is the
number of unit cells, equivalently the number of single-particle momentum
orbitals in the projected band, and $N_e$ is the number of electrons.  Thus
$N_e/N_k$ fixes the filling fraction, while $k$ labels the many-body momentum
sector.  The three comparisons below use the same interactions, energy
definitions, and band parameters as the corresponding panels of the main
text.

\begin{figure}[tbh]
    \centering
    \includegraphics[width=0.98\textwidth]
    {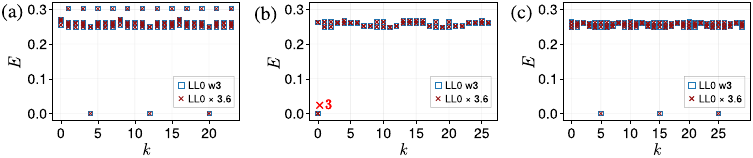}
    \caption{
    \textbf{Finite-size spectra at $\nu=1/3$ for the two-body interaction.}
    Panels (a)--(c) show clusters with
    $(N_k,N_e)=(24,8)$, $(27,9)$, and $(30,10)$, respectively.  Blue squares
    denote the w3 ideal-band spectrum and red crosses denote the LL0 spectrum
    rescaled by $3.6\approx M_2$.  As in Fig.~1 of the main text, the plotted energy is
    $E=(E_n-E_0)/[V_{\mathrm{int}}(d_s/\ell_B)^3]$, where
    $V_{\mathrm{int}}=e^2/(\epsilon\ell_B)$ and $d_s/\ell_B=0.01$.  The close
    spectral agreement, including the charge-neutral gap, persists across all
    three cluster sizes.
    }
    \label{fig:finite_size_nu13_two_body}
\end{figure}

\begin{figure}[tbh]
    \centering
    \includegraphics[width=0.98\textwidth]
    {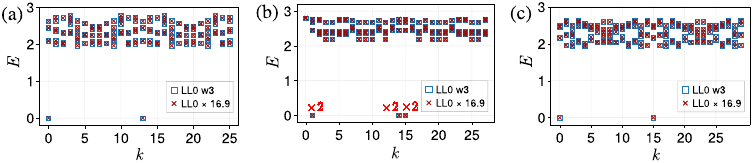}
    \caption{
    \textbf{Finite-size spectra at $\nu=1/2$ for the three-body Moore--Read
    interaction.}
    Panels (a)--(c) show clusters with
    $(N_k,N_e)=(26,13)$, $(28,14)$, and $(30,15)$, respectively.  Blue squares
    denote the w3 ideal-band spectrum and red crosses denote the LL0 spectrum
    rescaled by $16.9\approx M_3$.  The interaction, energy $E_{\mathrm{3\,body}}$,
    and energy zero are defined as in Fig.~3(a) of the main text.  The
    near-pointwise agreement of the spectra at each size shows that the
    three-body moment scaling is robust against changes of cluster size.
    }
    \label{fig:finite_size_nu12_three_body}
\end{figure}

\begin{figure}[tbh]
    \centering
    \includegraphics[width=0.98\textwidth]
    {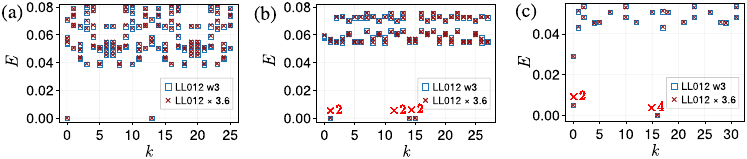}
    \caption{
    \textbf{Finite-size spectra at $\nu=1/2$ for the two-body interaction in
    the LL012-hybridized band.}
    Panels (a)--(c) show clusters with
    $(N_k,N_e)=(26,13)$, $(28,14)$, and $(32,16)$, respectively.  Blue squares
    denote the lattice-modulated LL012 w3 spectrum and red crosses denote the
    unmodulated LL012 spectrum rescaled by $3.6 \approx M_2$.  The interaction, energy
    $E_{\mathrm{2\,body}}$, and energy zero are defined as in Fig.~3(b) of the
    main text.  Although this three-LL-hybridized problem has appreciably
    larger finite-size effects, the largest cluster, $(N_k,N_e)=(32,16)$,
    retains a nonzero gap above the expected six-state Moore--Read ground-state
    manifold.  The modulated and rescaled unmodulated spectra nevertheless
    remain closely aligned.
    }
    \label{fig:finite_size_nu12_two_body}
\end{figure}

\FloatBarrier

Taken together, Figs.~\ref{fig:finite_size_nu13_two_body}--
\ref{fig:finite_size_nu12_two_body} show that the agreement is not specific to
one finite cluster.  Absolute gaps and the distribution of levels among
momentum sectors can display finite-size dependence, especially in the
three-LL-hybridized Moore--Read problem, but the overall spectral
renormalization remains highly accurate.  In particular, the two-body spectra
follow the $M_2$ scaling law, while the three-body parent Hamiltonian follows
its $M_3$ counterpart.  This size stability supports extending the moment
scaling of the excitation spectrum and gap toward the thermodynamic limit.

\end{document}